\documentclass[printer]{aa}  

\usepackage{graphicx}
\usepackage{txfonts}
\usepackage[colorlinks=false]{hyperref}
\usepackage{xcolor}
\usepackage{rotating}
\usepackage{pdflscape}
\usepackage[sort&compress]{natbib}
\bibpunct{(}{)}{;}{a}{}{,}

\newcommand{\um}{$\mu$m}

\newcommand{\kms}{km\thinspace s$^{-1}$}

\def\utw{\smash{\rlap{\lower5pt\hbox{$\sim$}}}}
\def\udtw{\smash{\rlap{\lower6pt\hbox{$\approx$}}}}

\def\Teff{\hbox{\it T$_{\rm eff}$}}

\def\Msun{\hbox{\it M$_\odot$}}

\def\Mbol{\hbox{\it M$_{bol}$}}

\def\Teff{\hbox{\it T$_{\rm eff}$}}

\def\K{\hbox{\it K}}

\def\Mk{\hbox{\it M$_{\rm K}$}}

\newcommand{\Ks}{{\it K$_{\rm s}$}}

\newcommand{\Aks}{{\it A$_{\it K_{\rm s}}$}}

\newcommand{\Av}{{\it A$_{\it V}$}}

\def\BCK{\hbox{\it BC$_{\it K}$}}
\def\BCKs{\hbox{\it BC$_{\it K_S}$}}

\def\simgr{\mathrel{\hbox{\rlap{\hbox{\lower4pt\hbox{$\sim$}}}\hbox{$>$}}}}

\def\vlsr{\hbox{V$_{\rm LSR}$}}

\begin{document}

   \title{Identification of late-type stars of class I. }

   \subtitle{Gaia DR3 Apsis parameters.}

   \author{Maria Messineo
          \inst{1}
           }

   \institute{ASTROMAGIC freelancer, Gartenstrasse 7,  14482 Potsdam, Germany\\
              \email{maria.messineo@protonmail.com}
             }

   \date{Received November 30, 2022; accepted January 13, 2023}

 
  \abstract
   {} 
  {Gaia DR3  GSP-Phot and GSP-Spec parameters of known 
  K- and M-type stars with luminosity class I 
  are examined and compared with  parameters collected from the literature,
  to assess their accuracy and their potential for stellar classification
  of unknown intrinsically bright late-types. 
  Gaia DR3  GSP-Phot and GSP-Spec parameters
  were generated by the Astrophysical Parameters Inference Software (Apsis).
  }
  {In the Gaia DR3 catalog, 
  there are about 40,000  entries with Apsis parameters similar 
  to those of known  red supergiants, RSGs, good parallaxes, 
  and infrared  2MASS and WISE data. 
  By using parallactic distances, 
  infrared photometry, and variability information, 
  only 203 new entries are found with luminosities and temperatures
  similar to that of known RSGs 
  and $G$-band amplitudes smaller than 0.5 mag.
  Their low-resolution BP/RP spectra are compared with an empirically made
  spectral library of BP/RP spectra of known bright late-type stars
  (C-rich, S-type, O-rich asymptotic giant branch stars (AGBs),  and RSGs) 
  to obtain their spectral types.} 
  {
  Among them, 15 S-type stars  are identified  by peculiar absorption features 
  due to ZrO and LaO visible in their BP/RP spectra, one S/C star, and
  nine C-rich stars by their  strong CN absorption bands.
  K- and M-types can be reproduced with an accuracy of two subtypes.
  20 new RSGs are confirmed, of which six  have bolometric 
  magnitudes brighter than those
  of the AGB limit: 2MASS J21015501+4517205, 2MASS J16291280$-$4956384,
  2MASS J10192621$-$5818105, 2MASS J20230860+3651450,
  2MASS J17084131$-$4026595, and 2MASS J16490055$-$4217328.
  The  flag for C-rich stars of the Gaia DR3 LPV pipeline
  is erroneously true for some  RSGs  and a visual inspection
  of the BP/RP spectra is mandatory. 
  }
  {}

   \keywords{stars: evolution --  
             stars: late-type --
          stars: supergiants -- 
          stars: massive --
          Galaxy: stellar content --
          technique: spectroscopic
   }

   \maketitle
%

\section{Introduction}

The term ``late-type stars'' indicate bright evolved stars 
cooler than $\approx 4500$ K. They divide into red supergiants (RSG)  and 
Asymptotic Giant Branch stars (AGB).
RSGs have masses from  $9$ to $40$ \Msun; 
they do not have a degenerated state in their core and  are burning He.
AGB stars have  masses $< 9$ \Msun,
and are characterized by two burning shells, 
an inner He shell and an outer H shell; 
the  core of CO is in an electron-degenerate state.
AGB stars experience complex interactions between 
the two burning shells and matter is dredge-up to the surface, 
thereby, changing the stellar chemistry.
AGBs  with masses from 1 to 5 \Msun\ evolve from an
O-rich atmosphere ([C/O] $< 1$) to a C-rich one ([C/O]$ < 1$).
S-type stars are  in a hybrid state, transiting toward a 
C-rich chemistry. Late-type stars are often also named with terms 
indicating their variability type, for example, Mira, semi-regular, irregular, 
long period variables (LPV), and large amplitude variables (LAV), 
or their envelope types (e.g., OH/IR stars, SiO masing stars).

In the inner Galaxy, uncertain distances and dust obscuration 
hamper the detection of RSGs. 
RSGs have ages from 8 to 30 Myr. 
AGBs have ages from 30 Myr to 12 Gyr, but
properties are similar to those of RSGs.
Currently,   about  600 Galactic stars
are probable  RSGs\footnote{Using 
the compilation of \citet[][]{skiff16} and \citet{messineo19},
203  RSGs mentioned in literature 
as possible members of stellar clusters and associations (independently of parallaxes)
were counted, 32 obscured RSGs in the proximity of the Galactic center, and 
other 394 isolated stars  with good Gaia DR3 parallaxes and 
luminosity in areas A and B.},
and to enlarge this number a careful analysis of 
dissimilarities between AGBs and RSGs is required. 
This is  an imperative requirement  to understand the history 
and morphology of the Milky Way, as
in a barred gravitational potential,
their spatial distributions are predicted to be 
different. 
It is misleading to talk about  initial mass functions of 
late-type stars and spatial distributions, 
without being able to distinguish RSGs from AGBs.

For a compilation of about 1,400 stars listed in the literature as 
class I K-type and M-type stars (O-rich stars) \citep{messineo19}, 
889 have good Gaia DR2 parallaxes and 966 in EDR3. 
With estimates of luminosity and effective temperature, \Teff,
the authors found a tail of 110 stars 
brighter than the AGB luminosity limit (\Mbol=$-7.1$ mag). 
Furthermore, at least 49\% of the stars in the sample are brighter 
than \Mbol=$-5.0$ mag and earlier than M4, 
which means they are more massive than 7 \Msun\ \citep{messineo19}.

The new DR3 release of 
Gaia\footnote{https://www.cosmos.esa.int/web/gaia/dr3}  
includes the first release of data from the two spectrographs on board
and variability data.
The new data may allow us to refine the Galactic catalog of RSGs.
The periods, Per, of RSGs range from 100  d to 1,000  d, 
similarly to Mira AGB stars, but Miras have larger amplitudes 
 \citep[Ampl $\ga 0.8$ in G-band, e.g.,][]{lebzelter22,messineo22}.

The  Gaia DR3 release (1.5 billion sources)  delivers  \Teff\ values, 
spectral types, and gravity of 470 million stars with 
low-resolution BP/RP spectra, which are computed with the
Astrophysical parameters inference system (Apsis) pipeline.
For 5.5 million stars, it also delivers parameters and chemical abundances
from the RVS spectra.
In principle, cool stars with [M/H] $< 0.5$ dex 
and log(g) $< 0.7$ cm s$^{-2}$ could be extracted from the 
Gaia DR3 catalog  as candidate RSGs. 
However, the parameters from Apsis are not optimized for 
AGB and RSG stars, because
circumstellar envelopes and compositional changes due to dredge-up
are not considered. 

In this work,  
parameters of known bright O-rich stars taken from the literature are compared  
with those estimated by Gaia DR3 (e.g., \Teff\ values, gravity, Ampl values),
and their BP/RP spectra are analyzed for classification purposes.
In Section \ref{sample} the sample of stars is presented.
The parameters from the Apsis 
pipeline are described in Sect. \ref{apsis}, and
the parameters from the Gaia variable catalog in Sect. \ref{variable}.
In Section \ref{classification}, molecular signatures to distinguish
O-rich stars from C-rich and S-type stars visible in the BP/RP spectra
are described.
In Section \ref{detection}, the Apsis parameters are used to extract 
a new sample of bright late-type stars, and their low-resolution BP/RP spectra
are analyzed to identify C-rich,  S-type,  O-rich AGBs, and RSGs.
Eventually, a summary is given in Sect. \ref{summary}.

\section{The Galactic sample of class I K-M stars} \label{sample}
The sample collected by \citet{messineo19} contains
1,406 stars which were reported at least once in the 
literature as K-M class I  stars.  
Six stars previously included were 
discarded\footnote{M. Messineo and A. G. A. Brown 2021, Zenodo technical note
$<$doi:10.5281/zenodo.4964818$>$.} and
121 new entries were added
from the updated catalog of \citet{skiff16} 
(version of 2020 in  the Centre de Données astronomiques de Strasbourg, CDS).
The number of sampled stars is 1,521.
The catalog is expected to be made of evolved bright late-type stars
(RSGs, O-rich AGBs, and red giant stars).

Among the stars, only 1,060 sources have good  
 Gaia DR3 parallaxes, $\varpi$, values,  where good means  
$\frac{\varpi}{\sigma_\varpi (est)}$ larger than 4. 
The external parallactic errors, 
$\sigma_\varpi$(est), are
calculated with the formula by \citet{mais21},  
which is a function of the $G$-band  magnitude and of the 
normalized unit weight error (RUWE).
The  $\varpi$ values  from the EDR3 to DR3 release are unchanged.
The Gaia EDR3 distances from \citet{bailerjones21} are here assumed.

The compiled infrared catalog  
(with  the Two Micron All Sky Survey (2MASS), 
 Deep near-infrared survey of the Southern Sky (DENIS), 
 the Midcourse Space Experiment (MSX), 
 the Wide-field Infrared Survey Explorer (WISE), 
 Galactic Legacy Infrared Midplane Survey Extraordinaire (GLIMPSE), 
and MIPSGAL measurements), the spectral types, 
the estimates of interstellar extinction, 
and bolometric magnitudes are  as described in  
\citet[][and references therein]{messineo19}.
The authors inferred the stellar \Teff\ values by assuming
the temperature scale by \citet{levesque05},
and the \Mbol\ values using the dereddened  \Ks\ magnitudes
and the bolometric corrections \BCKs\ of \citet{levesque05},
as well as by estimating the total flux enclosed under 
the infrared stellar energy distribution.
The interstellar extinction in \Ks-band, \Aks, 
is calculated with the 2MASS $JHK$ magnitudes 
and the intrinsic colors for RSGs
by \citet{koornneef83}. 
The  stars in areas A \& B of
\citet[][]{messineo19} with BP/RP spectra have   \Aks\ values
lower than  0.72 mag or \Av=7.82 mag.
The selective extinction ratios used are those given in \citet{messineo05}
for a power law with an index of $-1.9$.

In the following sections, late-type stars with different 
luminosity and \Teff\
ranges are mentioned as belonging to `areas' of the luminosity versus 
\Teff\ diagram, 
following the definition of  
\citet{messineo19}\footnote{A=Stars with \Mbol\ above the AGB limit (\Mbol$\ga-$7.1 mag);
B=Stars with $-5<$\Mbol$<-7.1$ mag earlier than M4 (  \Teff\ $ \ga 3548$ K);
C=Stars with $-5<$\Mbol$<-7.1$ mag later than an M4 (  \Teff\ $ \la 3548$ K);
D=Stars with $-3.6<$\Mbol$<-5$ mag along the observed distribution of  massive giants and
E=Stars with $-3.6<$\Mbol$<-5$ mag redder than that;
F=Stars with \Mbol$>-3.6 $ mag.}.
Typical Galactic RSGs are found to  populate areas A and B
(stars brighter than \Mbol$<-5$ mag and with 
spectral type not later than M4).

\section{Apsis parameters of class I K-M stars}
\label{apsis}

The Gaia DR3 release presents two sets of spectroscopic
parameters, those from the GSP-Phot module (low-resolution spectra)
and those from the GSP-Spec module (high-resolution spectra).

\subsection{GSP-Phot}
\label{GSP-Phot}

The GSP-Phot module releases  \Teff, gravity, and metallicities
from the low-resolution BP/RP spectra for 283 (out of 1,060) stars included
in the sample.
This number decreases to 159 when considering only the 713
stars with $\frac{\varpi}{\sigma_\varpi{\rm (est)}} > 4$ and 
luminosity estimates above the tip of the red giant
branch \citep[i.e., located in areas A, B, C, D, E of 
the color-magnitude diagram, CMD,][]{messineo19}.

For cool evolved stars, the Apsis GSP-Phot module uses 
models from the PHOENIX or MARCS libraries to find the 
best low-resolution matching spectrum and to obtain  
\Teff, gravity (log(g)), and total metallicity ([M/H]) 
\citep[][and references therein]{fouesneau22}.
The fit is based on the shape of the average spectrum and
stars are all assumed to be nonvariable. 
However, average spectra-based analysis of Mira AGB stars that often 
have optical brightness variations larger than 8 mag seems like 
a gross simplification. Furthermore, the fit strongly depends
on the treatment of interstellar extinction, 
which is self-derived from the Gaia data.
On the contrary, G-band variations of RSGs are typically smaller 
than 0.5 mag.

The  goodness-of-fit score for the adopted library spectrum 
(logposterior\_gspphot) and  the goodness-of-fit score 
for models with  two unresolved components  (logposterior\_msc)  
have low values. A lower value indicates a lower-quality fit.
The logposterior\_gspphot values range from $-31,895$ to $32,502$. 
The logposterior\_msc values range from $-1.09e7$   to   $-85,378$, while
\citet{fouesneau22} suggest to use values $> -1,000$. 
In conclusion,  due to large variability and quality of the fit, 
the Gaia DR3 Apsis pipeline performance for the BP/RP spectra
of bright late-type stars is  poorer than for the bulk
of low luminosity cool  stars.

\subsubsection{Temperatures and extinction}

\begin{figure*}
\begin{center}
\resizebox{0.48\hsize}{!}{\includegraphics[angle=0]{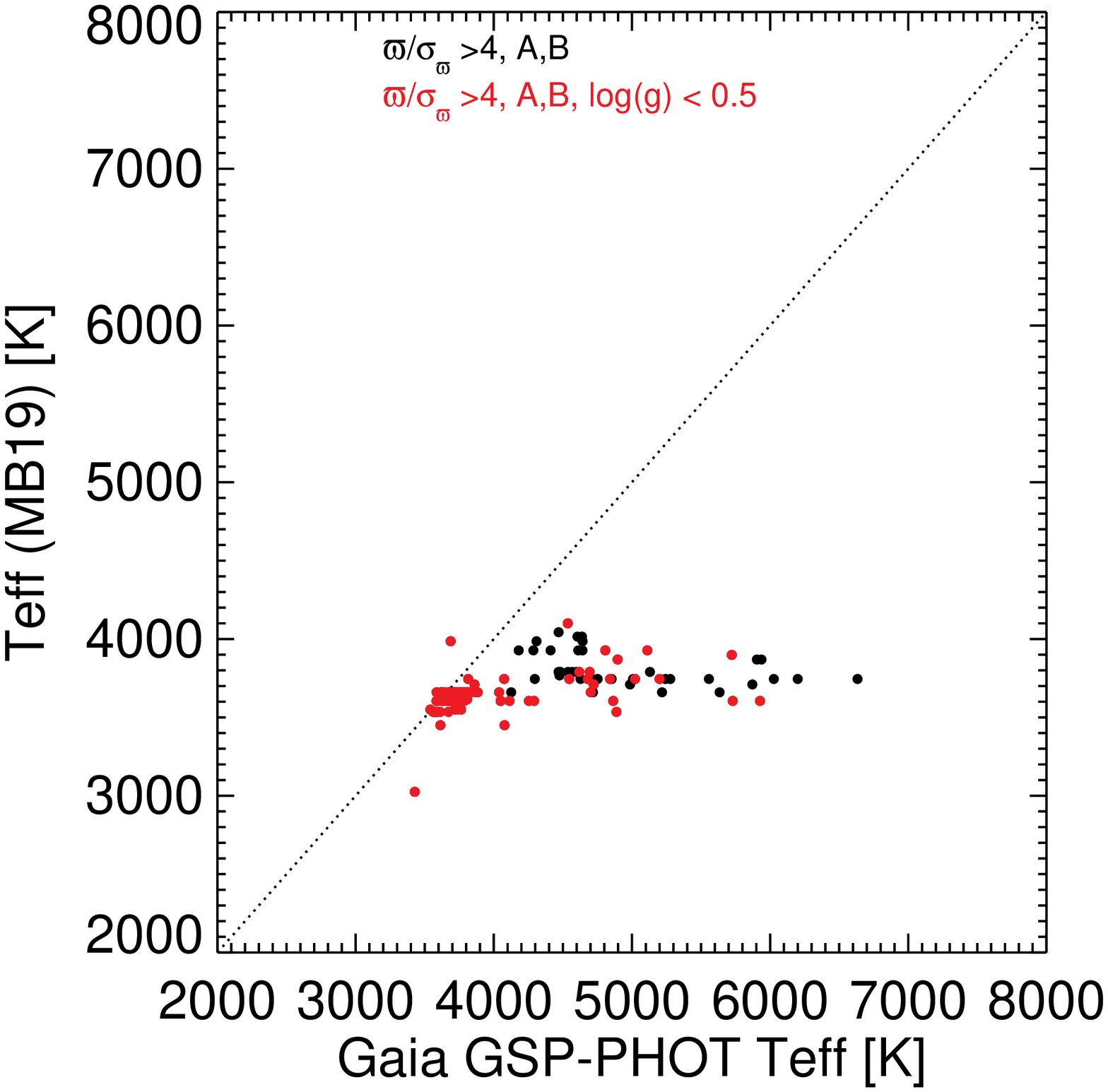}}
\resizebox{0.48\hsize}{!}{\includegraphics[angle=0]{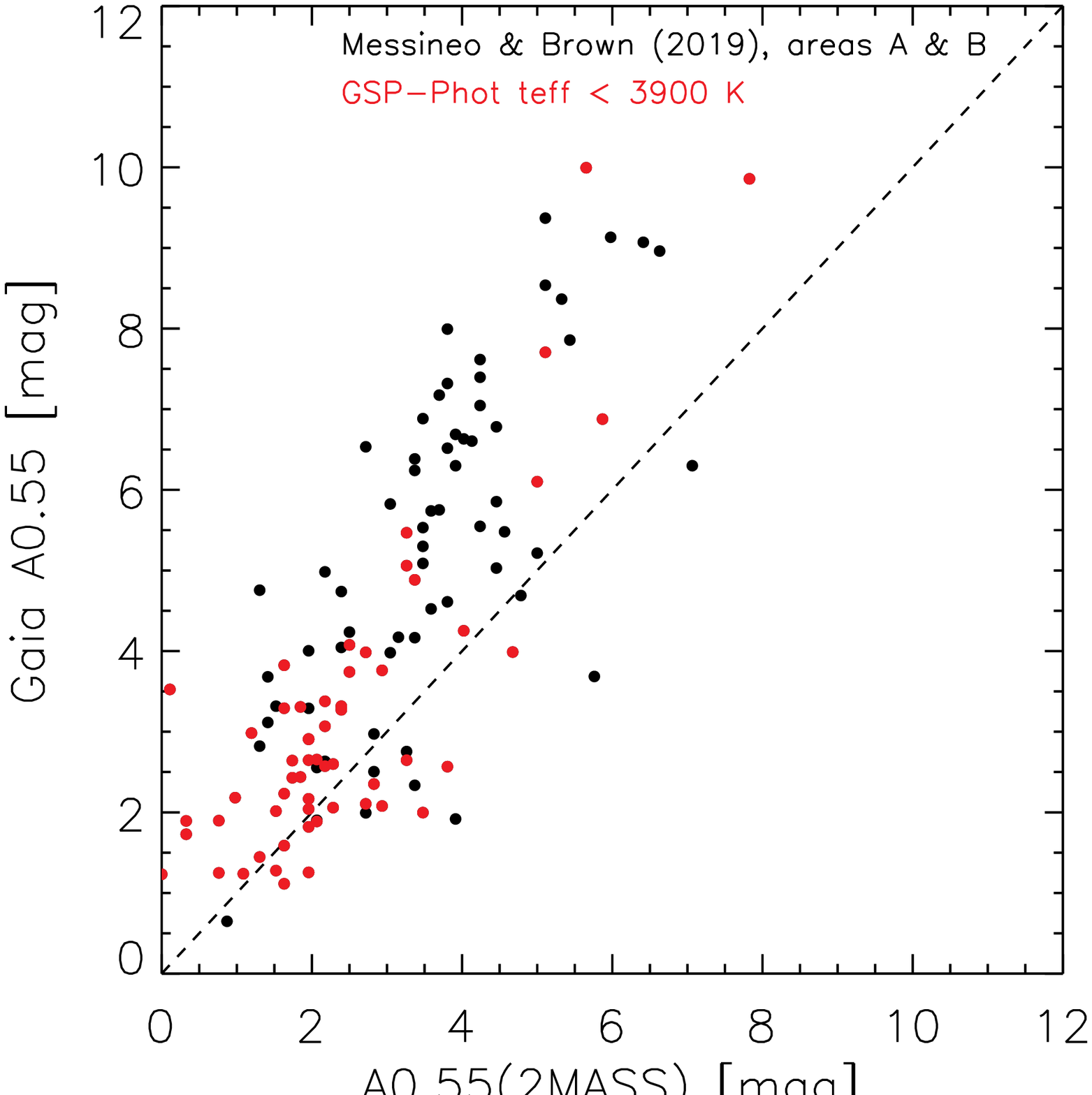}}
\end{center}
\caption{   \label{temp:comp} 
{\it Left panel:}
The  \Teff\ values adopted from literature by \citet{messineo19} vs.
the \Teff\ from the Gaia DR3 BP/RP spectra. Data points in red indicate Log(g) $< 0.5$.
{\it Right panel:} Gaia  GSP-Phot A0 (0.55 nm) values versus 
the A0 values (\Aks/0.092) inferred with 2MASS $JHK$ measurements
and the extinction law of \citet{messineo05} with an index of $-1.9$. 
}
\end{figure*}

In Figure \ref{temp:comp}, for the  159 stars
above the tip of the red giant branch, the GSP-Phot \Teff\ values  are compared 
with those adopted in \citet{messineo19} by assuming an initial class I. 
For the bulk of low gravity stars (log(g) $< 0.5$), 
there is a good agreement between the literature \Teff\ values and 
the new GSP-Phot \Teff\ with a mean delta of 105 K and a standard deviation
of 82 K (57 out of 82 objects). 
These are mostly M-type located at  extinction  below \Av =$2$ mag.
With increasing  extinction (distance) stellar \Teff\ values
may appear higher, because an overestimated extinction 
yields a warmer temperature.
Indeed, the Gaia self-derived extinction values appear higher
than those  inferred from infrared by \citet{messineo19},
roughly by a factor 1.5. 
In Figure \ref{temp:comp}, the Gaia A0 parameter
that is the line-of-sight monochromatic extinction  at 541.4 nm
inferred by the GSP-Phot pipeline \citep{andrae22} using the law of 
\citet{fitzpatrick99}, is plotted versus 
the $A_{0.55}$=\Aks/0.092 mag of \citet{messineo05}.
A different way of reading this bias is that
the bright M-type supergiants ($\sigma$(Temp)=72 K) delivered by the 
GSP-Phot module  are mostly located at low extinction, as shown 
in Fig.\ \ref{temp:comp}.

For lower-mass stars, a similar degeneracy between temperature 
and extinction is documented in Fig. 7 by \citet{andrae22}. 

\subsubsection{Gravity}

\begin{figure*}
\begin{center}
\resizebox{0.48\hsize}{!}{\includegraphics[angle=0]{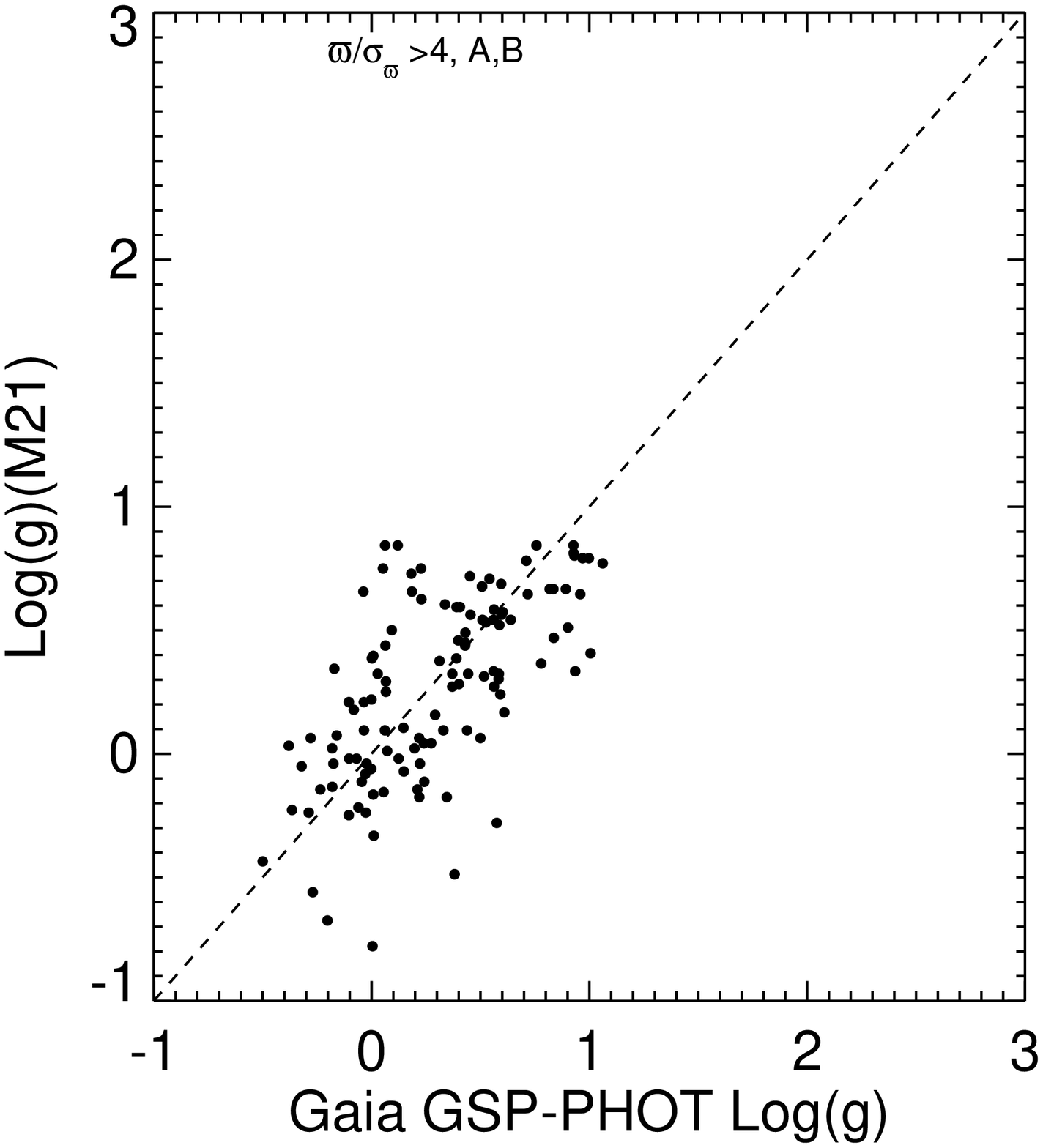}}
\resizebox{0.48\hsize}{!}{\includegraphics[angle=0]{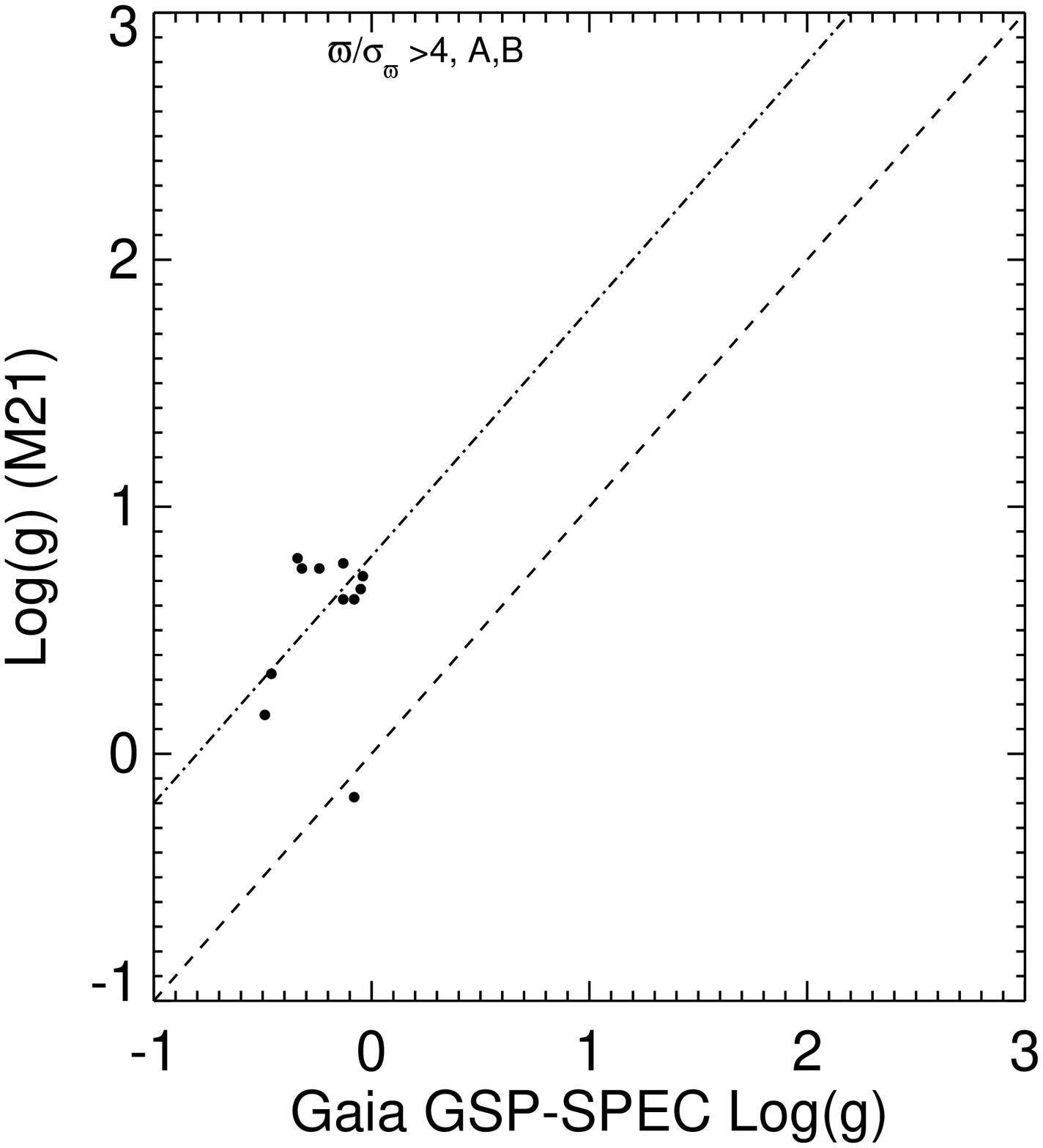}}
\end{center}
\caption{ \label{logg} {\it Left panel:} Log(g) from the luminosity  relation  
derived by \citet{messineo21} vs. GSP-Phot log(g).
{\it Right panel:} Log(g) from the luminosity  relation  
 vs. GAIA GSP-Spec log(g). Only data points with good distance
 and from areas A and B of \citet{messineo19} are plotted.
 The dashed line is the equity line, while
 the dashed-dotted line has been shifted by $-0.8$ dex.
}
\end{figure*}

In Figure \ref{logg}, the GSP-Phot log(g) values of stars with luminosities
above the tip  of the red giant branch are
compared with those estimated using the relation between log(g)
and luminosity by \citet{messineo21}.
There is a mean shift of $-0.05$ and $\sigma$=0.36.

The gravity of known RSGs ranges from $-0.7$ to 0.7 \citep{messineo21}.
A total of 114 (72\%) stars, out of the 159 sampled stars above 
the tip of the red giant branch  and with available Gaia BP/RP spectra, 
have GSP-Phot log(g)$ < 0.7$. 

\subsubsection{Metallicity}
The metallicity of  known RSGs ranges from $-0.5$ to 0.5, 
as inferred from values collected from the
literature \citep{messineo21}.
There are 148 GSP-Phot entries  with
[M/H] $> -0.5$ dex matching
 stars from \citet{messineo19} brighter than the tip 
 of the red giant branch and, 
 which constitute 93\% of the available  matches, and
this percentage remains  93\% in areas A and B.

\citet{fouesneau22} warn about systematics in the GSP-Phot [Fe/H], 
which, therefore, should only be used as indicative.
Indeed, for low gravity stars, it appears that the metallicity 
estimated with the GSP-Phot module
is poorly determined, and,
for members of open clusters, offsets up to  1 dex are measured
\citep{andrae22}. \\
For 23 bright late-type stars with  both GSP-Phot [Fe/H] values and 
[Fe/H] values from the literature, and with $\frac{\varpi}{\sigma_\varpi} >4$, 
an average difference of 0.1 dex is obtained with  $\sigma$=0.25 dex, 
with a  maximum difference of 0.68 dex (see Fig.\ \ref{Mtotspec}).

\subsection{GSP-Spec}
The Apsis GSP-Spec module performs the analysis  of a normalized 
spectrum and, therefore, the measured quantities are  independent 
of the stellar continuum. 

Parameters inferred from spectra taken with the Gaia high-resolution 
spectrograph (R=11,500) were retrieved for 336 catalog stars, 
325 of which are of good parallaxes ($\frac{\varpi}{\sigma_\varpi} >4$).
However, they are mostly giants (class F of \citet{messineo19}).
Only 72 stars  are brighter than the tip of the  red giant branch.
The retrieved entries were selected to have  the first nine digits 
of the GSP-Spec quality flag smaller than unity, while the 10, 11, 12 
digits were required to be null \citep{recio22}. 
This means   $\Delta$\Teff$ < 500$ K, 
$\Delta$log(g)$ < 1$ dex, and $\Delta$[M/H]$ < 0.5$ dex.

\subsubsection{Temperature}
In Figure \ref{tempSPEC}, for the 72 data points brighter 
than the tip of the red giant branch, \Teff\ values listed 
in \citet{messineo19} are compared with those estimated by 
the Gaia GSP-Spec module. There is an average difference 
of 119 K with $\sigma$=124 K.

\subsubsection{Gravity}
Seventeen data points with good quality GSP-Spec log(g) are 
located in areas A and B of \citet{messineo19}, and
9 have luminosity-inferred log(g) $< 0.5$ dex; 
their average difference between their GSP-Spec log(g) 
and the luminosity-inferred log(g) is 
$-0.5$ dex with  $\sigma$=1.65 dex. 
In Figure \ref{logg}, it appears that the GSP-Spec log(g) 
values are systematically lower.

\subsubsection{Metallicity}
The GSP-Spec [M/H] value is a measure of the stellar total
metallicity (Fe and $\alpha$ elements) and the GSP-Spec [$\alpha$/Fe] 
is the corresponding $\alpha$  enhancement. On the basis of chemical 
Galactic evolution studies, it is usually assumed 
[$\alpha$/Fe] =0 for [M/H] $> 0$ dex, 
[$\alpha$/Fe] =+0.4 for [M/H] $< -1$ dex,
and [$\alpha$/Fe] =+0.4$\times$[M/H] for $ -1 <$ [M/H] $<$ 0  dex.
The grid of the synthetic spectra  was   centered
on these canonical values \citep{recio22}.

\begin{figure*}
\begin{center}
\resizebox{0.32\hsize}{!}{\includegraphics[angle=0]{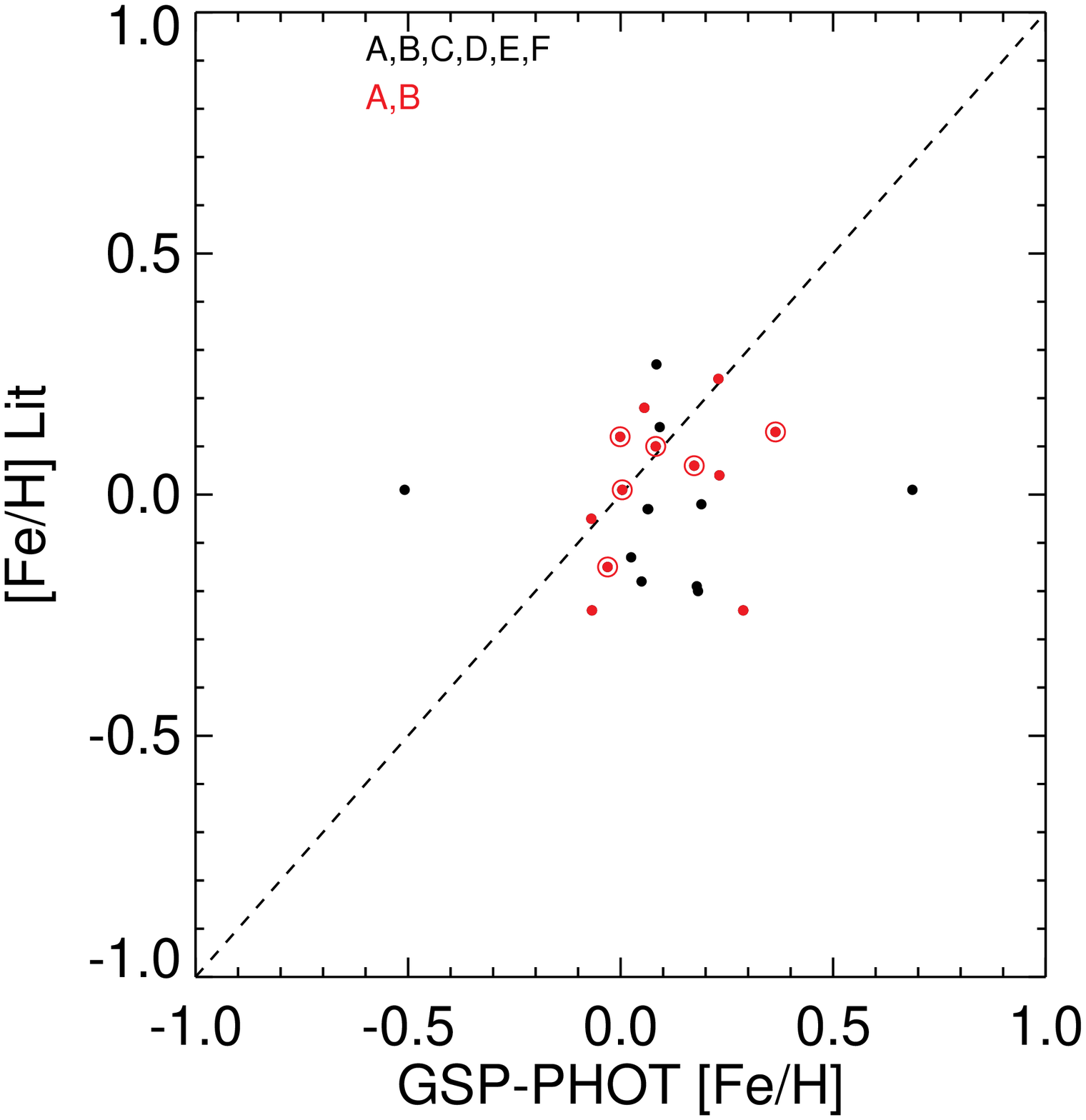}}
\resizebox{0.32\hsize}{!}{\includegraphics[angle=0]{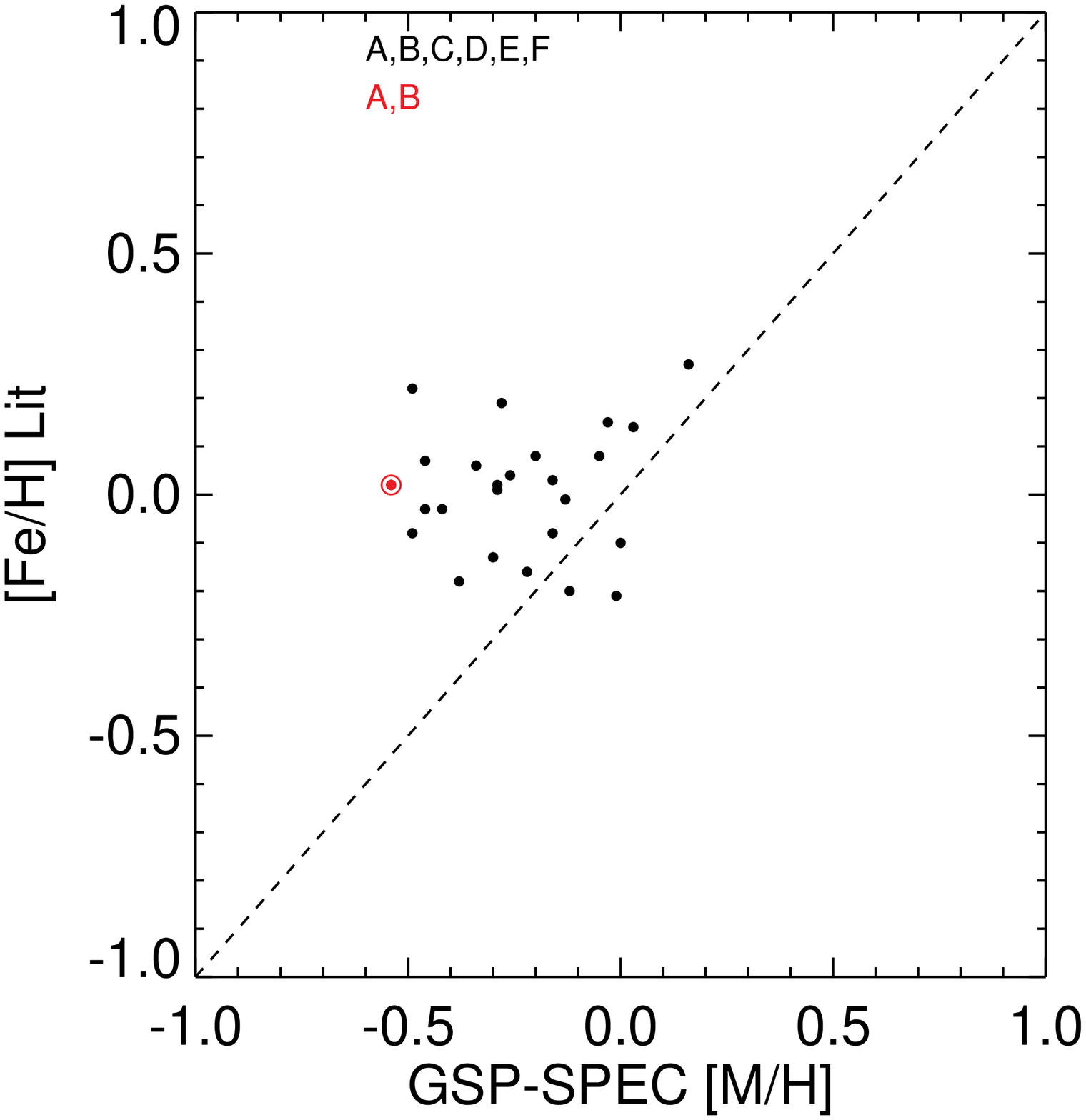}}
\resizebox{0.32\hsize}{!}{\includegraphics[angle=0]{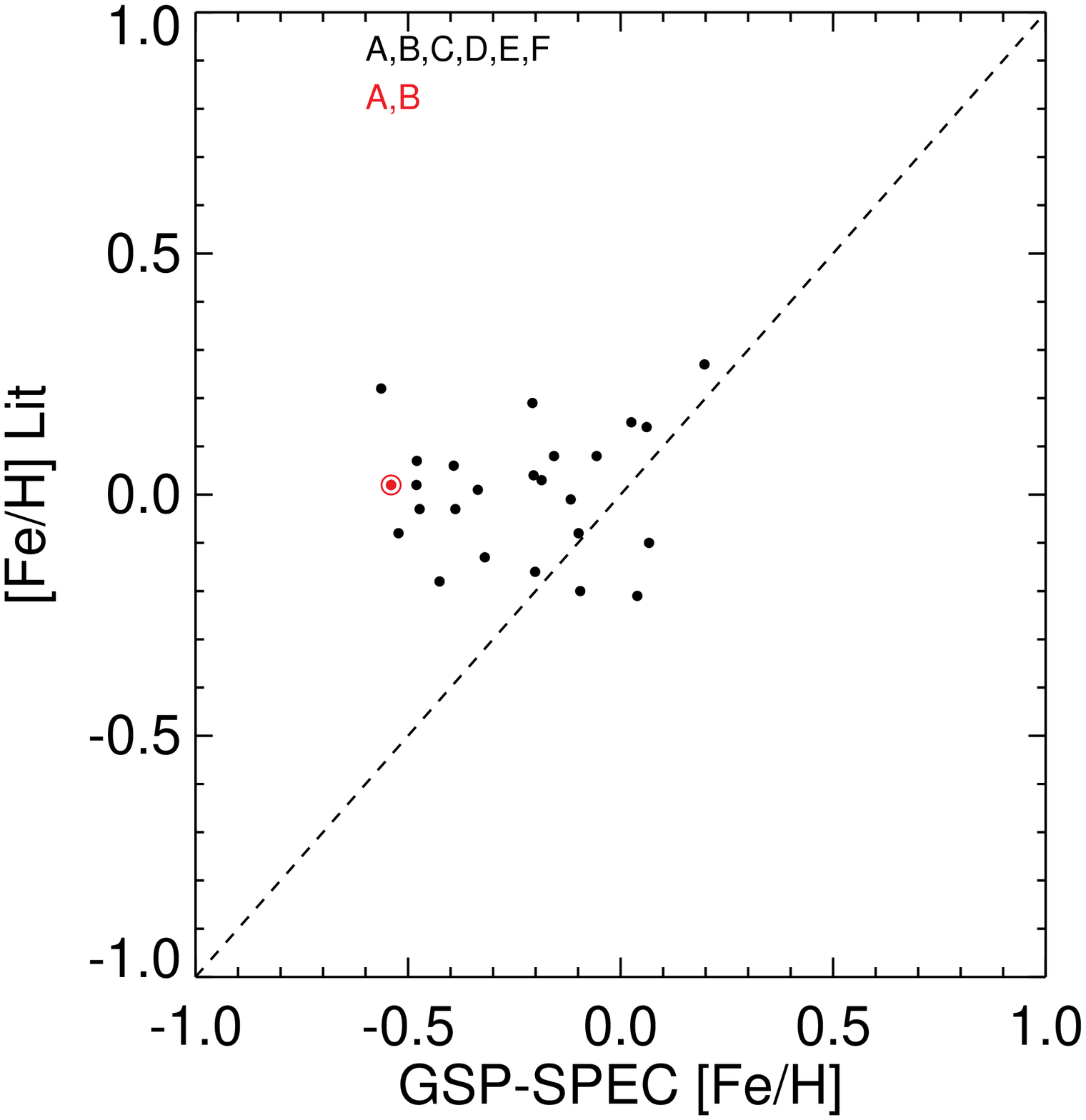}}
\end{center}
\caption{ \label{Mtotspec} 
[Fe/H] values from spectroscopic published 
works vs. the [Fe/H] values inferred by the  GSP-Phot and GSP-Spec module.
The red filled circles are data points located in the areas A and B of the 
luminosity-\Teff\ diagram of \citet{messineo19}.
Encircled  data points  are from the works of \citet{davies10} 
and \citet{gazak14}.
{\it Left panel:} [Fe/H] values from the literature vs.
those  inferred from GSP-Phot module.
{\it Middle panel:} [Fe/H] values from the literature vs.
the [M/H] values inferred from GSP-Spec.
{\it Right panel:} [Fe/H] values from the literature vs.
the [Fe/H] values inferred from GSP-Spec [M/H] and GSP-Spec [$\alpha$/Fe]
using the formula of \citet{ferraro99}.
}
\end{figure*}

\begin{figure}
\begin{center}
\resizebox{0.99\hsize}{!}{\includegraphics[angle=0]{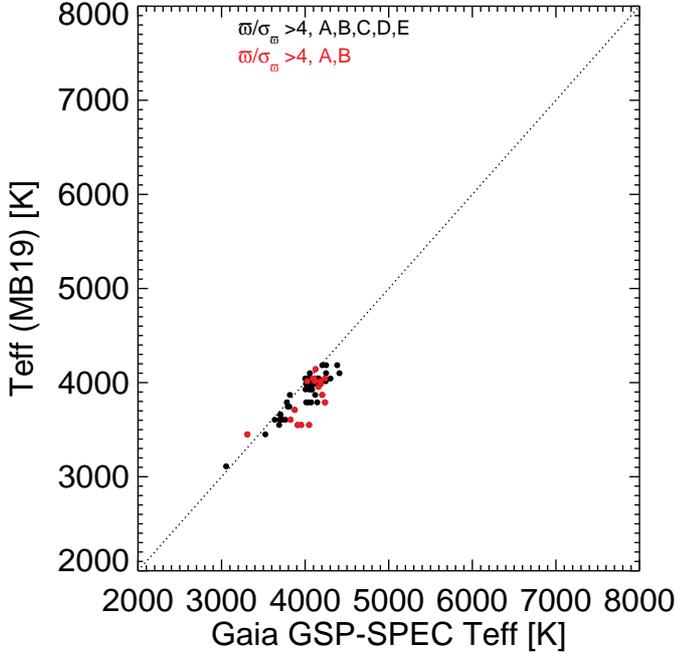}}
\end{center}
\caption{ \label{tempSPEC} 
 \Teff\ values from \citet{messineo19}
vs. the \Teff\ values inferred from the GSP-Spec module.
Only stars with luminosity above the tip of the red giant brach, 
and with  $\frac{\varpi}{\sigma_\varpi (\rm est)}$ larger than 4 are plotted.
Red points mark stars with luminosity inferred log(g) $< 0.5$ dex.  }
\end{figure}

In the sample, there are 25 data points  with available
GSP-Spec [M/H] values as well as  [Fe/H] values from the literature.
Estimates of [Fe/H] are obtained from GSP-Spec [M/H]
and GSP-Spec [$\alpha$/Fe] with the  Eq. (1) of \citet{ferraro99},
and, eventually, compared with the literature values in Fig.\ 
\ref{Mtotspec}. An average difference of $-0.24$ 
is measured and $\sigma$=0.25  dex; the maximum deviation is 0.78 dex.
From Figure \ref{Mtotspec}, however, it appears that the differences
are systematically bigger when GSP-Spec [Fe/H] decreases.
Indeed, for GSP-Spec [M/H] $> -0.2$ dex the  average difference 
of $-0.04$ dex and $\sigma$=0.16  dex; 
while for GSP-Spec [M/H] $< -0.2$ dex the  average difference of 
$-0.40$ dex and $\sigma$=0.18  dex.
It appears that the GSP-Spec module does not deliver super-solar 
metallicity for bright late-type stars.

\begin{figure*}
\begin{center}
\resizebox{0.49\hsize}{!}{\includegraphics[angle=0]{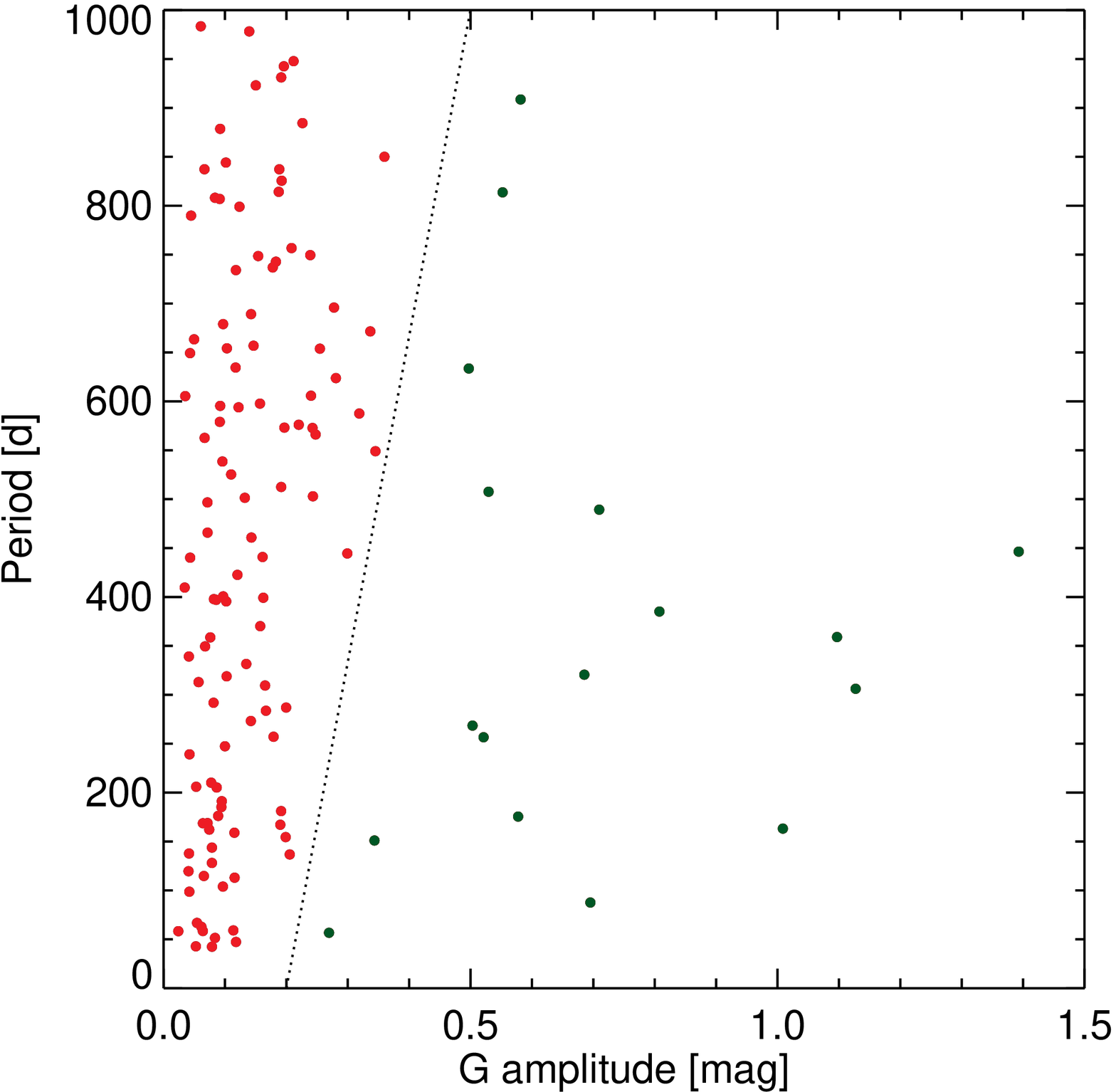}}
\resizebox{0.49\hsize}{!}{\includegraphics[angle=0]{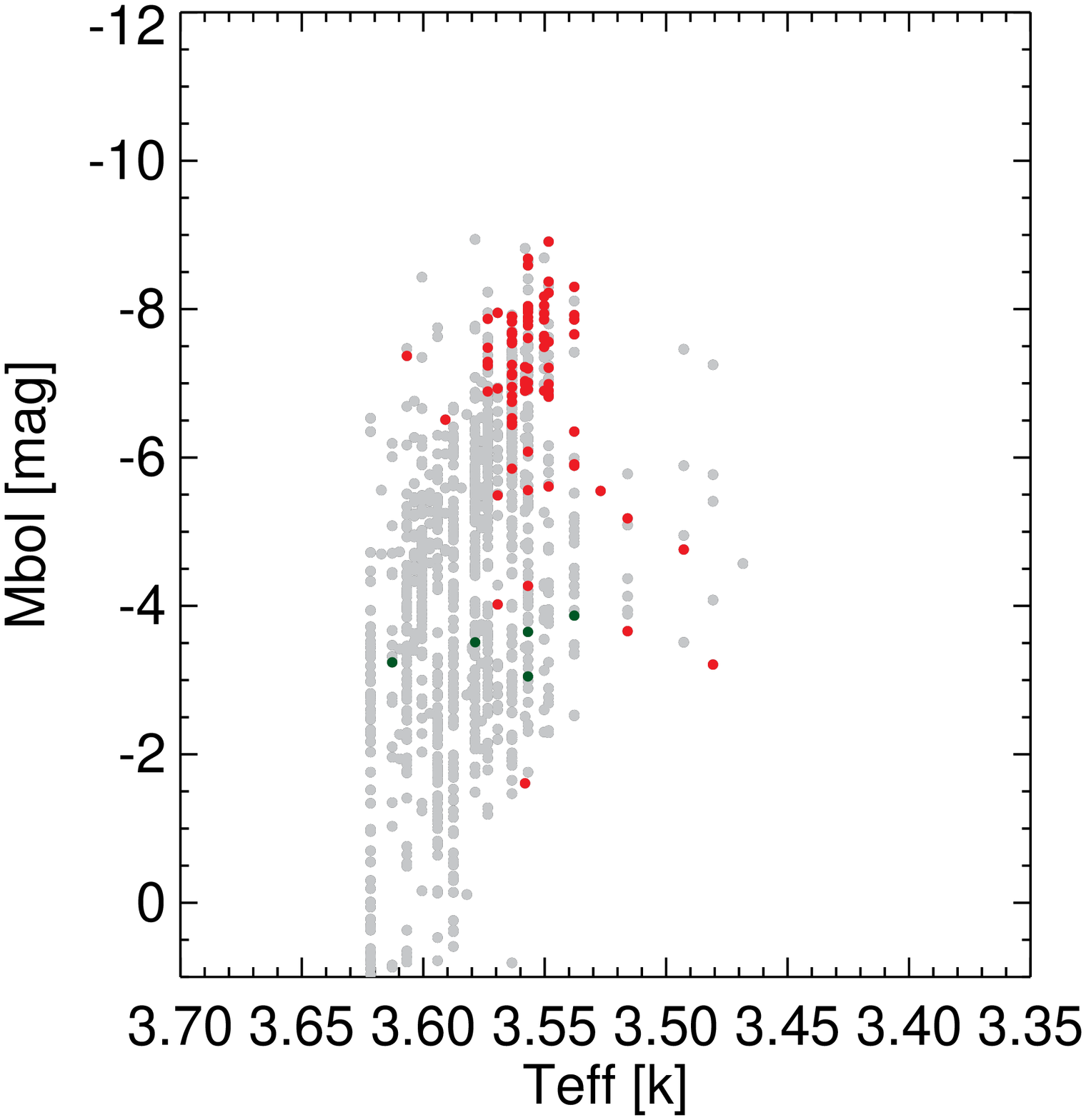}}
\end{center}
\caption{\label{per_ampl} {\it Left panel:} 
Per values  vs.  Ampl values of the Gaia variables. 
The dotted line is drawn by eye to separate the outliers (dark green) 
from the bulk of the distribution (red).
{\it Right panel:} Bolometric magnitudes vs. \Teff\ values
of  late-type stars from \citet{messineo19} are marked 
 with gray circles. Red and dark green symbols are those with measured
 Gaia Per values, as in the left panel.}
\end{figure*}

\begin{figure*}
\begin{center}
\resizebox{0.49\hsize}{!}{\includegraphics[angle=0]{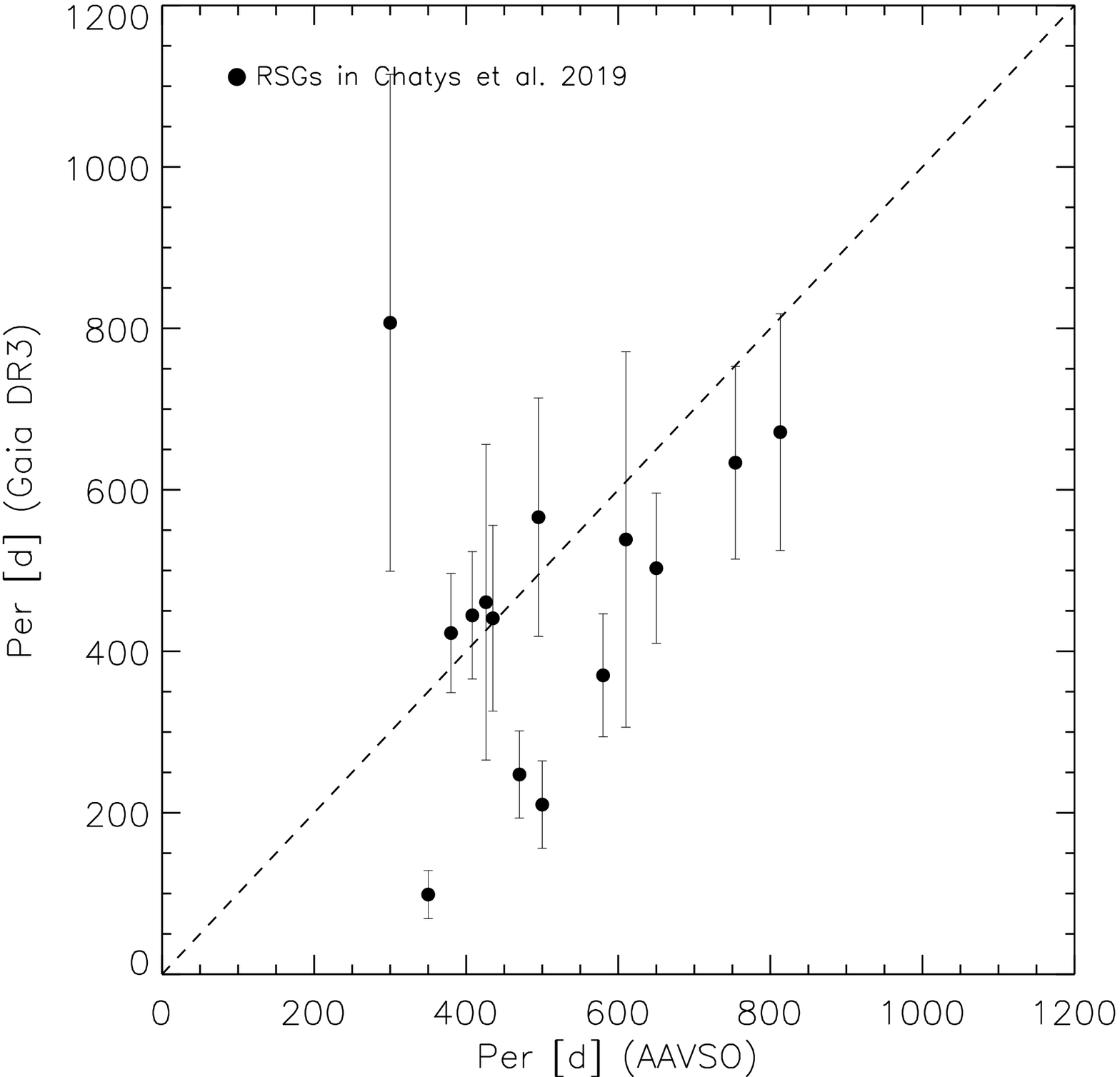}}
\resizebox{0.49\hsize}{!}{\includegraphics[angle=0]{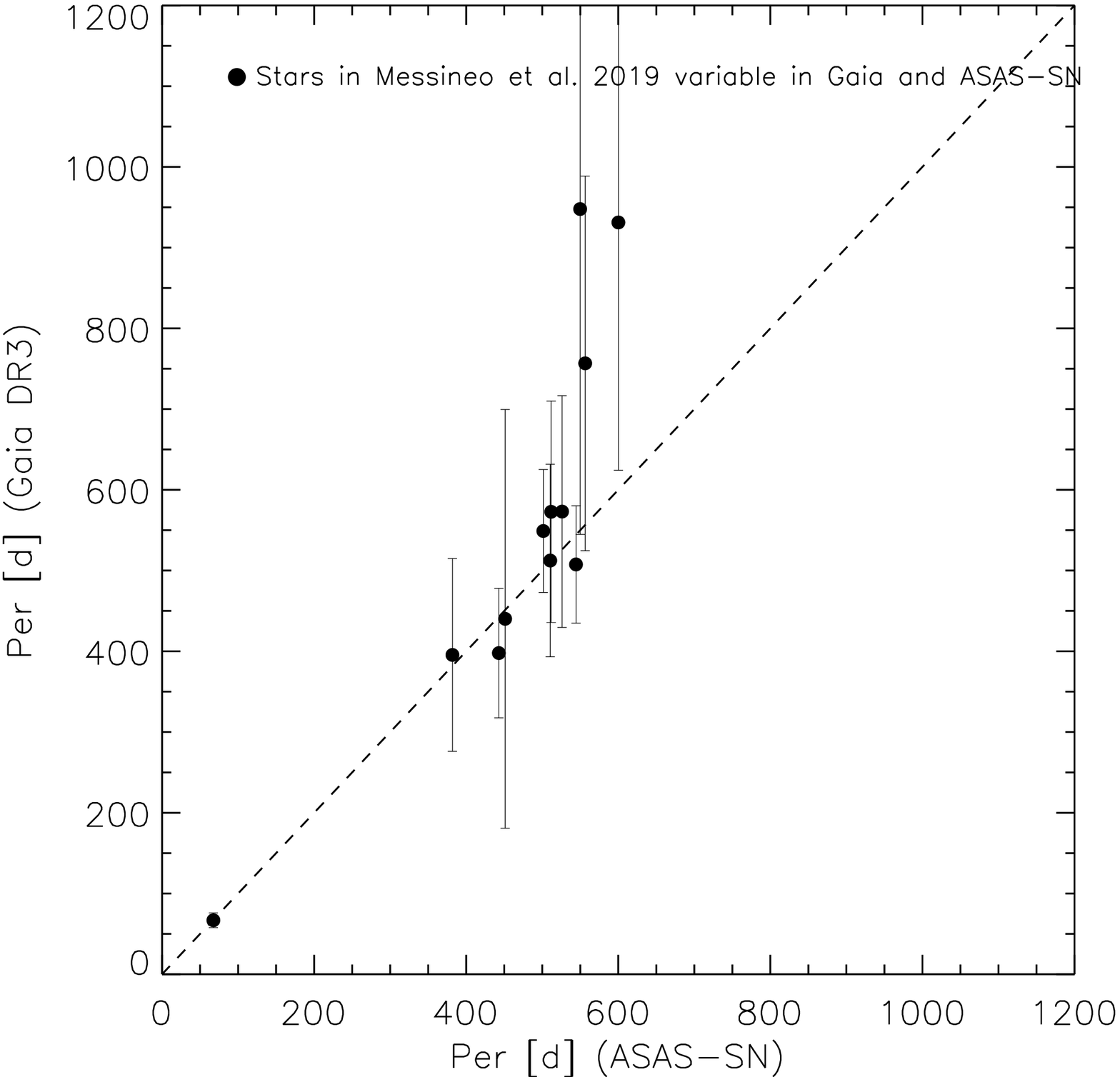}}
\end{center}
\caption{ \label{per_chathys-gaiaDR3}  \label{per_gaia_asas.eps} 
{\it Left panel:} Fundamental mode Per values of   RSGs from 
\citet{chatys19} vs. the Gaia DR3 Per values. 
{\it Right panel:} Gaia DR3 Per values of  12 late-type stars from 
\citet{messineo19} vs. Per values from the ASAS-SN survey. 
\citet{chatys19} do not include errors.
The ASAS II/366/catalog in Vizier does not contain errors.
The Gaia DR3 LPV pipeline table includes only frequency
and errors on the frequency.
  ${\rm Per} =\frac{1}{\rm frequency}.$
Errors on the Per values should be directly determined 
as the widths  of peaks in the periodogram \citep{kiss06}.
The plotted error bars for the Gaia Per values are
$\frac{1}{\rm frequency}-\frac{1}{\rm frequency+frequency\_error}$.}   
\end{figure*}

\begin{figure}
\begin{center}
\resizebox{0.99\hsize}{!}{\includegraphics[angle=0]{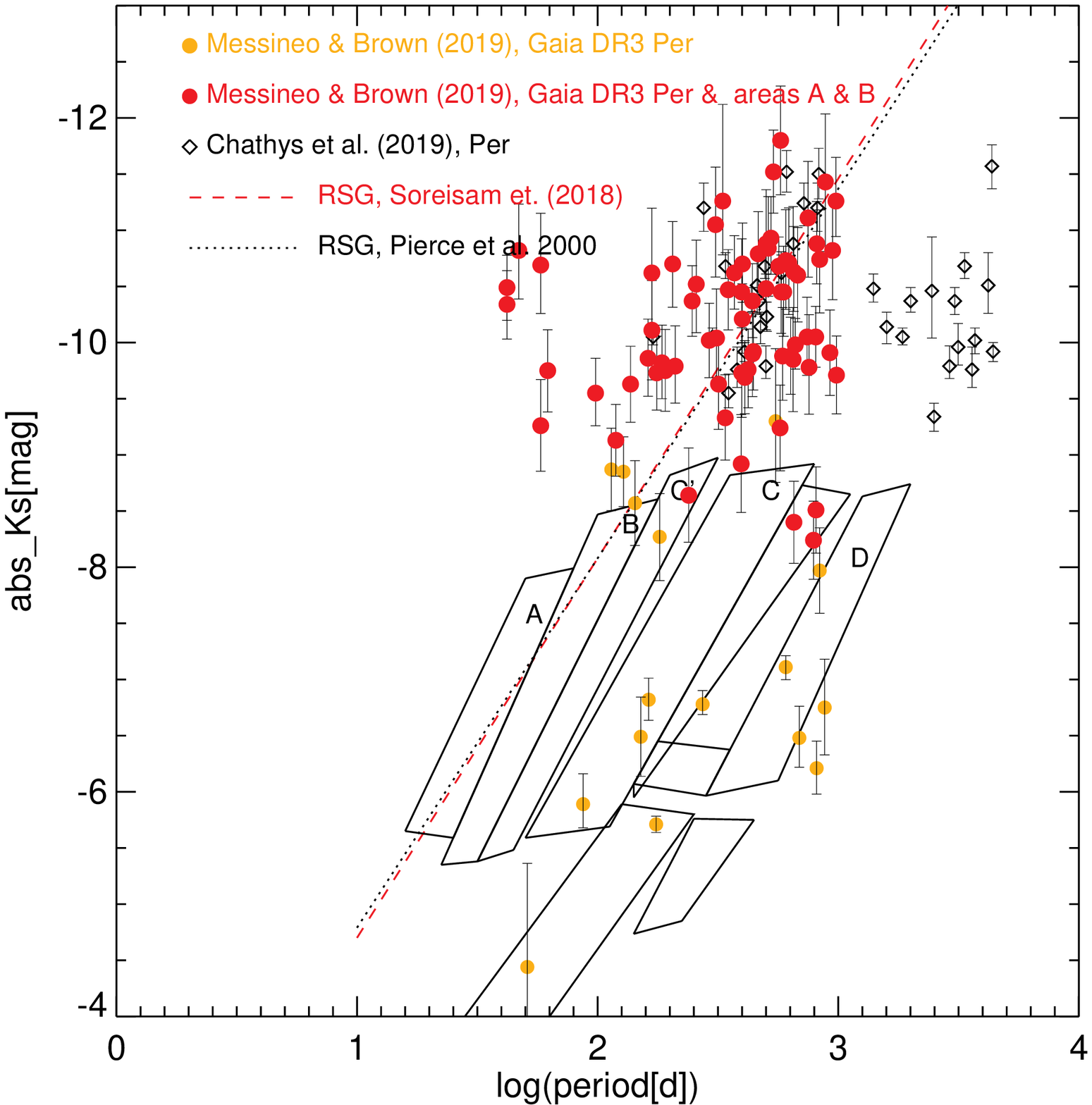}}
\end{center}
\caption{ \label{per_lum_uno}  
\Mk\ magnitudes vs. Gaia DR3 Per values of late-type stars
in the catalog of \citet{messineo19}  (red circles).
RSGs with periods from \citet{chatys19} are 
displayed with open diamonds.
For comparison, the  sequences of variable giants (A,B,C,C',D)
from \citet{riebel10} are also shown (black boxes).
} 
\end{figure}

\section{Amplitudes of class I K-M stars}\label{variable}
Among the 1,060 catalog stars with good parallaxes,
there are period estimates for 87 long-period variables \citep{lebzelter22}, 
of which 82 variables are located  among the 713 stars brighter than 
the tip of the red giant branch and 71  in areas A \& B of 
\citet{messineo19}.

The Ampl values of variable stars is an important parameter,
which allows us to separate variable RSGs and Mira AGBs. 
Typically, RSGs have Ampl $<0.8$ in G-band, 
while Mira AGBs from $\approx 0.8$  to 8 mag \citep[e.g.,][]{lebzelter22,messineo22}.
In Figure  \ref{per_ampl}, the Gaia DR3 Per values are plotted versus
the Gaia DR3 Ampl values. Per values range from about 42  to 983 d.
The  Ampl values mildly correlate with the  Per values
and are mostly smaller than $G$ Ampl $ < 0.5$ mag.
In the \Mbol\ versus \Teff\ diagram, the periodic variables
are  marking the brightest end of 
the stellar distribution. 
A few larger amplitudes are also present ($G $ Ampl $> 0.5$ mag), those 
belonging to low luminosity AGB stars; 
only four stars have a $G$ Ampl $> 1$ mag, which is
typical of Miras \citep{lebzelter22}.

As described in \citet{chatys19} and \citet{kiss06}, 
periodic variable RSGs
may present two  Per values, a short Per value (the fundamental mode)
from 200 to 1000 d and a long secondary period (LSP, Per $>$1000 d).
Only the short Per values can be detectable with Gaia data.

The All-Sky Automated Survey for Supernovae (ASAS-SN) 
surveyed the entire sky at optical wavelengths down to 17 mag 
for four years \citep{jayasinghe18}, and reported more than 66,000 new 
variables. 
When matching the catalog of KM-type stars \citep{messineo19} with 
that of ASAS-SN, 194 variables are identified and periods for 43 stars. 
Only 12 of those are already reported in the Gaia DR3 catalog 
of variables (mean difference is 83.d with $\sigma$=146 d).
The  differences between the nine  Per values smaller than 700 d 
have $\sigma$=37 d, as shown in Fig.\ \ref{per_chathys-gaiaDR3}.
The three longest Per values have a difference 
up to 397 d.

In Gaia DR3, there are period determinations for 14 stars 
from the sample of \citet{chatys19}, which have data from 
the American Association of Variable Star Observers 
(AAVSO)\footnote{\url{https://www.aavso.org/}}  
covering more than 50 years.  
The mean difference between the Gaia DR3 Per values  and the 
fundamental Per values of \citet{chatys19} is 54 d and 
the standard deviation, $\sigma$, is 201 d,
 as plotted in Fig.\ \ref{per_chathys-gaiaDR3}.
Typically, the fundamental Per values of RSGs range from 200 to 1,500 d.

In conclusion, with the current 2.8-year baseline, 
it appears that the Gaia DR3 Per values of bright late-type stars 
have an accuracy of 200 d.
The diagram of \Ks\ values versus  periods is shown 
in Fig.\ \ref{per_lum_uno}.

Matches to the stars
in \citet{messineo19} were  searched in the 
OGLE database\footnote{Vizier catalog I/244A, Optical Gravitational Lensing Experiment (OGLE) 
General Catalog of Stars.I. by 
\citet{szymanski96}  and catalog J/AcA/63/21, 
VI light curves of Galactic LPVs by \citet{soszynski13}, 
in the Wide Angle Search for Planets (WASP) time series \citep[][]{pollacco06},
and in the Kepler databases  \citep{pollacco06,jenkins10} 
without success\footnote{The NASA Exoplanet Archive is available at 
\url{https://exoplanetarchive.ipac.caltech.edu/}}.
The K2 input catalog includes the 2MASS J06090134+2352227 and
2MASS J17462538-2612011 sources \citep[e.g.,][]{huber16,howell14}, 
with an observed light curve  for the latter target.
2MASS J17462538-2612011 (\Mbol=$-5.280$ mag, RUWE=1.7) is reported as
an eclipsing  binary (K5Ib+A) with Per $>$ 35 d  
and composite spectrum at the maximum light by \citet{popper48};
indeed, it shows an anomaly in its motion \citep{kervella22}.
However, it does not have a Gaia BP/RP spectrum or RVS spectrum,
and it is neither flagged as a Gaia variable. }

\section{Identification of C-rich, O-rich, and S-type stars}
\label{classification}
The sample of \citet{messineo19} 
is made of a collection of stars taken from the literature
and  spectroscopically identified as K- and M-type stars of class I, therefore, with 
O-rich chemistry. Besides a few possible errors, 
the  stars in \citet{messineo19} are  O-rich by design.

C-rich stars are  identified by the Gaia DR3 LPV pipeline by 
locating  specific spectral features  due to   
 CN, CO$_2$, C$_2$, TiO, and VO molecular bands 
(astrophysical\_parameter.spectraltype\_esphs) 
\citep{lebzelter22}. The method is based on the 
fixed distance of specific peaks.
The DR3 flag for C-rich stars given by the Gaia DR3 LPV pipeline
erroneously marks several known RSGs (see Table \ref{Cerrore});
their BP/RP spectra are shown in Appendix  A.

\begin{table*}
\caption{\label{Cerrore} Stars among the $\approx 1,500$ O-rich stars 
in the catalog of \citet{messineo19} that are flagged as C-rich 
by the LPV pipeline. A visual inspection of their BP/RP spectra 
does not confirm  their C-rich nature.}
\begin{tabular}{llrll}
\hline
\hline
Alias & NAME DR3 & Area$^{(a)}$ & Spec.Type $^{(b)}$ & Inspection \\
\hline
              IRC~+60091   &    465401835863237504 &  A & M2      Iab&\\
CD-61~3575~~~~~~~~~~~~~~   &   6055427493901807104 &  A & M2        I&\\
IRAS~12563-6100~~~~~~~~~   &   6055686944283937408 &  A & M4     I-II&\\
                  KW~Sgr   &   4063462206570029312 &  A & M1.5        I&\\
               IRC-20409   &   4069136201903321600 &  A & M3& \\
                  PER085   &    523056686572432768 &  B & M2      Iab& \\
HD~12014~~~~~~~~~~~~~~~~   &    507785054178237824 &  B & K1&\\
HD~115921~~~~~~~~~~~~~~~   &   5868597894009411328 &  B & K4.5   Iab-Ib\\
IRAS~14599-5941~~~~~~~~~   &   5876489879290620416 &  B & M3.2Ib-II&\\
                   W~Per   &    460621159304139008 &  B & M4.5        I&\\
AS~381~~~~~~~~~~~~~~~~~~   &   2055181605358559872 &  B &  K0 I-II+B1[e] &  emission\\
                  PER287   &   2198645646339039360 &  B &  M2      Iab&\\
                  PER572   &    444883608855838720 &  E & M4    Ib-II &\\
                   MZM22   &   4146602259265301632 &  E & M0        I&\\
\hline
\end{tabular}
\begin{list}{}
\item $^{(a)}$ Area designates a specific region of the Luminosity versus 
\Teff\ diagram \citep{messineo19}.
In particular, area A includes stars brighter than the AGB limit.\\
$^{(b)}$ Spectral types are collected from the literature by \citet{messineo19}.
\end{list}
\end{table*}

\label{classS}

Following the Appendix of \citet{deangeli22}, the BP/RP spectra
of bright K-M stars from \citet{messineo19} 
were retrieved and the GaiaXPy code was run to  extract a wavelength function.
For comparison, a library of known C-rich stars 
and  S-type stars  was compiled and their BP/RP spectra 
were similarly extracted  \citep[e.g.,][]{keenan80,ake79,stephenson76}.

The spectra of M-type AGBs and RSGs are dominated by absorption due to TiO bands.
The  BP/RP spectra of   11 late S-type  stars (e.g., S4/5 and S8/5) present 
three distinct  absorptions, 
at an average wavelength of $652.9$ nm with a standard deviation $\sigma=1.6$ nm,  
at $790.7$ nm with $\sigma=4.4$ nm,  
and $939.3$ nm with $\sigma=2.1$ nm.
The strong feature centered at $\approx940$ nm is  due to ZrO bands 
\citep[e.g.,][]{wing72,rayner09,messineo21}.
The spectral regions around 
740 and 790 nm is characterized by two LaO
absorption bands 
\citep{keenan50,macconnell00}\footnote{Recently, the two LaO band-heads at 740.6 and 791.4 nm 
were measured in the laboratory by \citet{gaft22}}.
LaO molecular bands appear in the S4 type, and are strong 
in the S5 type and later types \citep{keenan80}.
The absorption around 653 nm is weak and is due to two  ZrO molecular  bands
from 647.4-647.9 nm to 650.8-651.2 nm \citep[e.g.,][]{evans89,chen22}.
The BP/RP spectra of  eight C-rich present  three featuring minima
at $704.8$ nm with  $\sigma$=1.0 nm,  at $803.3$ nm
with  $\sigma$=1.5 nm, and  at $932.8$ nm with   $\sigma$=1.5 nm,
which are due to three strong CN molecular bands
\citep[e.g.][]{kraemer05}. 


The visual inspection of the BP/RP spectra  of 
stars from \citet{messineo19}  
confirms that there are  no  C-rich stars and neither S-type stars 
of class 4 or later (SX/4) among them.
Therefore, the stars listed in Table \ref{Cerrore}
and  flagged as C-rich stars by the Gaia DR3 LPV pipeline, are
false C-rich. Indeed, they appear KM-types.

Furthermore, an empirical library of BP/RP spectra of 
known K- and M-type RSGs \citep{messineo19} 
and metal rich O-rich AGB stars 
(from the sample described in  Sect. \ref{detection})
was built with a few  BP/RP spectra, as shown in Appendix A.
The  spectra were dereddened, using the infrared extinction values (\Aks)
determined from the 2MASS $JHK$ measurements,
and the extinction curve of \citet{cardelli89} extrapolated to the
near-infrared with a power law with an index of $-1.9$ \citep{messineo05}.
The spectra were normalized to the mean flux.
By cross-matching the dereddened BP/RP spectra 
with  the built  library of BP/RP spectra, 
a best matching reference spectrum was selected
for each bright  star in areas A and B of  \citet{messineo19}. 
The minimization was done with the summation of the differences between 
the two dereddened flux density vectors. The re-assigned
spectral types  are  within  $\approx 2$ spectral types 
from those collected from the literature ---
more precisely, $\pm 1$ spectral type for K-type stars,
$\pm 2.5$ spectral types for M0-M1 stars, $\pm 1$ spectral 
type for M2-M3 stars,
and $\pm 2$ spectral type for M4-M5 stars.

\section{New highly-probable RSGs with Apsis parameters and BP/RP spectra} \label{detection}

The  entire Gaia DR3 catalog is searched to extract stars 
with Apsis parameters 
similar to those of bright stars populating
areas A and B of \citet{messineo19}.

By selecting with the GSP-Phot parameters stars with
$-0.7<$ log(g) $< 0.7$ dex, $ -0.5<$[Fe/H] $<0.5$ dex, 
$2500<$\Teff$<3900$ K (M-type), 
$\frac{\varpi}{\sigma_\varpi{(\rm est)}} > 4$,
$\rm G_{\rm BP}-G_{\rm RP} >2$ mag 
(as RSGs in Messineo's catalog), and imposing also that
there are 2MASS and WISE matches, one counts 19,499 stars. 
Only M-type  stars are considered, because
for those the  GSP-Phot  \Teff\ values appear reliable.

When counting the Gaia GSP-Spec  entries with 
$-0.7<$ log(g) $< 0.7$ dex,   
$ -0.5<$[Fe/H] $<0.5$ dex,  
and $2500<$\Teff$<4500$ K (K- and M-types), 
$\frac{\varpi}{\sigma_\varpi{(\rm est)}}>4$, 
$\rm G_{\rm BP}-G_{\rm RP} >2$ mag, and imposing also that 
there are 2MASS and WISE matches, 22,470 stars are counted. 
The retrieved entries  have the first nine digits of the GSP-Spec 
quality flag smaller than unity, while the 10, 11, and 12 digits 
were required to be null \citep{recio22}.

2MASS $JHK$ photometric measurements were retrieved.
For each source, by assuming that they are candidate RSGs, 
a total extinction \Aks\ was estimated with the $JHK$
magnitudes, the intrinsic colors for K- and M-type 
supergiants of \citet{koornneef83},
and the selective extinction ratios of \citet{messineo05},
as described in \citet{messineo19}.
Absolute \Ks\ (\Mk) were estimated using the distances from \citet{bailerjones21},
and estimates of their bolometric magnitudes were obtained using the
\BCKs\ of \citet{levesque05}.

Among the 19,499 stars from the GSP-Phot selection,
160 stars are found to belong to class A or B, 
to have  $G$ Ampl $< 0.5$ mag (when determined from the LPV pipeline), 
and   not included in the catalog of  \citet{messineo19}.
Similarly, 50 stars were found among the 22,470 stars from the GSP-Spec selection,
with seven stars already included in the GSP-Phot selection.
In conclusion, by using the Apsis parameters and 2MASS photometry 
203 (160+50-7) new candidate RSGs from class A or B  are identified.
The  sample has  been selected based on
their Apsis parameters and may contain O-rich, C-rich, and S-type AGBs 
and RSGs. 
By contrast, the catalog of \citet{messineo19}, by 
design, is  made up of O-rich AGBs and RSGs 
(with chemistry known from literature).
When searching for stars from \citet{messineo19} which are included in the 19,499 
GSP-Phot selected stars,  23 stars is retrieved.

In the next sections, this number (203) will be further reduced to a dozen 
highly-probable RSGs by considering other constraints from the  BP/RP spectra,
intrinsic photometric colors, and luminosity.

\subsection{ Mid-infrared counterparts of the candidates}
For the sample of 203 near-IR bright luminous late-type stars,
mid-infrared measurements were retrieved from the
MSX, WISE, GLIMPSE, and MIPSGAL surveys.
MSX matches were available for 146 sources,
WISE matches were available for all 203 sources, 
GLIMPSE matches for 37 sources, and MIPSGAL for 27 sources.

\subsection{Spectroscopic identification of S- and C-rich stars}
When searching on  the SIMBAD  astronomical database, 
ten of the new 203 Apsis selected bright stars 
are reported as C-rich stars,  15 as S-stars, and one as S/C star
(see Table \ref{knownSC}).
The  BP/RP spectra show clear signatures of late S-types for 11 stars, as displayed 
in Appendix  A. For example,  star 2162991061217892736 is 
marked as an S in  the SIMBAD  astronomical database,
but its spectrum is matched with an M2 star  because it is  most likely an early S star, 
SX/(1-2).
All, but one, known C-rich stars are  flagged by the Gaia DR3 LPV pipeline, 
and confirmed with a visual inspection of their BP/RP spectra.
Cya86  	(2MASS  J20220156+3634558, Gaia DR3 2057504976503308928) 
was listed as a C-rich stars by \citet{aaronson90}, however,
Gaia data do not support such type; its spectrum resembles that of a K-type star
(see Table \ref{knownSC} and Appendix A).

\subsection{Photometric identification of S-type, C-rich, and O-rich stars}
\label{Cphot}

\begin{table*}
\caption{\label{knownSC} List of C-rich and S-type stars found among the 203 selected 
Apsis bright late-type stars}
\begin{tabular}{llllllrr}
\hline
\hline
ID                 & Gaia eye Type  & Gaia LPV-pipe C & SIMBAD&Comment\\ 
                   & \\
\hline
 1869327345293745408  &         S  &     false  &         S &  \\
 2054730187112773888  &         S  &      $..$  &         S &  \\
 2162991061217892736  &        (M3) &      $..$  &         S & S(1-3)/(1-3)$^*$ \\
 2164275668781737088  &         S  &     false  &         S &  \\
 3109665212615470080  &         S,poor  &     false  &       M3S &  \\
  392259848477966464  &         S  &      $..$  &S2.5-5/6-7 &  \\
 5233074194539855360  &         S  &     false  &      S6-8 &  \\
  525500041926758912  &         S  &     false  &         S   \\
 5326839648494951168  &         (M3) &      $..$  &        S?  & S(1-3)/(1-3)?\\
 5515454717849642624  &         (M1) &     false  &      S2*2  & S(1-3)/(1-3) \\
 5593221113697014144  &        (M5S?)&     false  &         S  & S5/(1-3)\\
 5613437666506594816  &         S  &      $..$  &         S   \\
 5615401806591049600  &         S  &     false  &        S6   \\
 5929367695531204352  &         S  &     false  &      S7,7   \\
  999737853862776576  &         S  &     false  &        S!   \\
 2999161373443493504  &       S/C  &      true  &       S/C   \\
  182625449700126336  &         C  &      true  &     C-N5+   \\
 2002465559026804480  &         C  &      true  &         C   \\
 2057504976503308928  &        K4  &      $..$  &        (C)& C not confirmed   \\
 2199232613746829312  &         C  &      true  &         N   \\
  272850583600304640  &         C  &      true  &         C   \\
  461731459888465664  &         C  &      true  &      C5,3   \\
  513069513222015360  &         C  &      true  &         C   \\
 5716487091710504064  &         C  &      true  &         C   \\
 5718419826993558016  &         C  &      true  &         C   \\
 5863457058660466688  &    C,poor  &      true  &         C   \\
\hline
\end{tabular}
\begin{list}{}
\item $^*$ Early S-type following the definition by \citet{keenan80}.
\end{list}
\end{table*}

\begin{figure*}
\begin{center}
\resizebox{0.45\hsize}{!}{\includegraphics[angle=0]{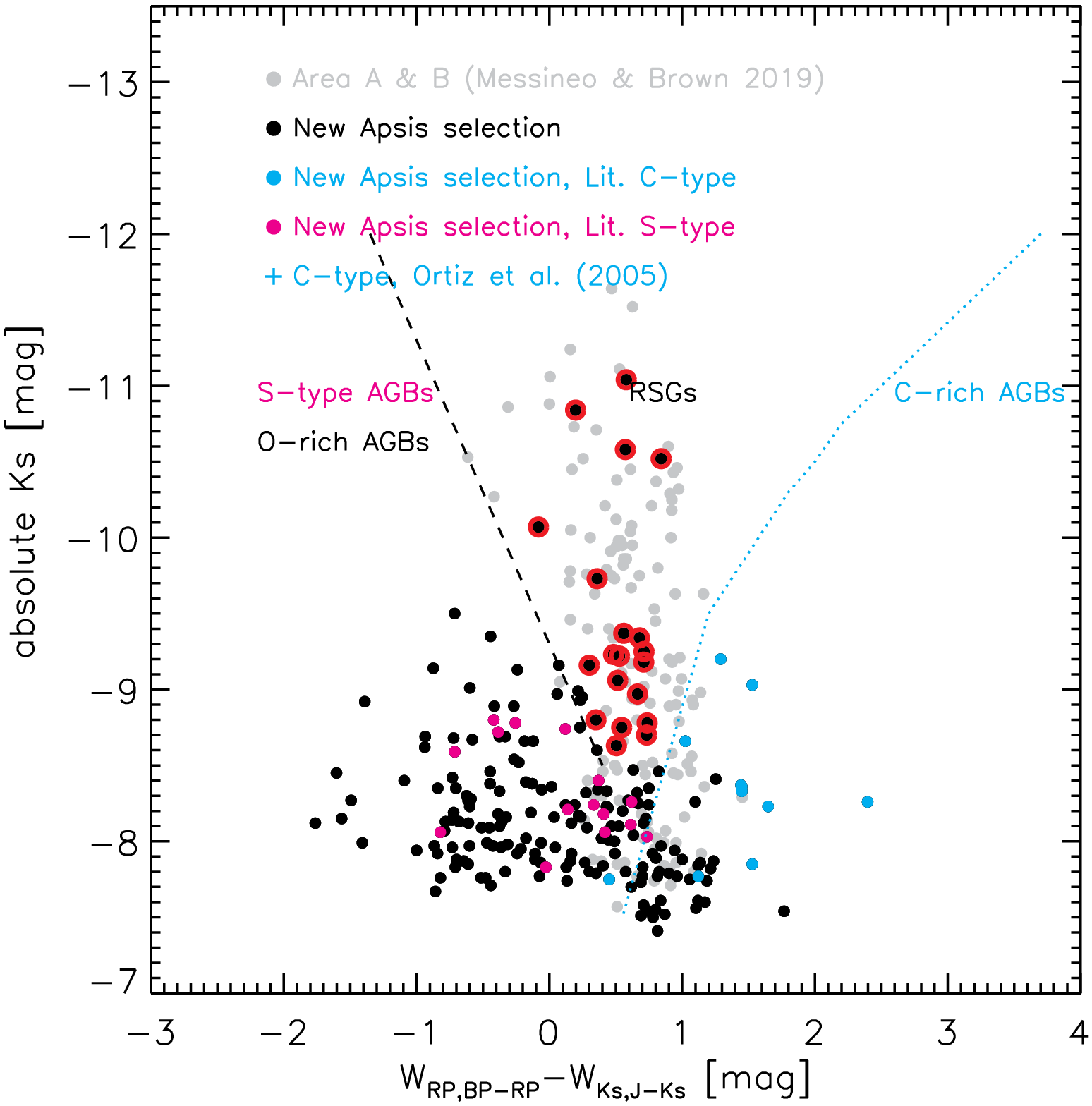}}
\resizebox{0.45\hsize}{!}{\includegraphics[angle=0]{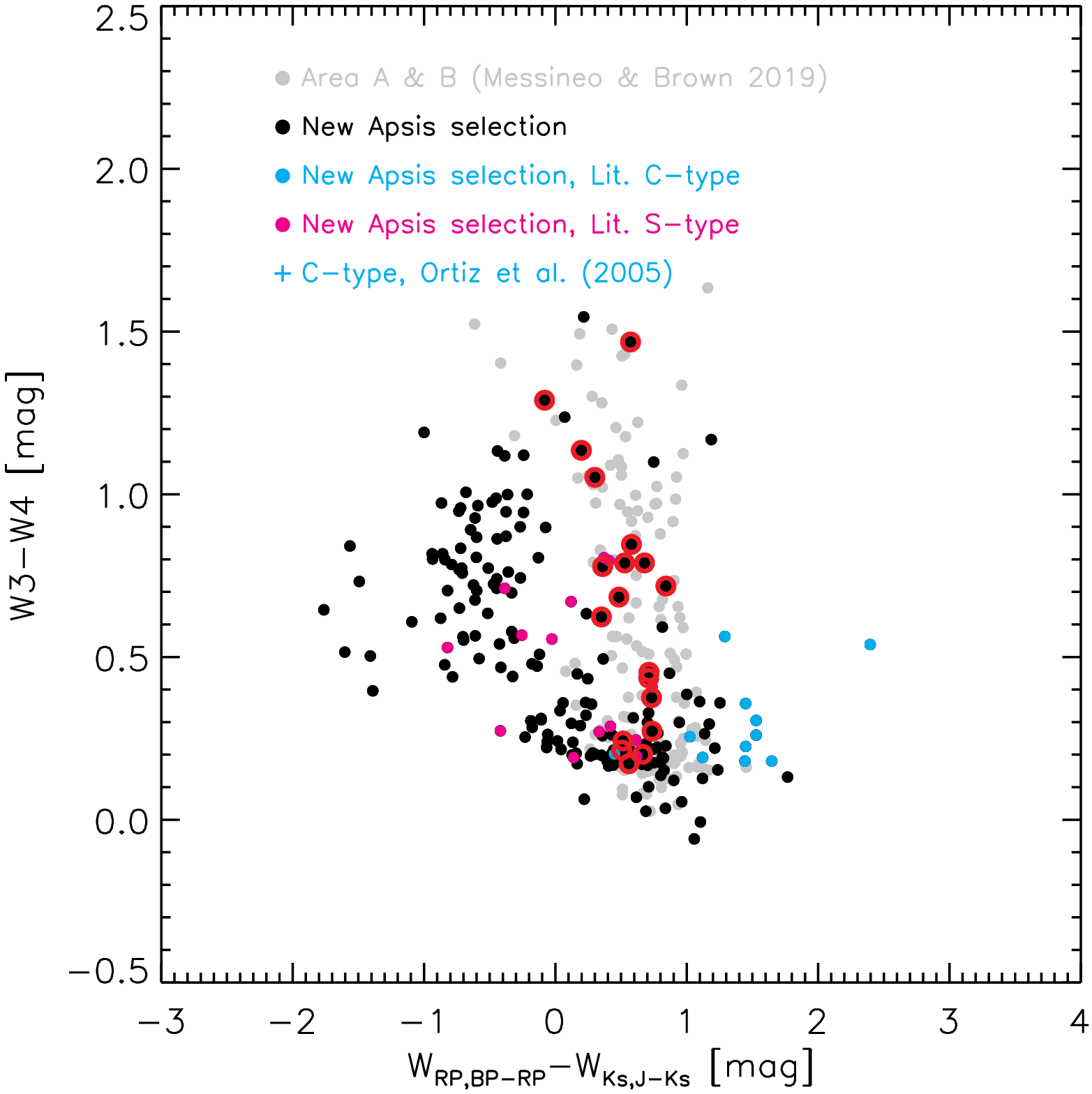}}
\end{center}
\caption{ \label{colorcarbon} 
{\it Left panel:}  \Mk\ values vs. W$_{\rm RP,BP-RP}-$W$_{\rm K_S,J-K_S}$.
Gray filled circles mark  class I stars from areas A \& B of \citet{messineo19}.
Black-filled circles indicate the newly Apsis selected bright late-type stars, in magenta
those reported in the literature as S-type stars, 
and in cyan those reported as C-rich stars. 
The long-dashed line separated the bulk of AGB stars from the sequence of
known RSGs \citep{messineo19},
which lyes in a vertical sequence between color 0 and 1 mag.
The C-rich stars are redder  than the roughly traced 
cyan-dotted curve \citep{abia22}.
{\it Right panel:} W$_{\rm RP,BP-RP}-$W$_{\rm K_S,J-K_S}$ 
vs. WISE [W3-W4] colors. Symbols are as in the  left panel.
}  
\end{figure*}

\begin{figure}
\begin{center}
\resizebox{0.99\hsize}{!}{\includegraphics[angle=0]{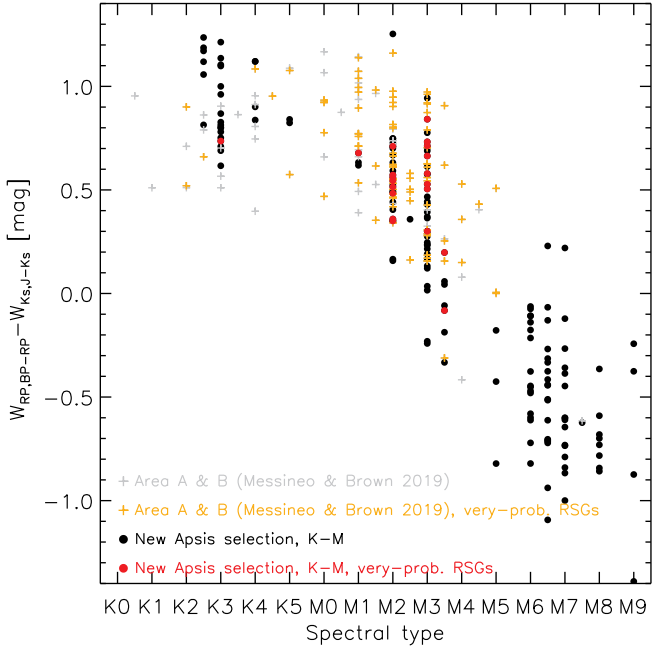}}
\end{center}
\caption{ \label{dimostra} Spectral types inferred from the BP/RP spectra vs. 
the W$_{\rm RP,BP-RP}-$W$_{\rm K_S,J-K_S}$ colors of the 203 Apsis selected bright late-types
(black-filled circles). Red-filled circles indicate highly-probable RSGs. 
For comparison, spectral types and W$_{\rm RP,BP-RP}-$W$_{\rm K_S,J-K_S}$ colors of KM-type
stars in areas A or B of \citet{messineo19} are marked with plus signs; in orange
those highly-probable RSGs.}  
\end{figure}

There are  several photometric diagnostics available in 
literature to select C-rich stars and   S-type stars. 
For example, C-rich stars can be identified by their 
near- and mid-infrared colors (e.g., [W3-W4] vs. [\Ks-W3] and
[A-D] vs. [\Ks-A]), as described in  
\citet{messineo18} and \citet{ortiz05}.
Recently, \citet{abia22} has analyzed the Gaia photometry
 and 2MASS data of evolved bright late-type stars and found that
using the extinction-free 
color W$_{\rm RP,BP-RP}-$W$_{\rm K_S,J-K_S}$\footnote{
W$_{\rm RP,BP-RP}$=${\rm G_{\rm RP}}-1.3$($G_{\rm BP}-G_{\rm RP}$);
W$_{\rm K_S,J-K_S}$=$\K_S-0.686$($J-K_S$), 
as defined in \citet{abia22}.}
and the luminosity,
it is possible to identify C-rich stars (redder), 
mass-loosing AGB stars (bluer), and RSGs.
For the newly selected Apsis bright late-type stars,
the \Mk\ magnitudes are plotted 
versus the W$_{\rm RP,BP-RP}-$W$_{\rm K_S,J-K_S}$
colors in Fig.\  \ref{colorcarbon}.
The few known C-rich and S-type stars are marked.
For comparison, the brightest K- M-type stars of the \citet{messineo19}
catalog (areas A \& B) are also shown. 
In the \Mk\ versus W$_{\rm RP,BP-RP}-$W$_{\rm K_S,J-K_S}$ diagram,
the class I K- M-type stars appear located in a vertical sequence with a narrow 
color range (0.0-1.0 mag), which  extends to \Mk=$-12$ mag, as
predicted by \citet{abia22}.
In the diagram W$_{\rm RP,BP-RP}-$W$_{\rm K_S,J-K_S}$
vs. WISE [W3-W4] color, two sequences are also visible.
One sequence is traced by the class I KM-type stars 
(areas A \& B) of \citet{messineo19},
and the other by late-type O-rich and S-type AGB stars.
The sequences appear merged at W$_{\rm RP,BP-RP}-$W$_{\rm K_S,J-K_S}$= 0.5-1 mag 
and [W3-W4]=0.2-0.4 mag, but  they split into two separated sequences 
for redder [W3-W4]. The two sequences have  
different  W$_{\rm RP,BP-RP}-$W$_{\rm K_S,J-K_S}$ 
colors, but similar ranges of [W3-W4] color.
This suggests  differences in the surface \Teff\ values/spectral types.
Indeed, a correlation appears between the  W$_{\rm RP,BP-RP}-$W$_{\rm K_S,J-K_S}$ 
colors and the spectral types, as shown in Fig.\ \ref{dimostra}.
Spectral types were derived by crossmatching the BP/RP spectra
with those of known RSGs \citep{messineo19} and known late-type AGB stars
present in the sample (as described in Sect. \ref{classS}).
The vertical sequence of class I stars is made up of stars earlier 
than M5,
while most of the low-luminosity AGBs have spectral type from M5 to M9.
Note that the [W3-W4] colors have not been dereddened;
at these wavelengths the obscuration is negligible with respect to the range
of [W3-W4] colors spanned by the stars. The \Aks\ values of stars
in areas A and B of \citet{messineo19} range from 0 to 0.7 mag,
and, therefore,  A(W3) is below 0.38 and A(W4) is below 0.28 mag \citet{messineo05}.
This implies that the reddening E(W3-W4) is below 0.1 mag.
The [W3-W4] color  is likely indicating a different types of envelopes (mass loss), as
seen for the \Ks-W4 \citep[e.g.][]{messineo17}.

\subsection{A sample of very-likely RSGs }
\label{cutdescribe}
From a sample of about 40,000 stars with Apsis parameters compatible 
with those of
RSGs, 203 bright stars are found to populate   areas A and B of the
luminosity versus \Teff\ diagram of \citet{messineo19}.
When removing the C-rich stars and S-type stars, 
and retaining the stars brighter 
than \Mk = $-8.5$ mag and bluer than the dashed line
\Mk=$-9.5$+$2.0\times$W$_{\rm RP,BP-RP}-$W$_{\rm K_S,J-K_S}$
which is drawn in Fig.\ \ref{colorcarbon},
only 20 RSGs remain (out of 203 bright stars populating areas A and B),
which are listed in Table \ref{tablenewrsg}.
The cut removes the  strong contamination of O-rich AGBs. 

When applying the same cut to the class I K-M stars from areas A and B 
of \citet{messineo19}, a sample of 312 RSGs out of 486 stars is obtained,
or 87 RSGs out of 133 stars when considering those 
with available BP/RP spectra. Therefore, the new 20 RSGs represent
an increase of 23\% of the currently known Gaia RSGs with BP/RP spectra.

\begin{landscape}
    \begin{table}
\vspace*{+1cm}
\caption{\label{tablenewrsg}Twenty new Galactic RSGs.}
\centering
\begin{tabular}{llllllllllll}
\hline
\hline
Gaia DR3 ID & 2MASS ID & \Teff & \Aks\ & DM  & \Mbol(1)&\Mbol(2)& rangeG &Ampl&Per&Sp. Type (here) & Sp. Type (lit)\\
            &       &   [K] & [mag] &[mag]& [mag]   & [mag] &[mag]&[mag] &[d]&\\
\hline
 2162696838801402624  &           21015501+4517205  & 3706.136  & 0.38 $\pm$ 0.21 &13.58  $^{ 0.27  }_{-0.26 }$&-8.28  $^{-0.45  }_{ 0.45 }$&-8.41  $^{-0.36  }_{ 0.36 }$&  0.46 &   0.10 & 571.40 &        M3 I &        $..$  \\
 5940602264810935296  &           16291280-4956384  & 3619.762  & 0.43 $\pm$ 0.19 &12.43  $^{ 0.36  }_{-0.45 }$&-8.01  $^{-0.48  }_{ 0.55 }$&-8.15  $^{-0.43  }_{ 0.51 }$&  0.35 &    $..$&    $..$&      M3.5 I &        M5  \\
 5255557489361493504  &           10192621-5818105  & 3705.174  & 0.64 $\pm$ 0.21 &12.65  $^{ 0.31  }_{-0.28 }$&-7.83  $^{-0.48  }_{ 0.46 }$&-7.98  $^{-0.40  }_{ 0.37 }$&  0.53 &   0.20 & 818.00 &        M2 I &     M4/S?  \\
 2057885446186660992  &           20230860+3651450  & 3672.156  & 0.70 $\pm$ 0.23 &12.43  $^{ 0.32  }_{-0.28 }$&-7.74  $^{-0.52  }_{ 0.50 }$&-7.84  $^{-0.42  }_{ 0.39 }$&  0.22 &    $..$&    $..$&        M3 I &  M1/M4/M5  \\
 5966134707191684608  &           17084131-4026595  & 3715.311  & 0.56 $\pm$ 0.15 &12.17  $^{ 0.43  }_{-0.44 }$&-7.32  $^{-0.54  }_{ 0.54 }$&-7.49  $^{-0.45  }_{ 0.45 }$&  0.58 &    $..$&    $..$&      M3.5 I &        $..$  \\
 5967949038813123328  &           16490055-4217328  & 3807.621  & 0.65 $\pm$ 0.23 &12.29  $^{ 0.30  }_{-0.31 }$&-7.04  $^{-0.50  }_{ 0.50 }$&-7.15  $^{-0.41  }_{ 0.41 }$&  0.15 &    $..$&    $..$&        M2 I &        $..$  \\
 2162838224802765312  &           21052536+4609193  & 3746.285  & 0.21 $\pm$ 0.01 &13.40  $^{ 0.21  }_{-0.19 }$&-6.61  $^{-0.22  }_{ 0.20 }$&-6.72  $^{-0.21  }_{ 0.19 }$&   $..$& n/v$^*$ &    n/v&        M1 &        em  \\
 2034031438383765760  &           19554232+3205492  & 3692.973  & 0.19 $\pm$ 0.19 &12.68  $^{ 0.16  }_{-0.16 }$&-6.59  $^{-0.44  }_{ 0.44 }$&-6.68  $^{-0.22  }_{ 0.22 }$&  0.11 &    $..$&    $..$&        M2 &M2/M4.5III  \\
 4280063935601354752  &           18370725+0305122  & 3747.560  & 0.25 $\pm$ 0.02 &13.33  $^{ 0.39  }_{-0.32 }$&-6.50  $^{-0.39  }_{ 0.33 }$&-6.60  $^{-0.39  }_{ 0.32 }$&   $..$& n/v &    n/v&        M2 &        M0  \\
 2166846155161277568  &           20555124+4726196  & 3758.865  & 0.32 $\pm$ 0.02 &12.98  $^{ 0.18  }_{-0.19 }$&-6.46  $^{-0.20  }_{ 0.21 }$&-6.55  $^{-0.19  }_{ 0.20 }$&   $..$& n/v &    n/v&        M2 &        $..$  \\
 4066271286955813120  &           18101469-2400211  & 3657.726  & 0.33 $\pm$ 0.17 &10.83  $^{ 0.16  }_{-0.13 }$&-6.45  $^{-0.34  }_{ 0.32 }$&-6.56  $^{-0.26  }_{ 0.24 }$&  0.13 &    $..$&    $..$&        M3 &        M2  \\
 4093888820012699776  &           18122069-2106308  & 3687.384  & 0.32 $\pm$ 0.21 &10.82  $^{ 0.15  }_{-0.14 }$&-6.45  $^{-0.39  }_{ 0.39 }$&-6.56  $^{-0.30  }_{ 0.30 }$&  0.14 &    $..$&    $..$&        M3 &     M1/M4  \\
 5964587629936987264  &           16565150-4341211  & 3669.180  & 0.54 $\pm$ 0.20 &11.24  $^{ 0.33  }_{-0.26 }$&-6.37  $^{-0.47  }_{ 0.42 }$&-6.46  $^{-0.41  }_{ 0.36 }$&  0.42 &    $..$&    $..$&        M3 &        $..$  \\
 3030313424084918528  &           07333169-1351089  & 3692.653  & 0.14 $\pm$ 0.02 &13.45  $^{ 0.23  }_{-0.22 }$&-6.29  $^{-0.24  }_{ 0.23 }$&-6.40  $^{-0.23  }_{ 0.22 }$&  0.21 &    $..$&    $..$&        M2 &      M2.5  \\
 5641052313347740928  &           08465133-3144492  & 4038.000  & 0.31 $\pm$ 0.01 &12.99  $^{ 0.12  }_{-0.15 }$&-6.27  $^{-0.13  }_{ 0.16 }$&-6.49  $^{-0.12  }_{ 0.15 }$&   $..$& n/v &    n/v&        K3 &        $..$  \\
 3371386878815821824  &           06333294+1829346  & 3697.617  & 0.20 $\pm$ 0.18 &12.62  $^{ 0.26  }_{-0.19 }$&-6.20  $^{-0.47  }_{ 0.43 }$&-6.27  $^{-0.28  }_{ 0.21 }$&  0.26 &   0.10 & 225.70 &        M3 &     M2/M4  \\
 4350721267200216576  &           16310076-0810201  & 3710.535  & 0.05 $\pm$ 0.16 &12.61  $^{ 0.16  }_{-0.17 }$&-6.04  $^{-0.38  }_{ 0.38 }$&-6.15  $^{-0.20  }_{ 0.21 }$&   $..$& n/v &    n/v&        M2 &        M0.5$^{**}$  \\
 5865178545956998272  &           13231269-6326447  & 3764.869  & 0.22 $\pm$ 0.02 &13.06  $^{ 0.20  }_{-0.16 }$&-5.99  $^{-0.21  }_{ 0.17 }$&-6.07  $^{-0.20  }_{ 0.16 }$&   $..$& n/v &    n/v&        M3 &        M4  \\
 6054417592496091520  &           12132666-6252136  & 3705.015  & 0.36 $\pm$ 0.01 &13.07  $^{ 0.30  }_{-0.24 }$&-5.99  $^{-0.30  }_{ 0.25 }$&-6.07  $^{-0.30  }_{ 0.24 }$&  0.12 &    $..$&    $..$&        M2 &        $..$  \\
 3428621063247110912  &           05465281+2452220  & 3684.000  & 0.24 $\pm$ 0.02 &12.88  $^{ 0.35  }_{-0.24 }$&-5.85  $^{-0.36  }_{ 0.25 }$&-5.96  $^{-0.35  }_{ 0.24 }$&  0.12 &    $..$&    $..$&        M3 &      M2/3  \\
\hline
\end{tabular}
\begin{list}{}
\item {\bf Notes:} \Teff\ is the effective temperature given by the GSP-Phot processing.\\
{\it M$_{bol}$(1)} is the bolometric magnitudes derived with the \BCK\ of \citet{levesque05}.\\
{\it M$_{bol}$(2)} is the bolometric magnitudes derived with all available near- and mid-infrared magnitudes, 
as in \citet{messineo19}.\\
rangeG is range\_mag\_g\_fov from Gaia DR3 gaiadr3.vari\_summary that is
the difference between the highest and lowest G FoV magnitudes.\\ 
Ampl and Per are the amplitude and period estimated from the Gaia DR3 LPV pipeline \citep{lebzelter22}.\\
Sp. Type (here) it is the spectral type inferred by comparison with a library of BP/RP spectra of known RSGs.\\
Sp. Type (lit) is the spectral type reported by \citet{skiff16}.\\
($*$) n/v= information about variability `NOT-AVAILABLE'.
($^{**}$) Spectral type is  from the online catalog by \citet{houk99}.\\
None of the new RSGs above listed has data in the WASP, Kepler, K2, and OGLE databases.
None of them shows anomalies in their motions due to companions \citep{kervella22}.
\end{list}
    \end{table}
\end{landscape}

Six newly selected stars have estimated \Mbol$<-7.1$ mag.\\
2MASS J21015501+4517205 is  on the Perseus arm at (l,b)=(86.59$^\circ$,$-$0.77$^\circ$). 
It was unsuccessfully searched for a 43 GHz SiO maser
by \citet{deguchi05}. In the present work, an M3 type is inferred.\\
2MASS J16291280-4956384 and 2MASS J17084131-4026595  seem to belong to the
Scutum-Crux arm with  (l,b)=($-$13.77$^\circ$, $-$0.13$^\circ$)  
and (l,b)=($-$25.49$^\circ$,$-$0.91$^\circ$), respectively.
2MASS J16291280$-$4956384 coincides with IRAS 16254$-$4950 (M3-5) which
has a strong 10 \um\ silicate feature in emission \citep{volk89}.
2MASS J17084131$-$4026595 coincides with IRAS 17051$-$4023.
In the IRAS LRS catalog, its spectrum  is 
characterized by a strong silicate feature in emission \citep{volk89}
and with the BP/RP spectrum an M3.5 I type  is derived.\\
%
2MASS J10192621$-$5818105 is located on the Sagittarius-Carina arm at
(l,b)=($-$75.95$^\circ$,$-$1.11$^\circ$) and  
coincides with IRAS 10176-5802. 
An SiO maser emission was detected at 86 GHz with a central velocity of
\vlsr=$-$10.0 \kms\ by \citet{haikala94}.
The source has an IRAS low-resolution spectrum that displays
the 10 \um\ silicate feature in emission \citep{olnon86}, which indicates 
an oxygen-rich envelope that is optically thin.
\citet{skiff16} reports an M4/S? type, while in this work 
luminosity and BP/RP spectrum suggest an M2 I.\\
%
2MASS J20230860+3651450 is on the local arm at (l,b)=(75.47$^\circ$, $-$0.21$^\circ$) 
and  coincides with IRC +40413, which is classified as an  M5 supergiant by 
\citet{grasdalen79}. The supergiant class is, however, not annotated in the SIMBAD 
database and  \citet{skiff16} reports  M1/M4/M5 types. Its BP/RP spectrum fits better
that of an M3 I star.\\
%
2MASS J20230860+3651450 coincides with IRAS 16454-4212 and 
it is most likely an M2 type. 
There is no other information on  the SIMBAD  database about this star. 

\begin{figure}
\begin{center}
\resizebox{0.99\hsize}{!}{\includegraphics[angle=0]{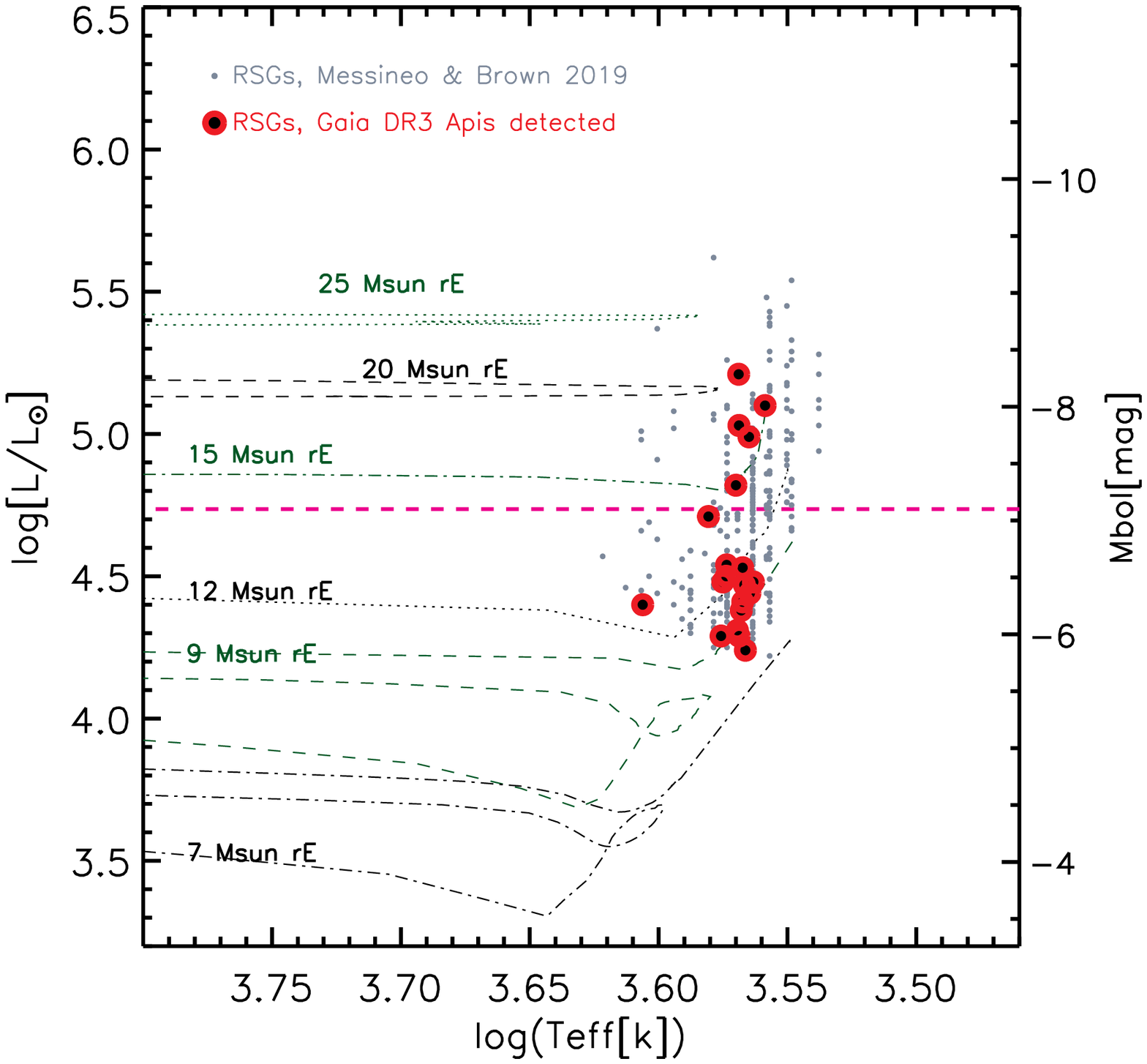}}
\end{center}
\caption{ \label{lumnew}  Luminosities vs. \Teff\ values of the 20 new very-probable RSGs
(filled-black circles,  encircled by red rings),
which were found analyzing stars with Apsis parameters. 
For comparison, the highly-probable RSGs among the class I stars listed by 
\citet{messineo19} are also plotted with gray-filled circles.
Tracks of rotating stars  at solar metallicity   from \citet{ekstrom12}
 are over-plotted; the black dotted–dashed curve marks a 7 \Msun\ track; 
the green long-dashed curve marks a 9 \Msun\ track; 
the black-dotted curve marks a 12 \Msun\ track; 
the green dotted–dashed curve shows a 15 \Msun\ track; 
the black long-dashed curve marks 
a 20 \Msun\ track; the top green-dotted line shows a 25 \Msun\ track.
The horizontal  magenta-dashed line indicated the AGB magnitude limit.
}
\end{figure}

\begin{figure}
\begin{center}
\resizebox{0.99\hsize}{!}{\includegraphics[angle=0]{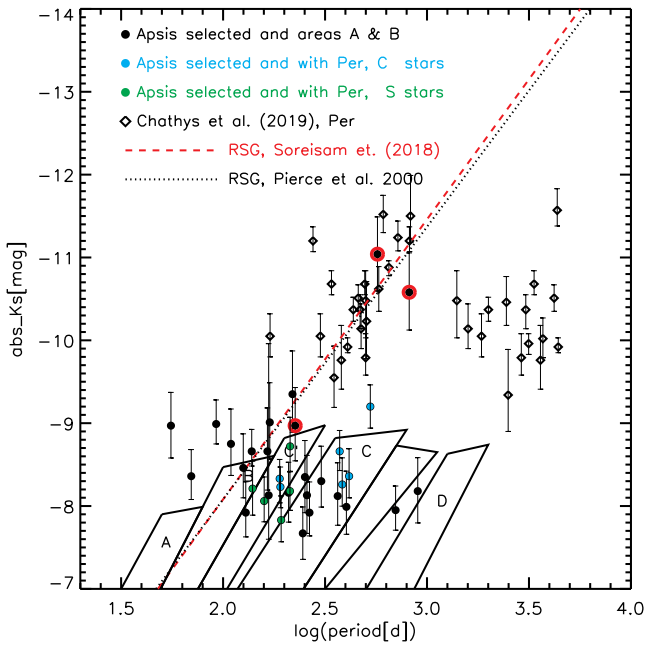}}
\end{center}
\caption{ \label{lumpernew} \Mk\ magnitudes vs. periods of the variables
in the newly selected Apsis sample (black-filled circles).
Encircled in red are the very-probable RSGs, and
in green are the targets reported  as
C-rich or S-type stars in  the SIMBAD  database. For comparison, diamonds mark the location
of well-known RSGs \citep{chatys19, messineo21}.
The red long-dashed line is the period-luminosity relation from \citet{soraisam18}
and the dotted line is that of \citet{pierce00}. 
The boxes indicate the loci of the 
giant sequences as described in \citet{riebel10}.}
\end{figure}

\label{cutinvarlum}
In Figure \ref{lumnew}, the luminosities of the newly detected 20 RSGs
are plotted against their \Teff\ values. 
The stellar tracks of rotating stars with solar metallicity 
and  masses from 9 to 25 \Msun\ successfully enclose
the range of luminosities of the new RSGs \citep{ekstrom12}.
In Figure \ref{lumpernew}, the \Mk\ magnitudes  
of the variables are plotted versus their 
Gaia DR3 Per values.
The two brightest stars (2MASS J21015501+4517205 and 
2MASS J10192621$-$5818105) of the sample are  
located on the period-luminosity sequence of  RSGs and have luminosity above that
of the AGB limit (\Mbol = $-$8.4 and $-$7.9 mag, respectively).

\section{Conclusive summary}
\label{summary}

Existing Apsis GSP-Phot and GSP-Spec parameters of known RSG stars
and bright AGB stars from the catalog of \citet{messineo19} are analyzed.
Only nine known RSGs have GSP-Spec parameters available from DR3,
while 79 RSGs have GSP-Phot parameters.
The BP/RP spectra of class I stars 
in \citet{messineo19} confirm that the sample is made 
of O-rich stars.  

The Apsis parameters  from the GSP-Phot 
and GSP-Spec  pipelines have been compared with parameters from the literature,
keeping in mind that the synthetic models used to fit the BR/RP spectra
are not suitable for mass-losing stars and do not
include the peculiar chemistry of the AGB envelopes.
Several biases have been  demonstrated. 

The GSP-Phot module analyzes the stellar energy distribution (SED) 
of the stars and absolute fluxes; it
suffers from  degeneracy between temperature 
and interstellar extinction (a higher  estimated extinction mimics a higher 
 \Teff) and is advised only for stars 
located at moderate extinction  \citep[\Av $<2$ mag,][]{andrae22}. 

The GSP-Spec module is independent of distances and extinction, based
only on the analysis of normalized spectra.
However, it fails to measure supersolar metallicity in RSGs.
There are only a few gravity values of good quality,
and it is advisable to recalculate them from the stellar luminosity.
The stellar \Teff\ values estimated by \citet{messineo19} 
agree with those released by the GSP-module  within 300 K, 
well within the selected quality flag ($\Delta$ \Teff =500 K).

The  Gaia DR3 database was searched 
for Apsis GSP-Phot and GSP-Spec parameters
similar to those of known RSGs. 
The initial  44,000 entries were reduced to  203
entries after having estimated the infrared luminosities, and to
20 entries when cutting the catalog at \Mk=$-8.5$ mag
and removing the bluest sources 
in the W$_{\rm RP,BP-RP}-$W$_{\rm K_S,J-K_S}$ colors,
which are very likely to be AGBs \citep{abia22}.
The BP/RP spectra of the new 20 RSGs display features
of M1-M3.5 stars, plus one K3.

Some C-rich stars and S-type stars were contaminating the built sample.
They were identified in the SIMBAD database and confirmed
with the \Mk\ versus W$_{\rm RP,BP-RP}-$W$_{\rm K_S,J-K_S}$ diagram
of \citet{abia22}, as well as with a direct inspection of the
morphology of their BP/RP spectra.
Indeed, their BP/RP spectra show absorption features
 due to ZrO and LaO molecules,
the most prominent being centered at 940 nm.


\appendix
\section{BP/RP spectra}

\begin{figure*}
\begin{center}
\resizebox{0.33\hsize}{!}{\includegraphics[angle=0]{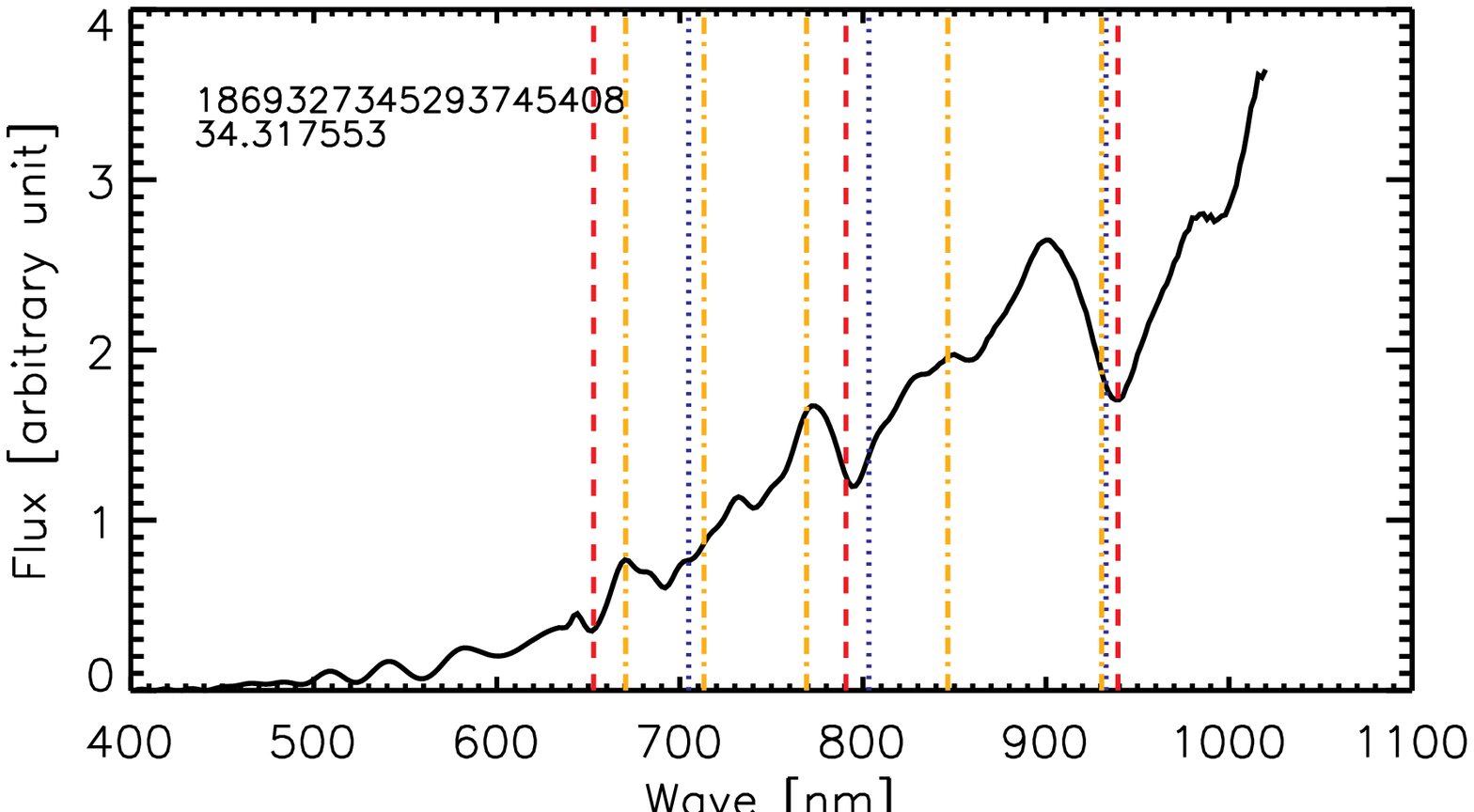}}
\resizebox{0.33\hsize}{!}{\includegraphics[angle=0]{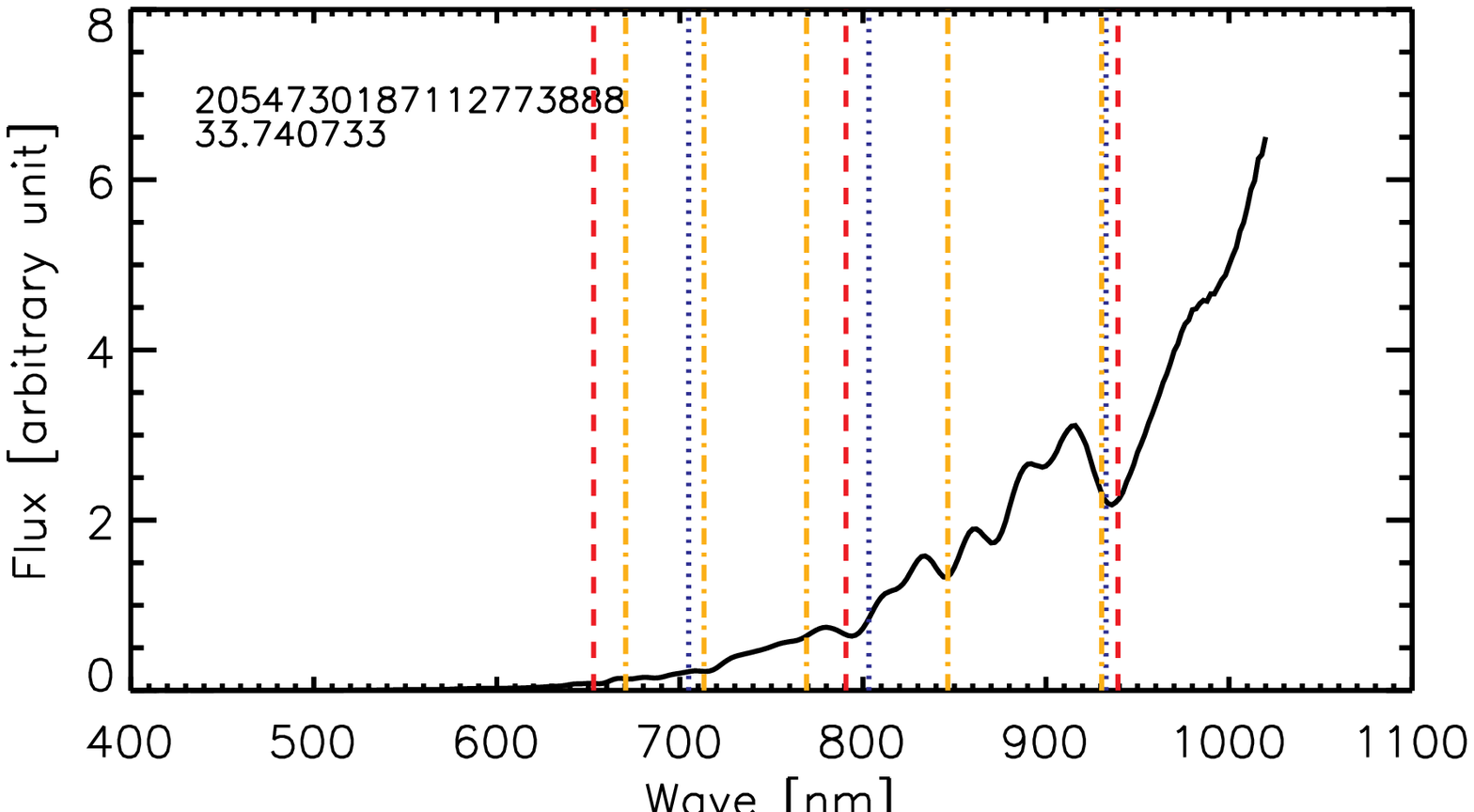}}
\resizebox{0.33\hsize}{!}{\includegraphics[angle=0]{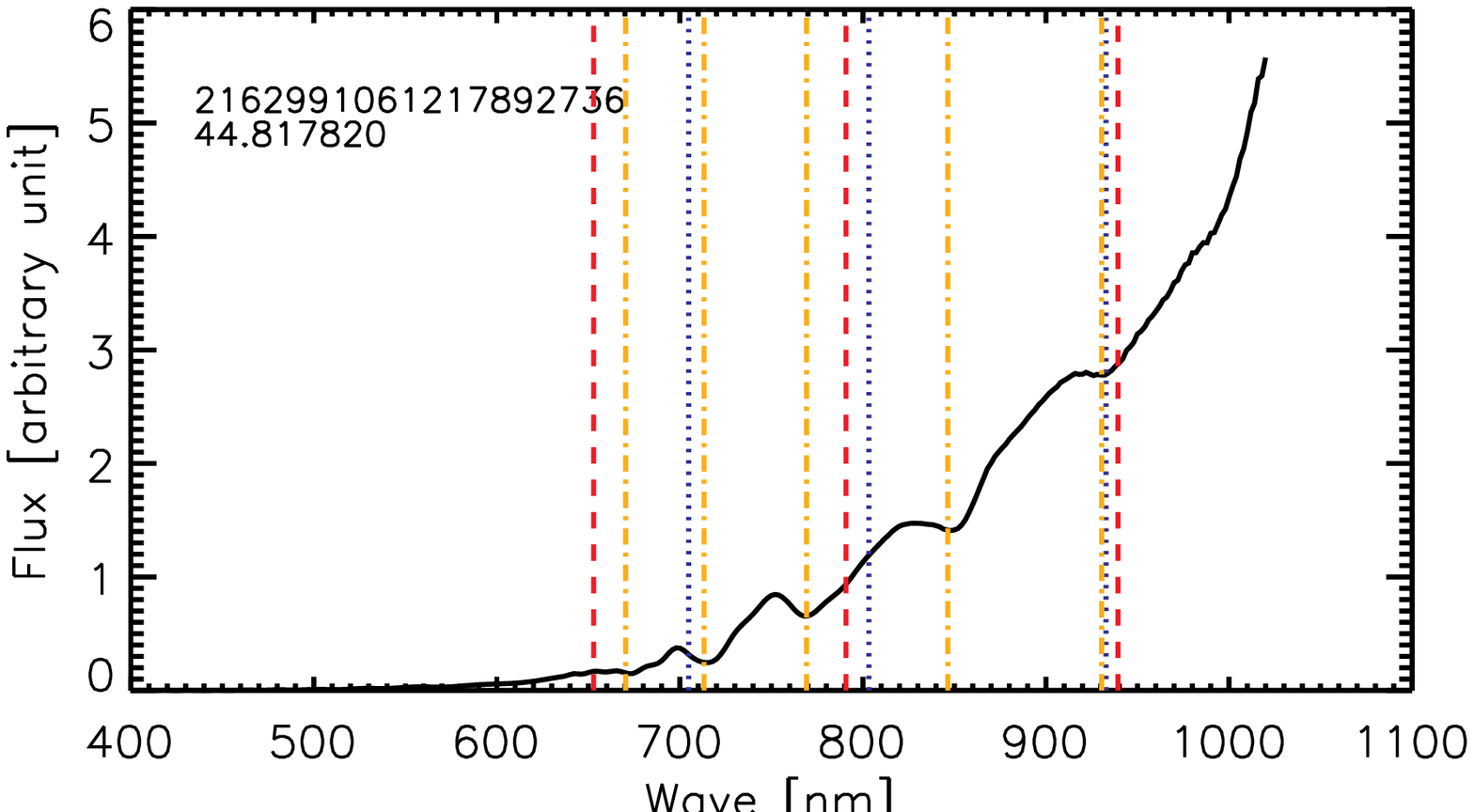}}
\resizebox{0.33\hsize}{!}{\includegraphics[angle=0]{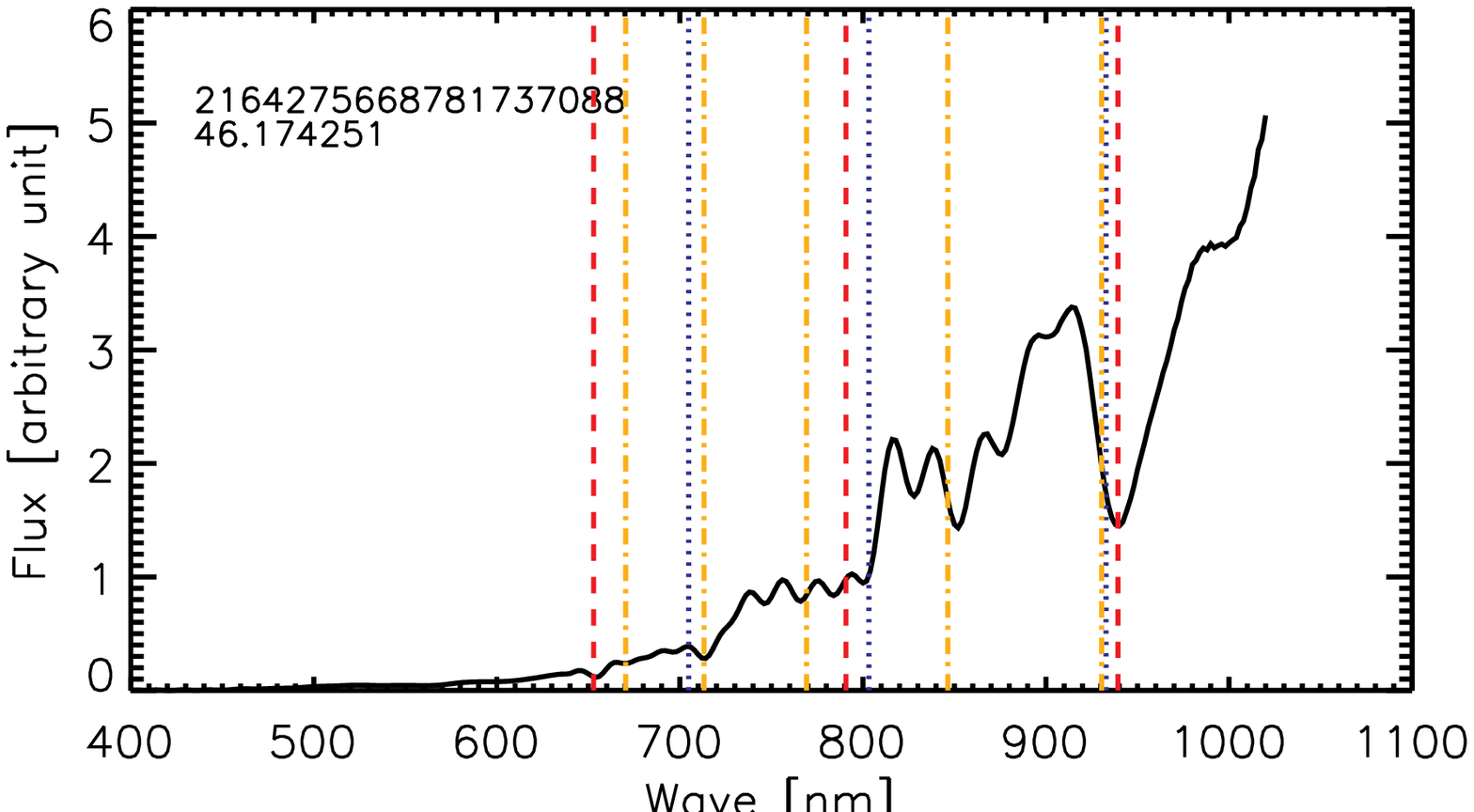}}
\resizebox{0.33\hsize}{!}{\includegraphics[angle=0]{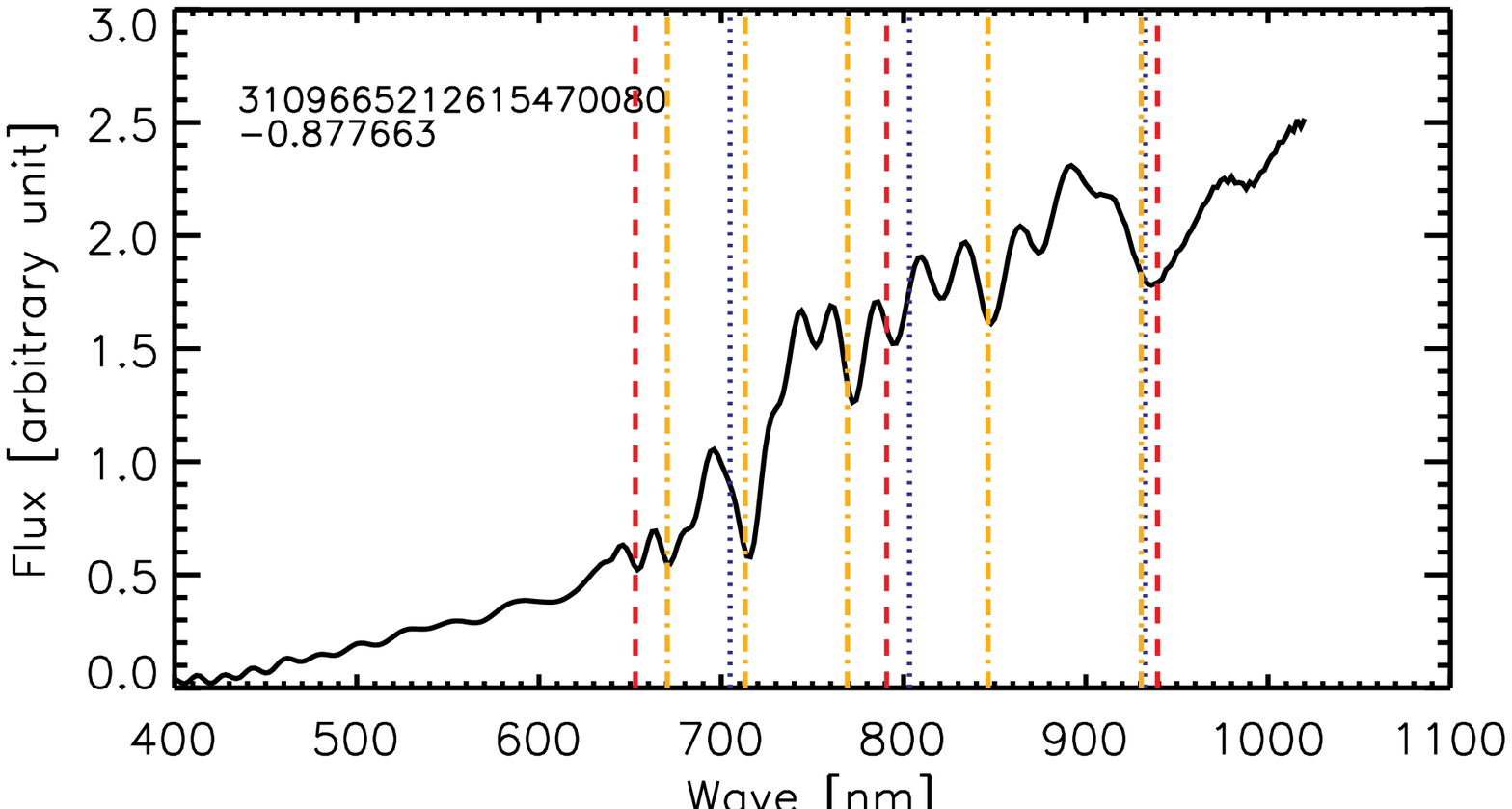}}
\resizebox{0.33\hsize}{!}{\includegraphics[angle=0]{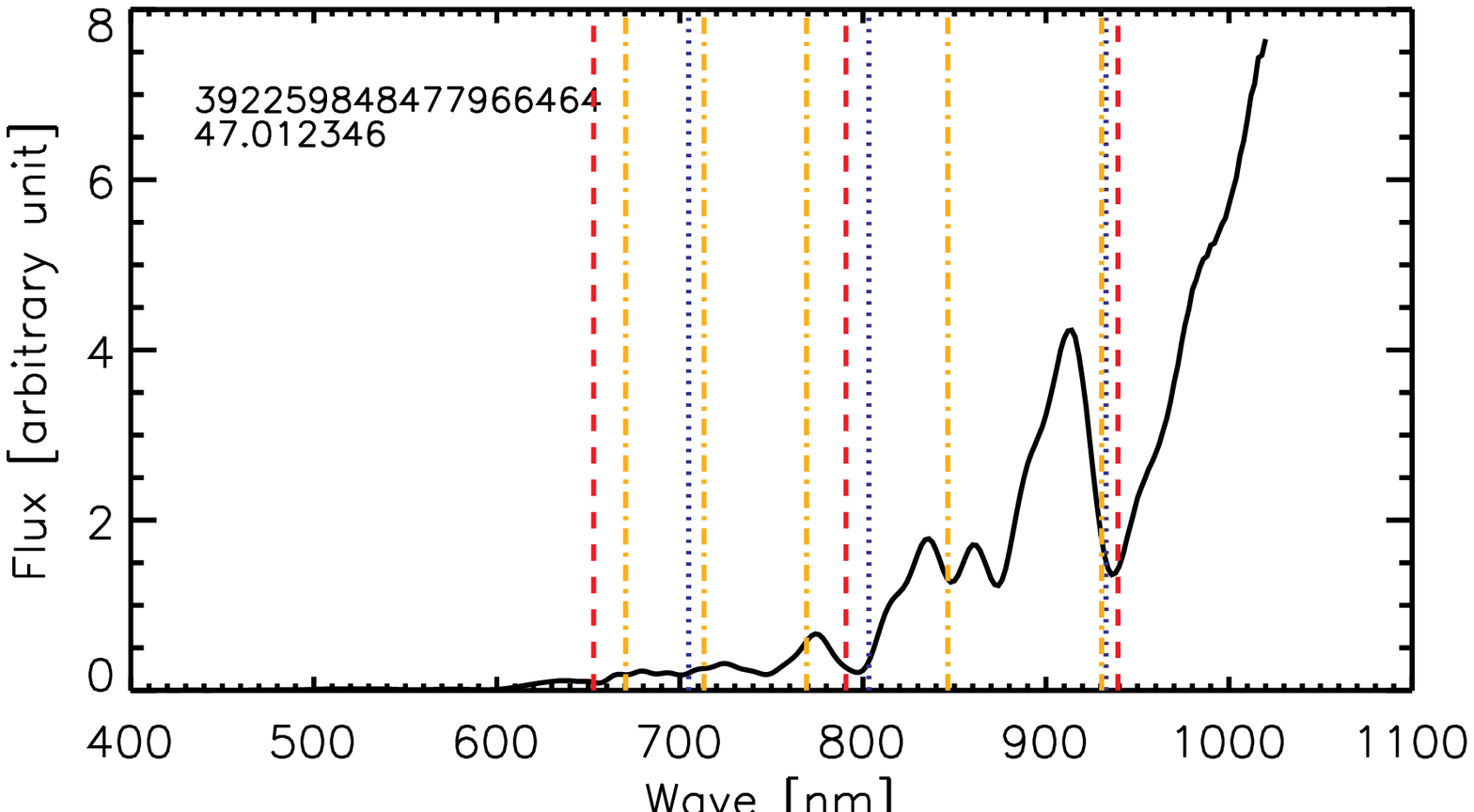}}
\resizebox{0.33\hsize}{!}{\includegraphics[angle=0]{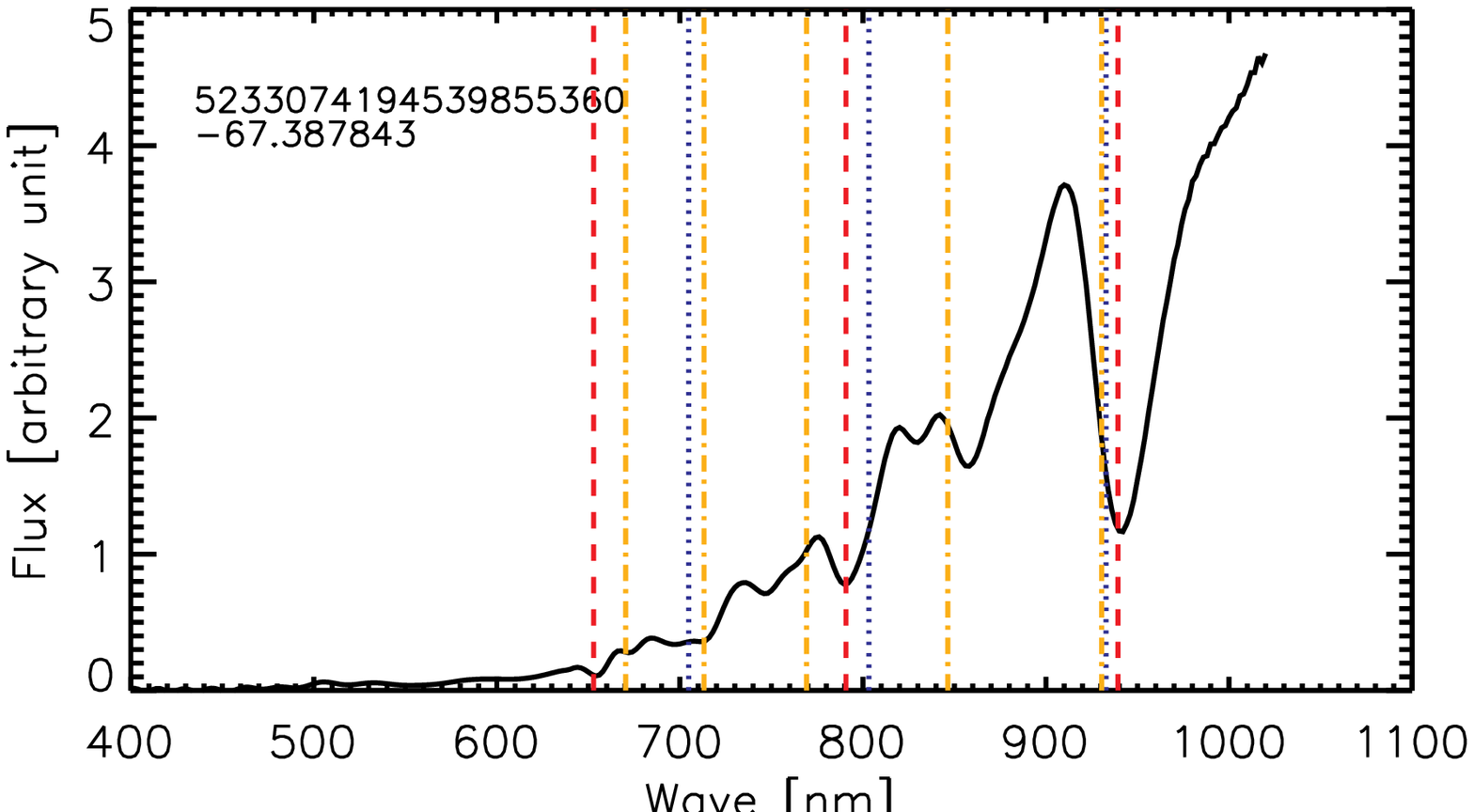}}
\resizebox{0.33\hsize}{!}{\includegraphics[angle=0]{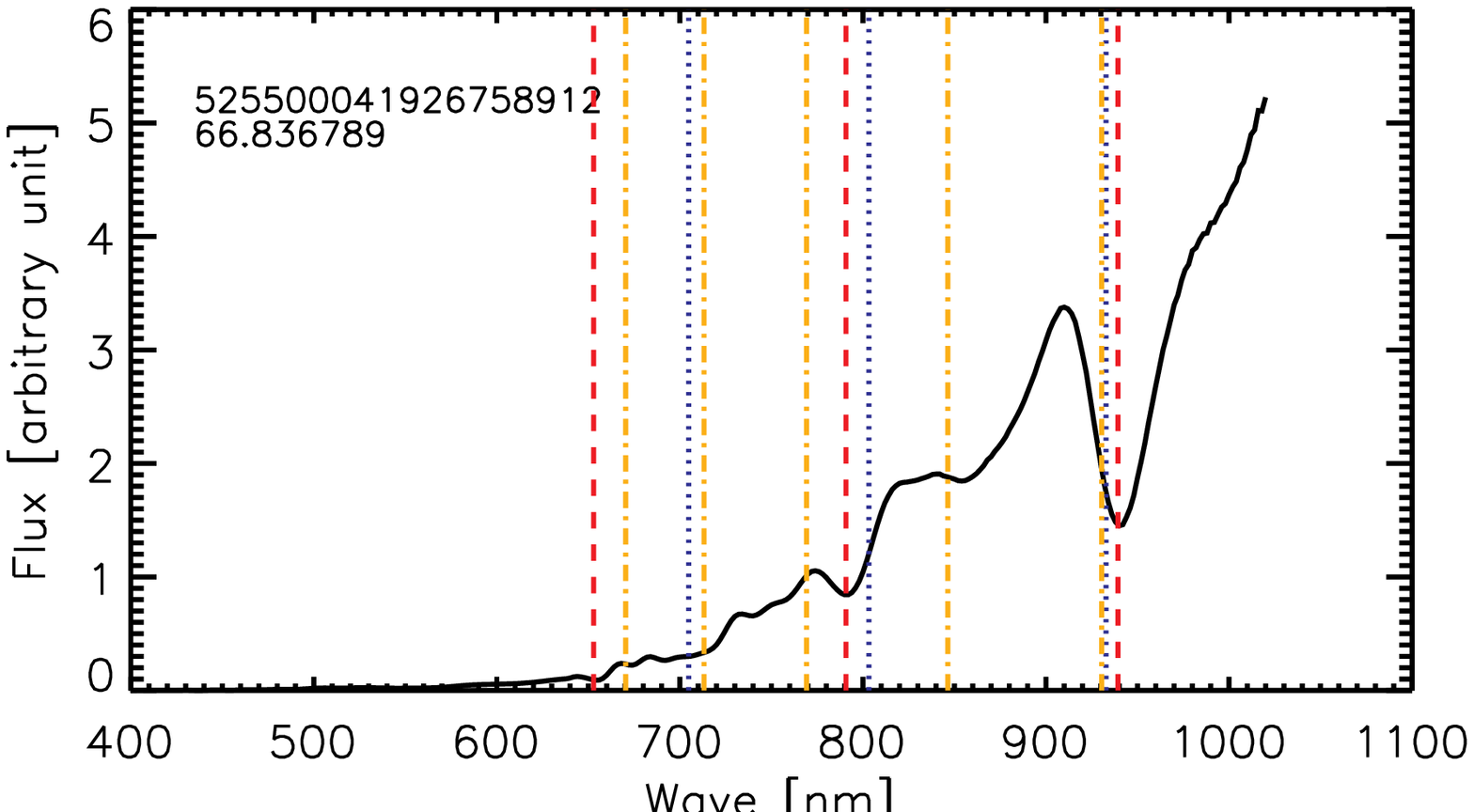}}

\resizebox{0.33\hsize}{!}{\includegraphics[angle=0]{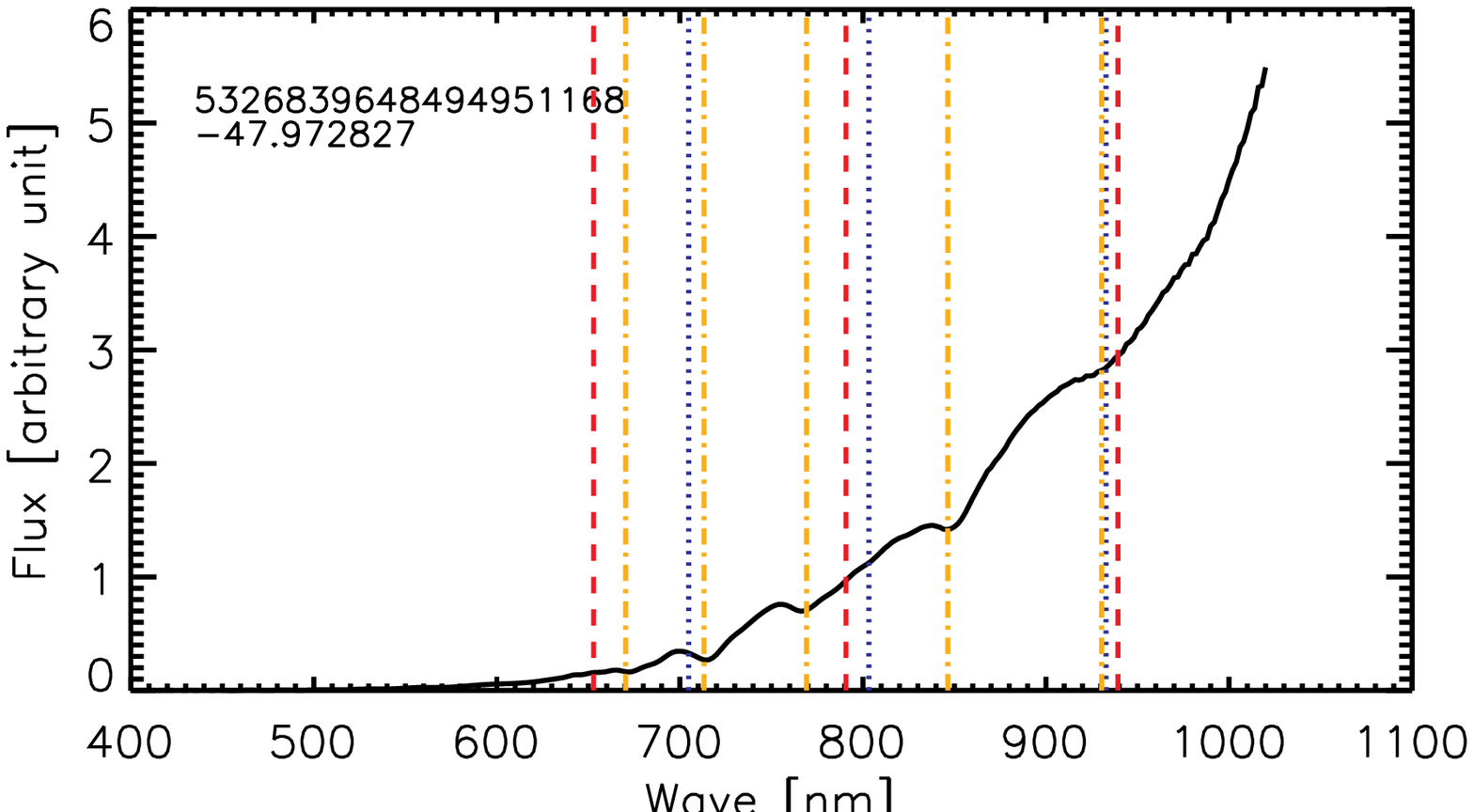}}
\resizebox{0.33\hsize}{!}{\includegraphics[angle=0]{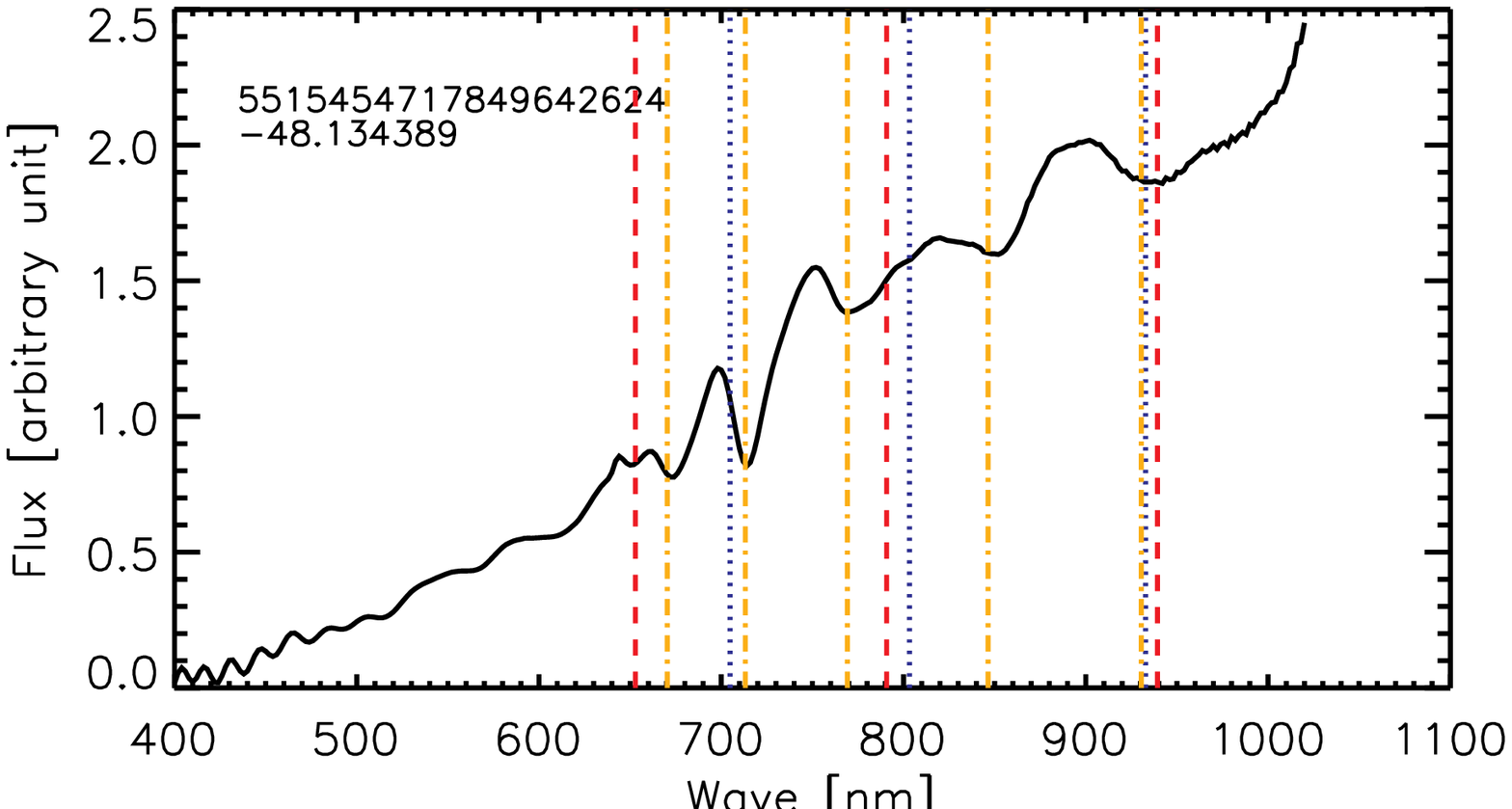}}
\resizebox{0.33\hsize}{!}{\includegraphics[angle=0]{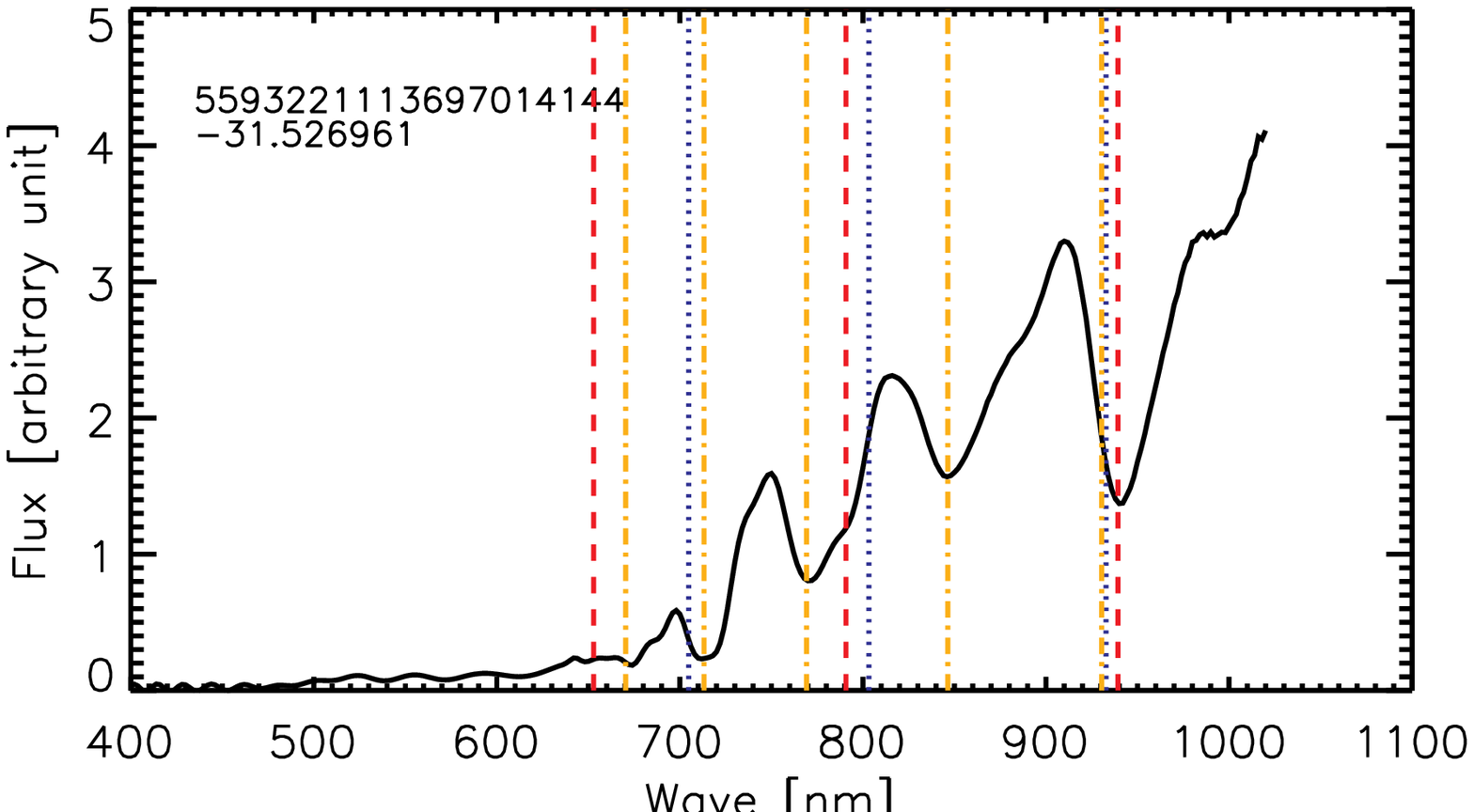}}
\resizebox{0.33\hsize}{!}{\includegraphics[angle=0]{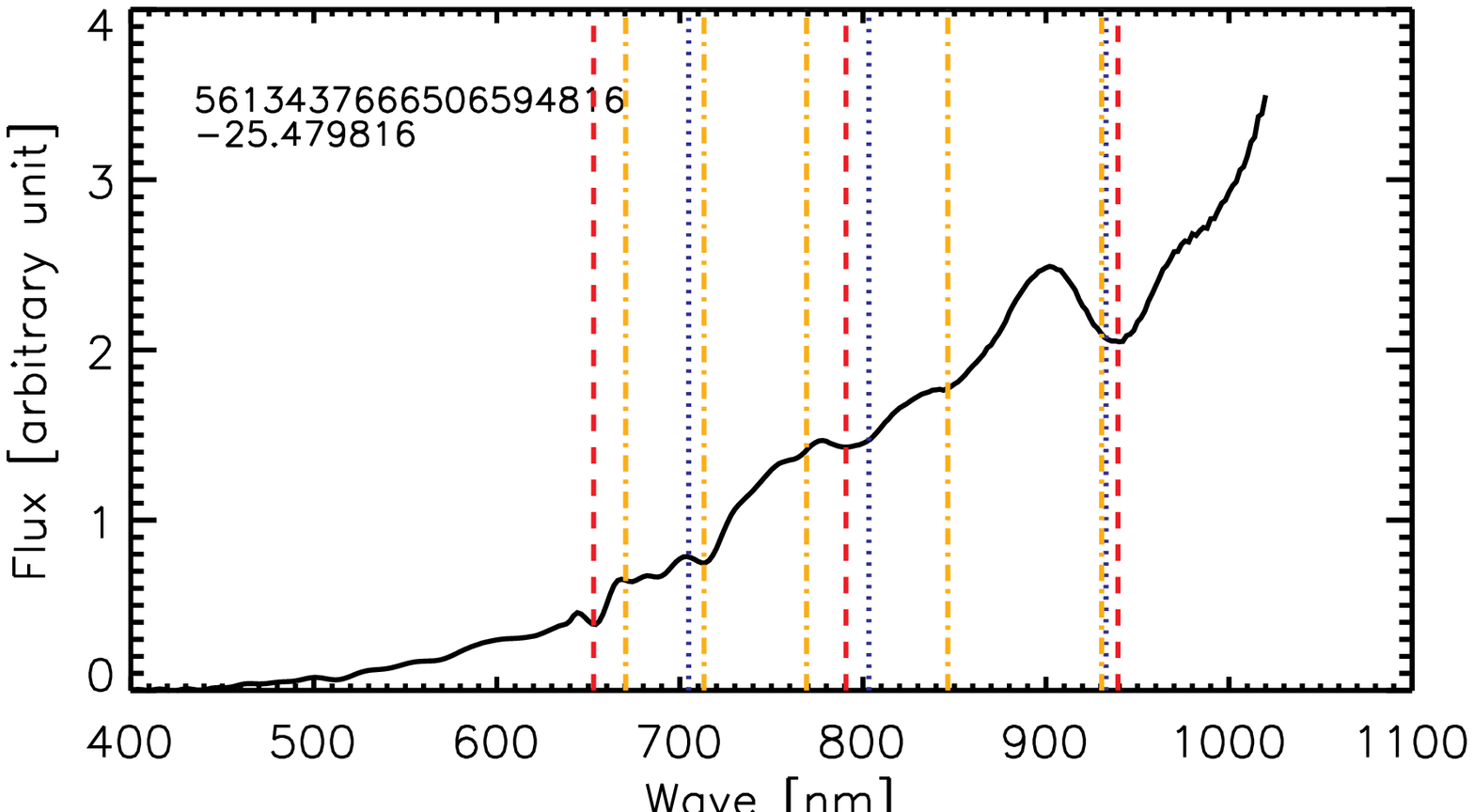}}
\resizebox{0.33\hsize}{!}{\includegraphics[angle=0]{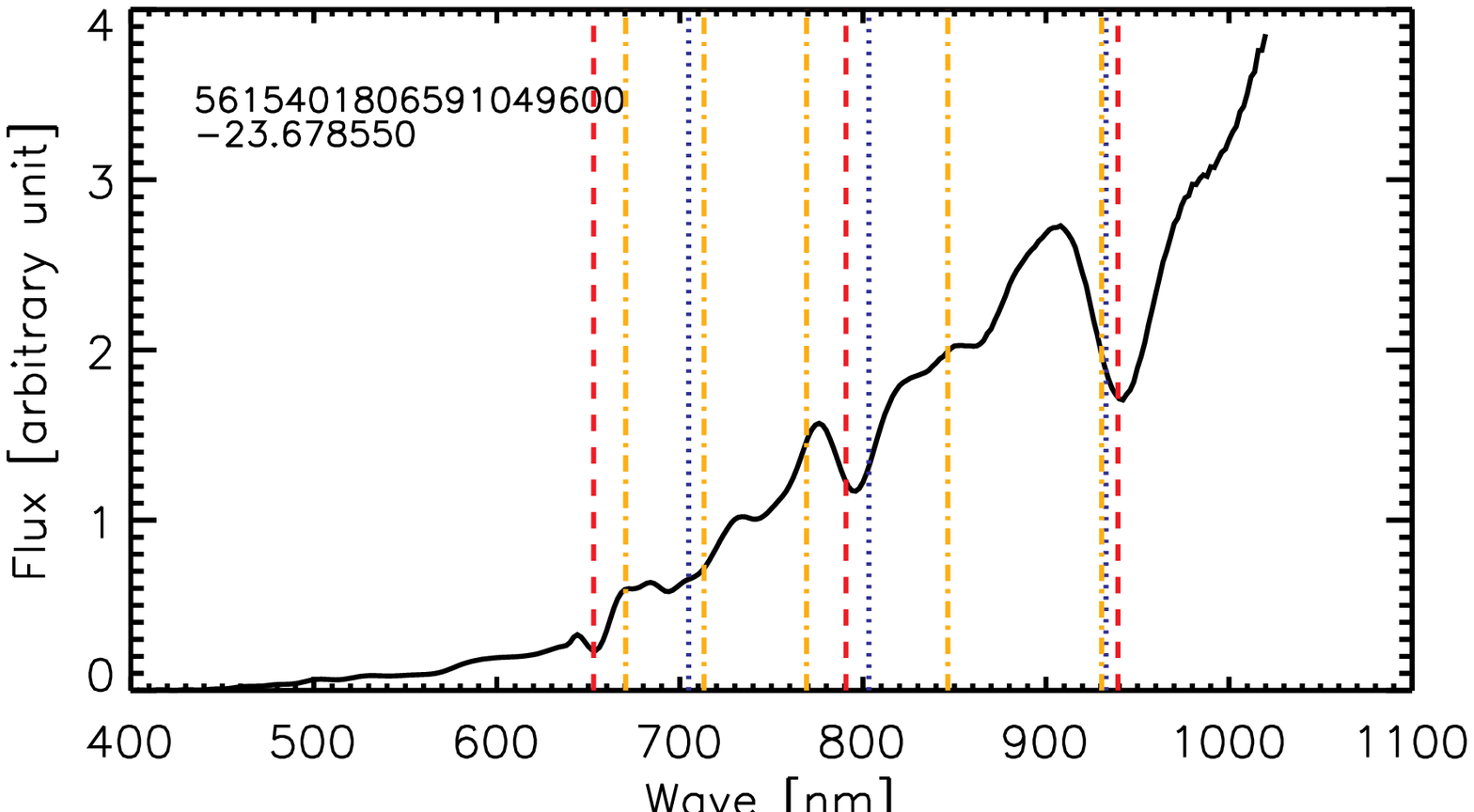}}
\resizebox{0.33\hsize}{!}{\includegraphics[angle=0]{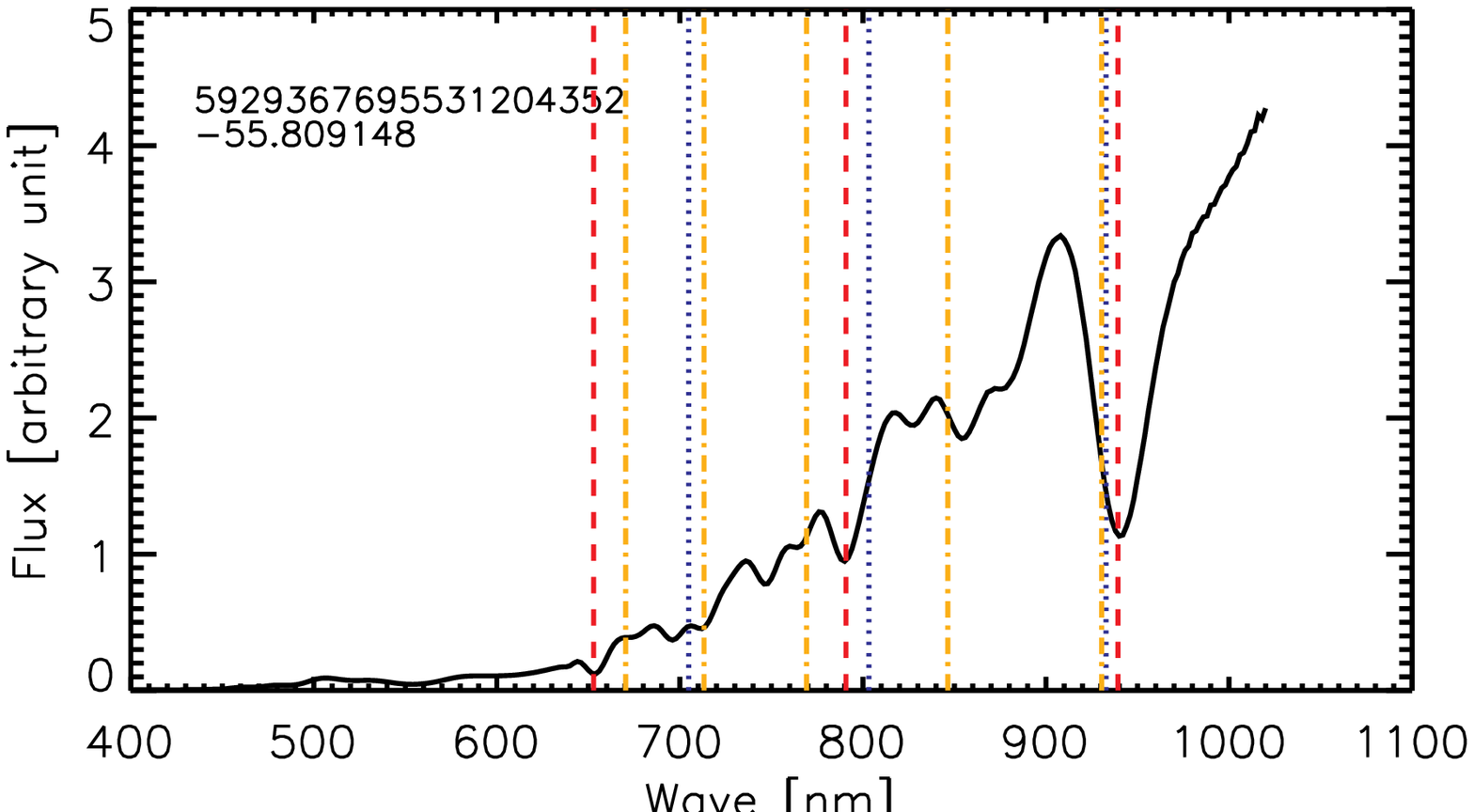}}
\resizebox{0.33\hsize}{!}{\includegraphics[angle=0]{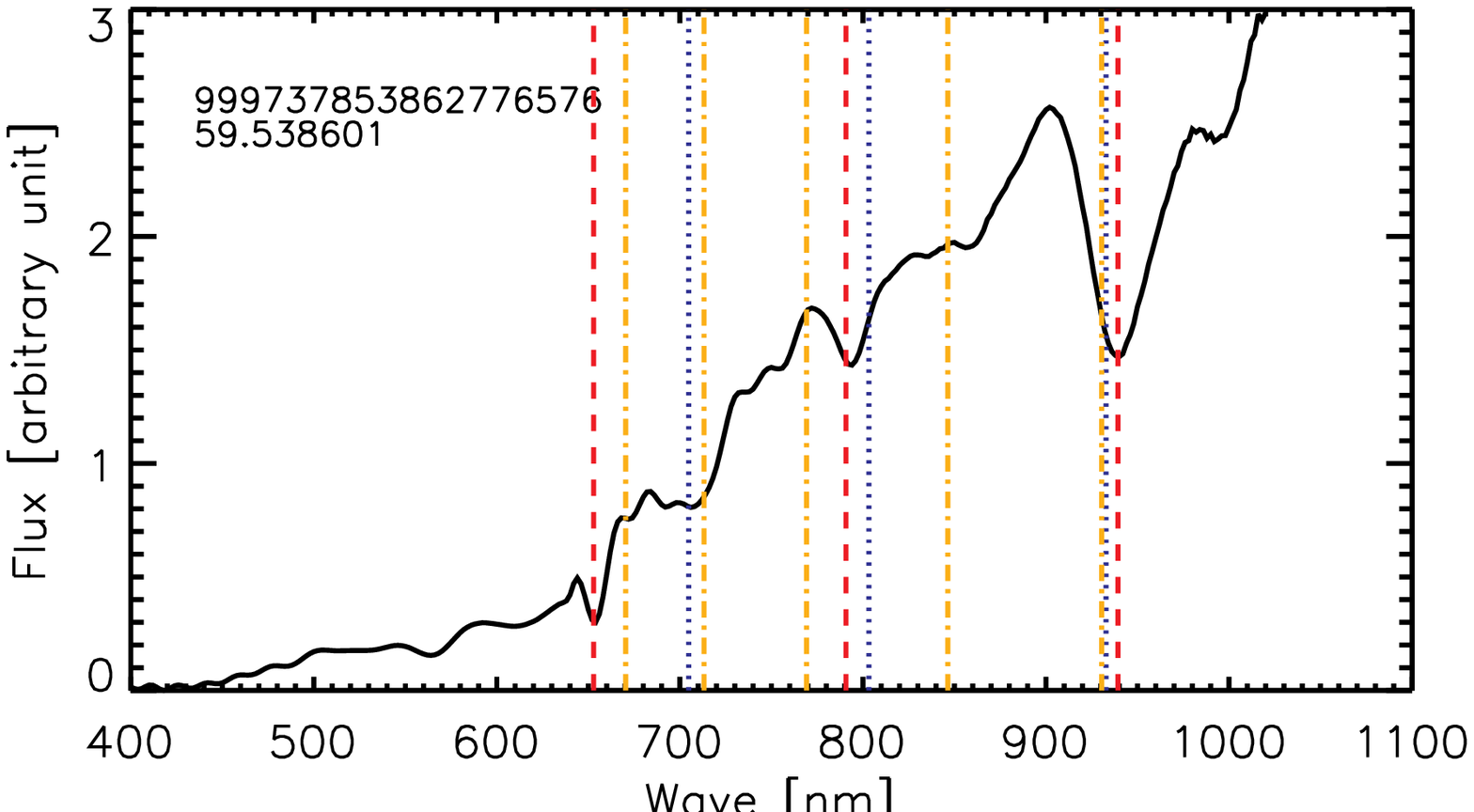}}
\resizebox{0.33\hsize}{!}{\includegraphics[angle=0]{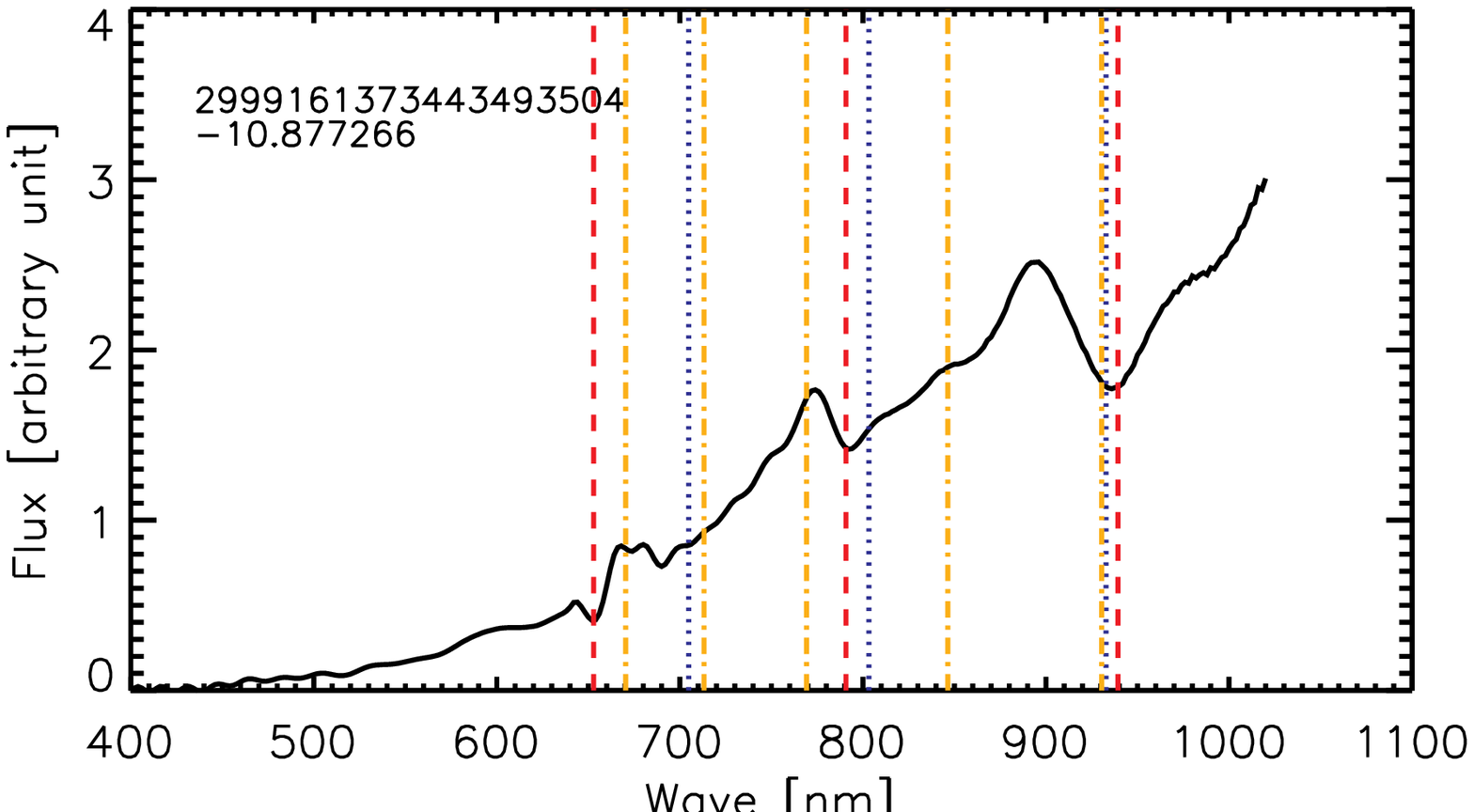}}
\end{center}
\caption{\label{Sstars} Gaia BP/RP spectra of  S-type stars in the
new selected  203 bright late-type stars in Sect. \ref{detection}.
The vertical red long-dashed lines  mark the locations of the three 
broad absorptions
seen in S-rich stars.
For comparison, the centroids of the absorption seen in C-type (blue) 
and  M1-M3 O-rich stars (orange) are also shown.
} 
\end{figure*}

\begin{figure*}
\begin{center}
\resizebox{0.33\hsize}{!}{\includegraphics[angle=0]{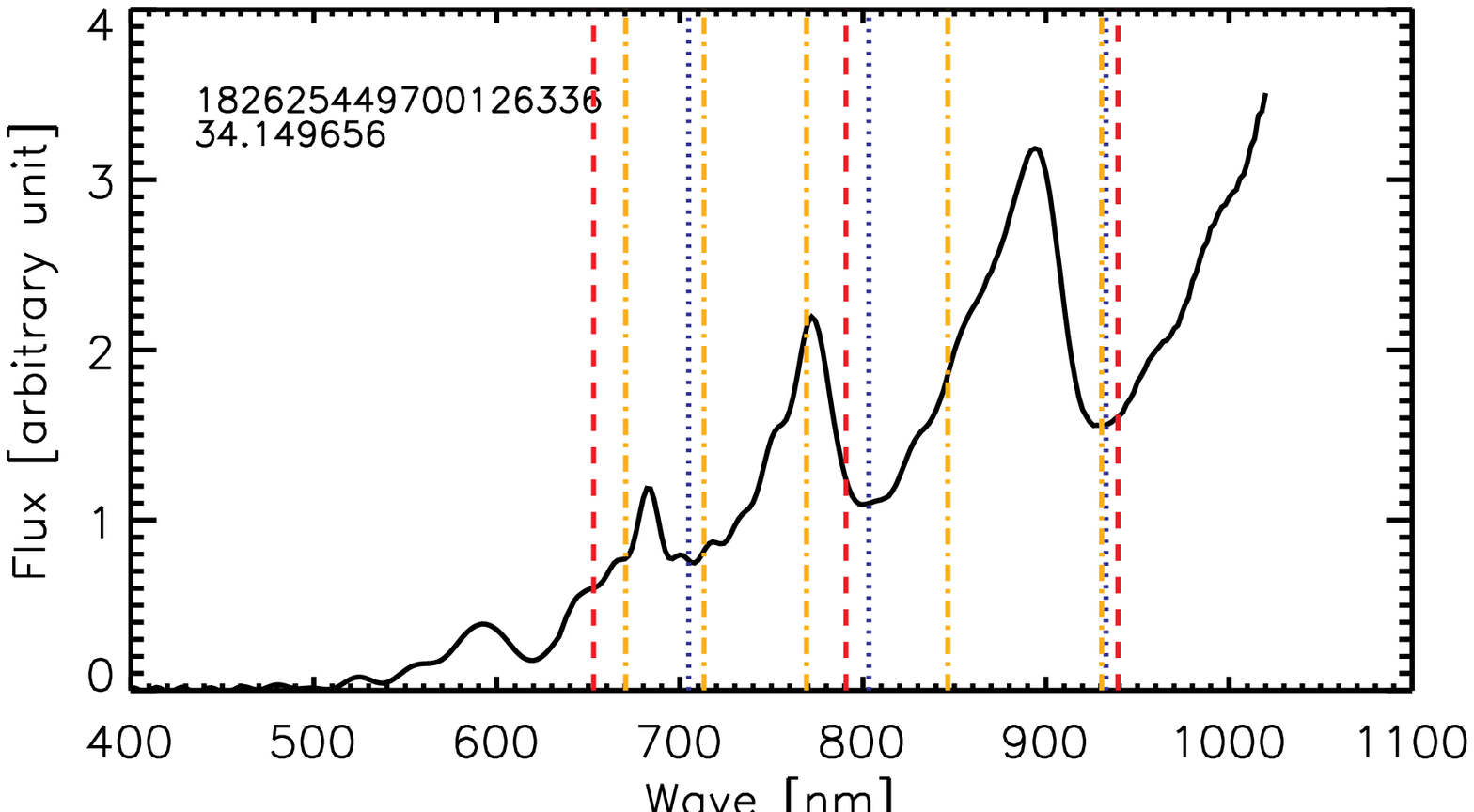}}
\resizebox{0.33\hsize}{!}{\includegraphics[angle=0]{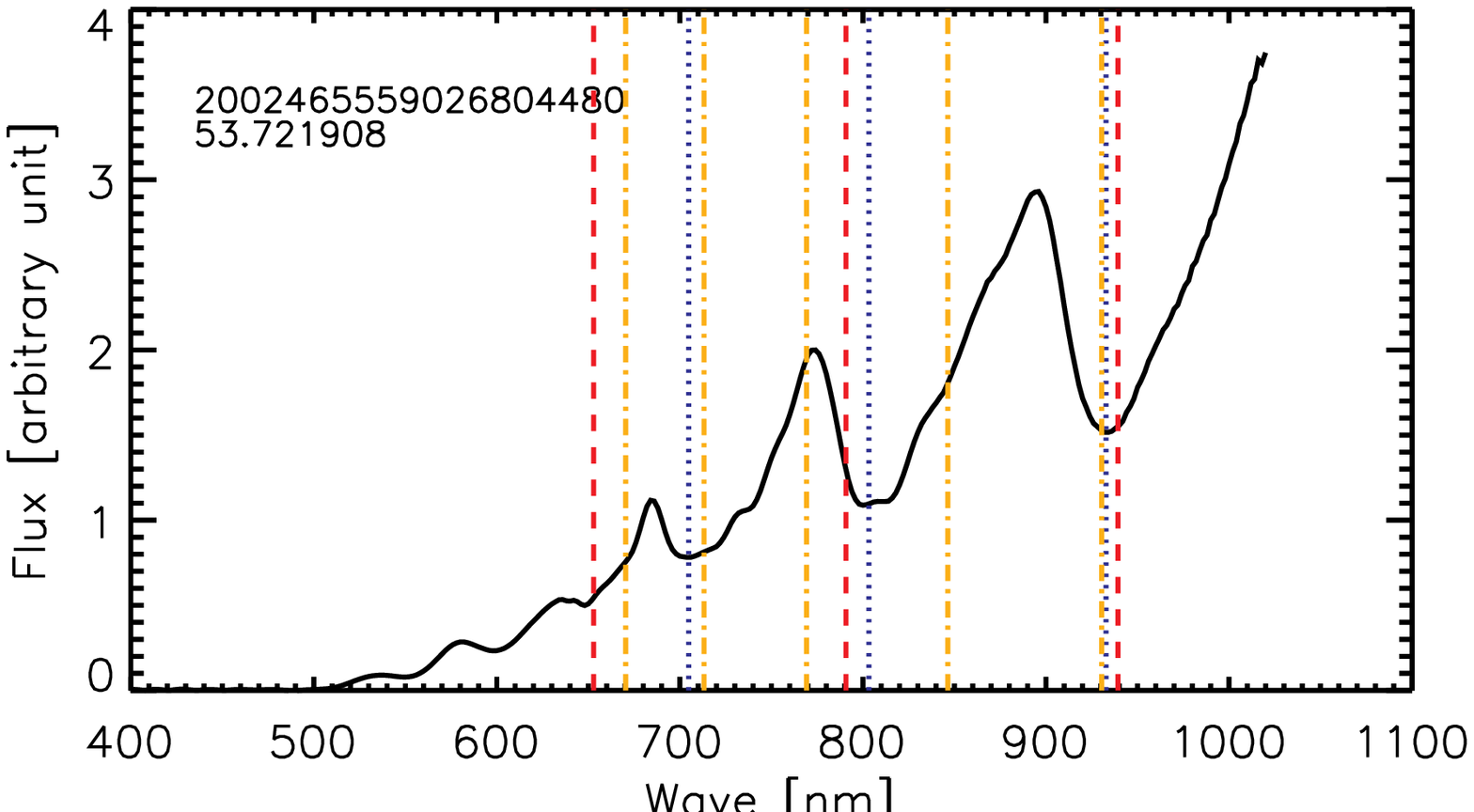}}
\resizebox{0.33\hsize}{!}{\includegraphics[angle=0]{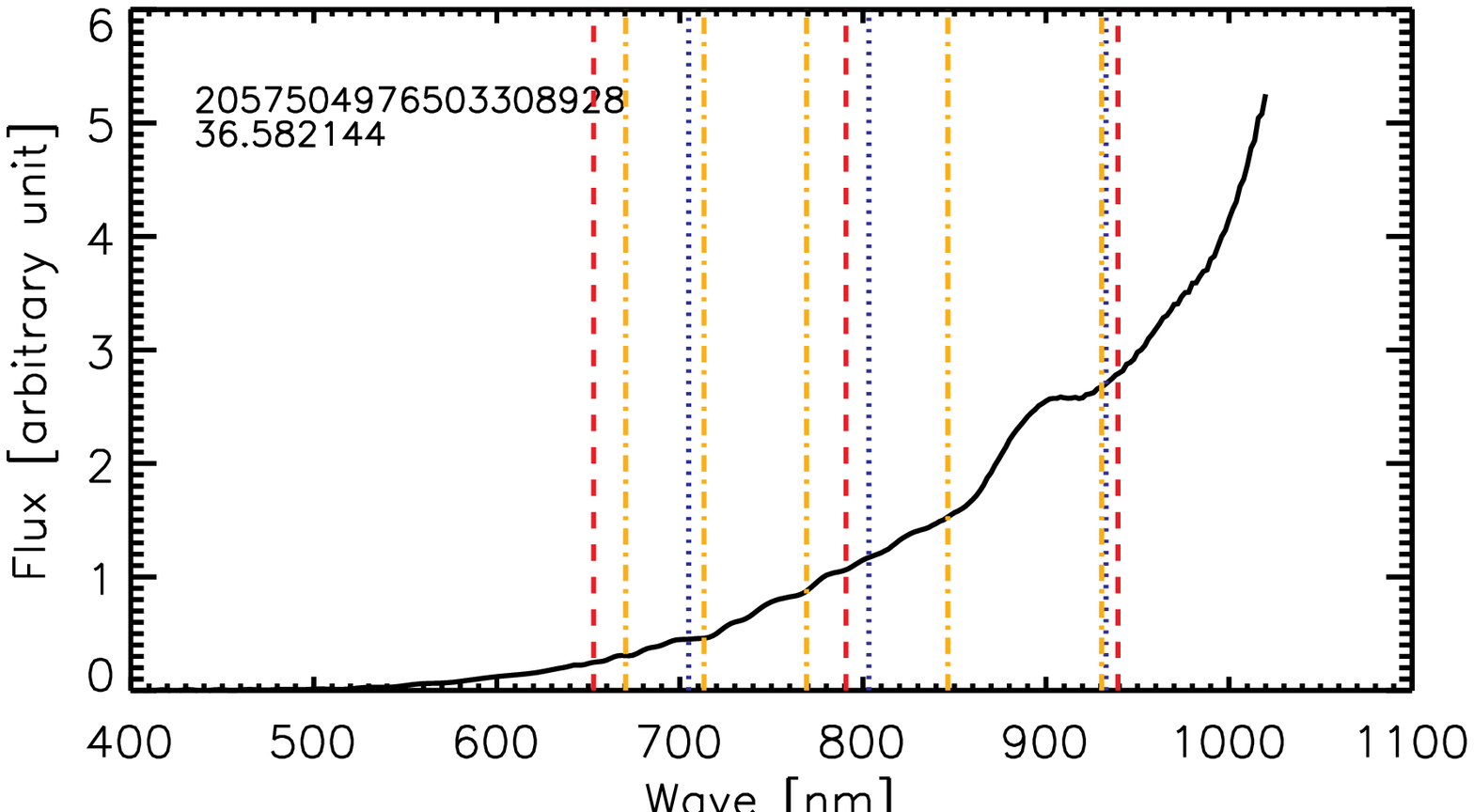}}
\resizebox{0.33\hsize}{!}{\includegraphics[angle=0]{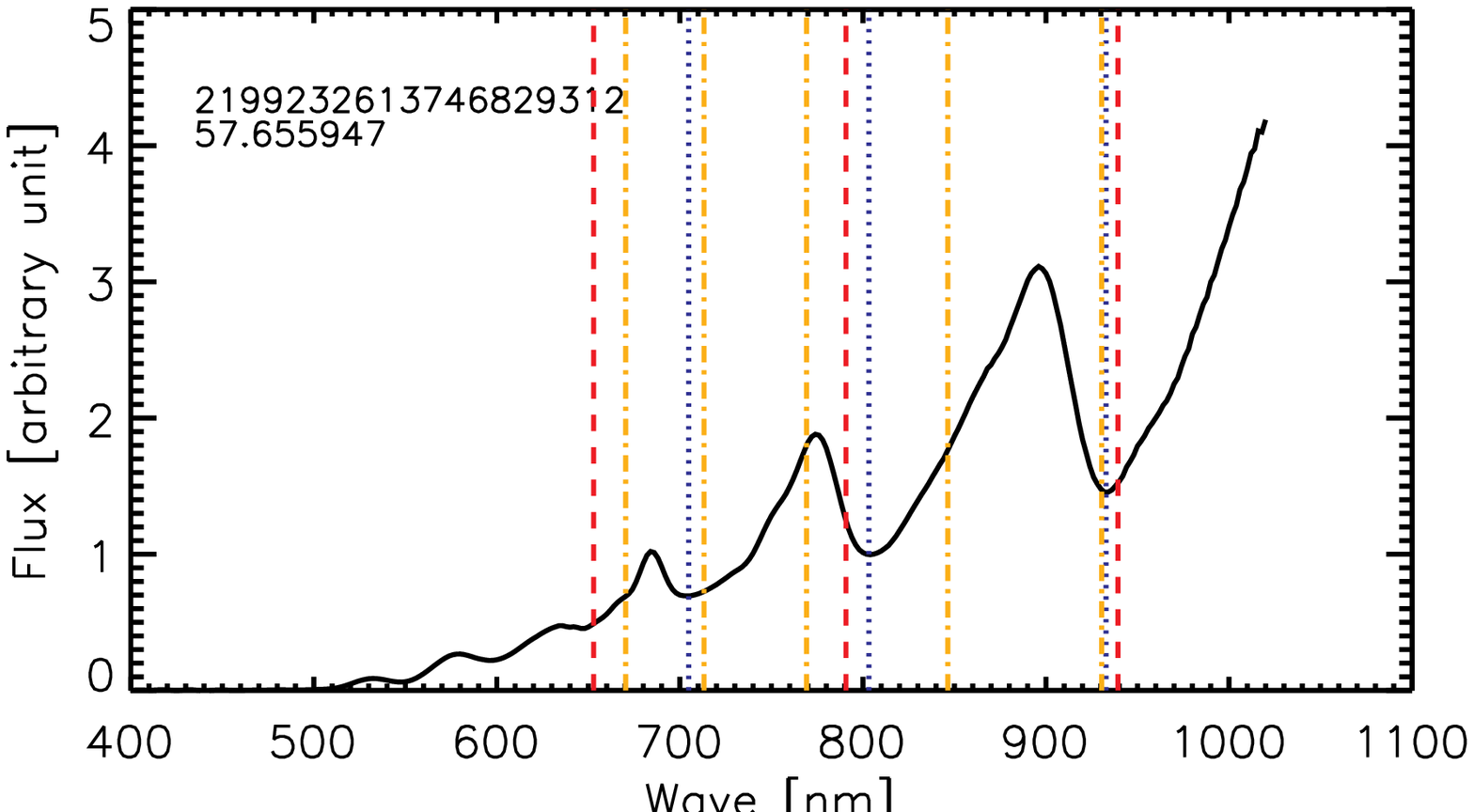}}
\resizebox{0.33\hsize}{!}{\includegraphics[angle=0]{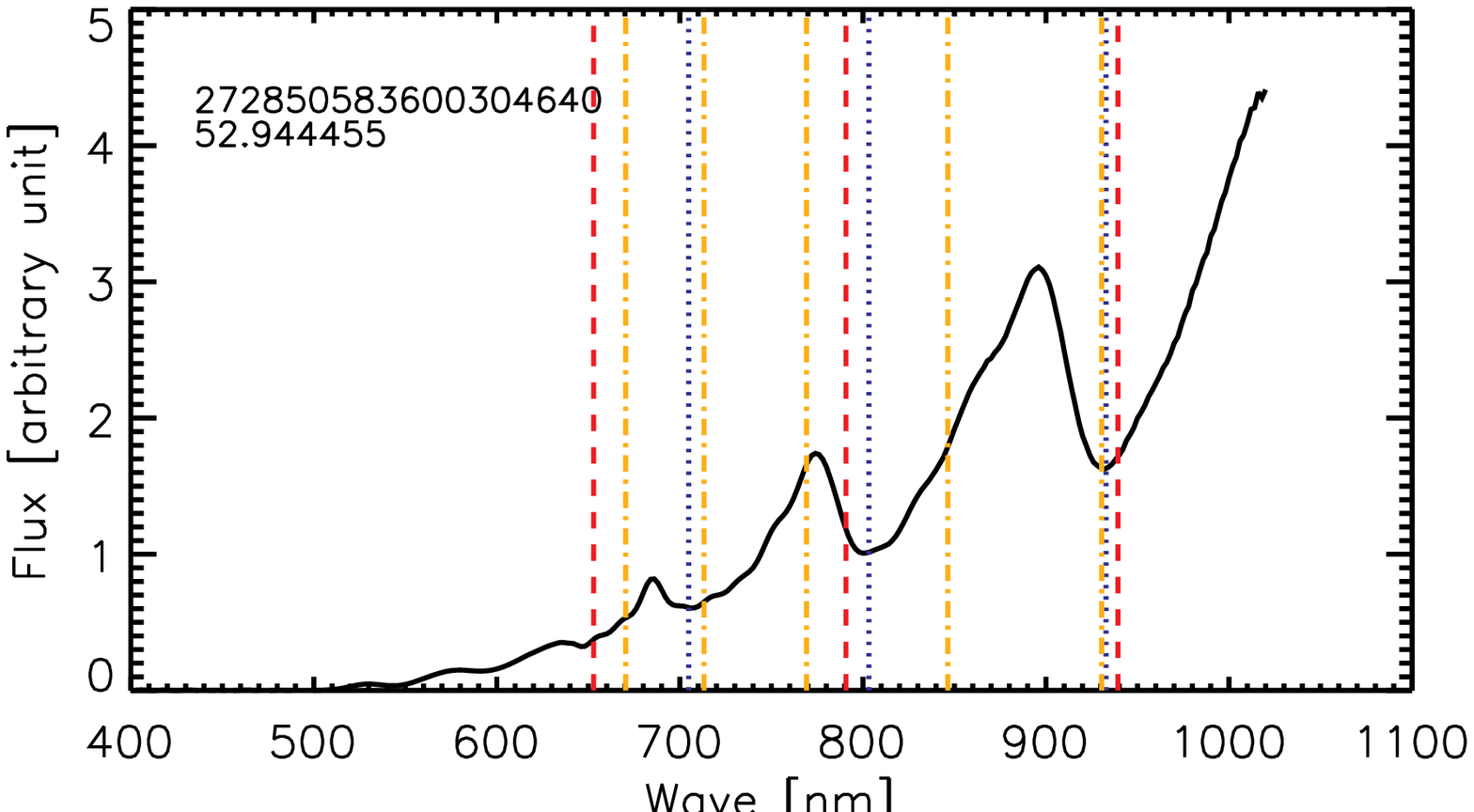}}
\resizebox{0.33\hsize}{!}{\includegraphics[angle=0]{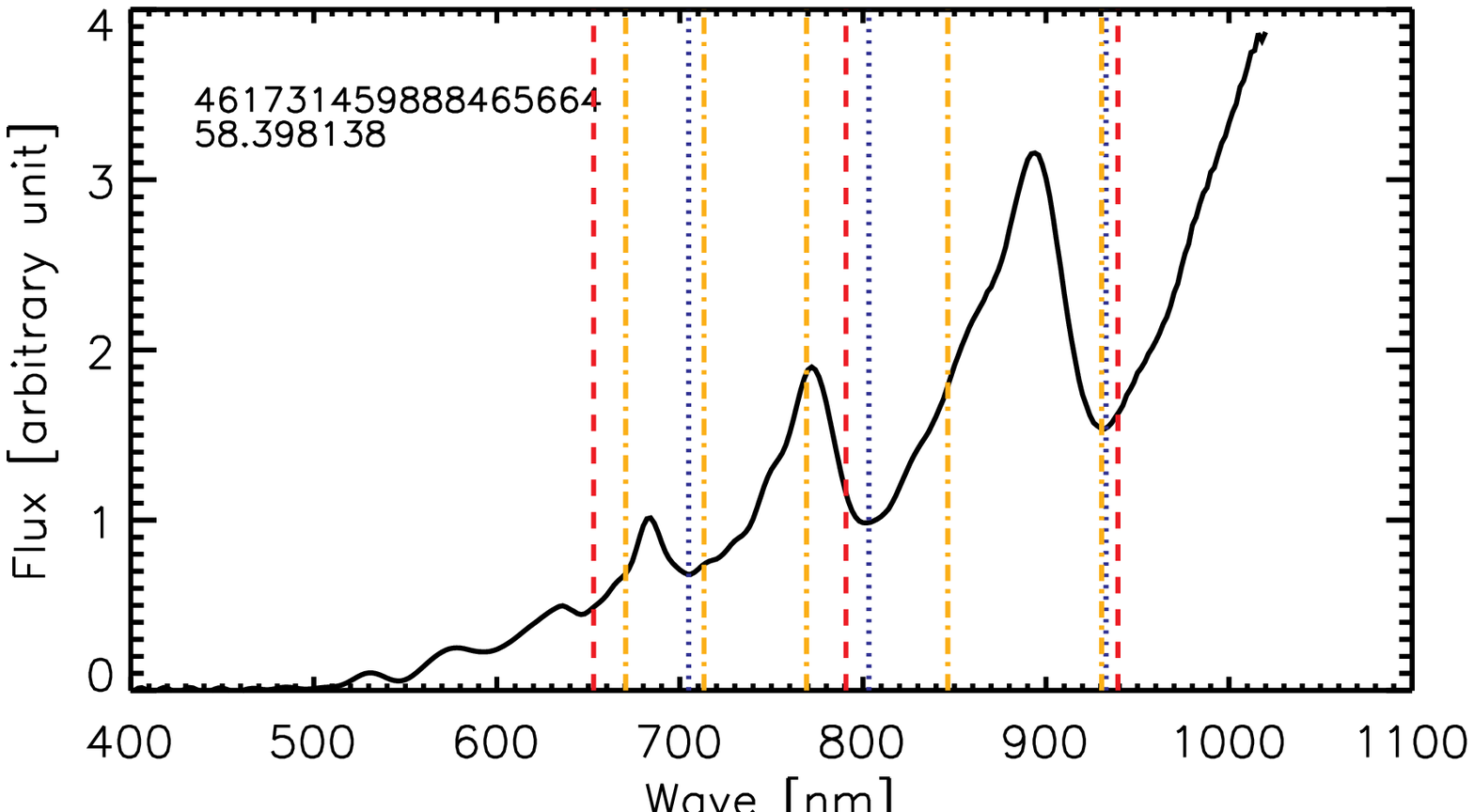}}
\resizebox{0.33\hsize}{!}{\includegraphics[angle=0]{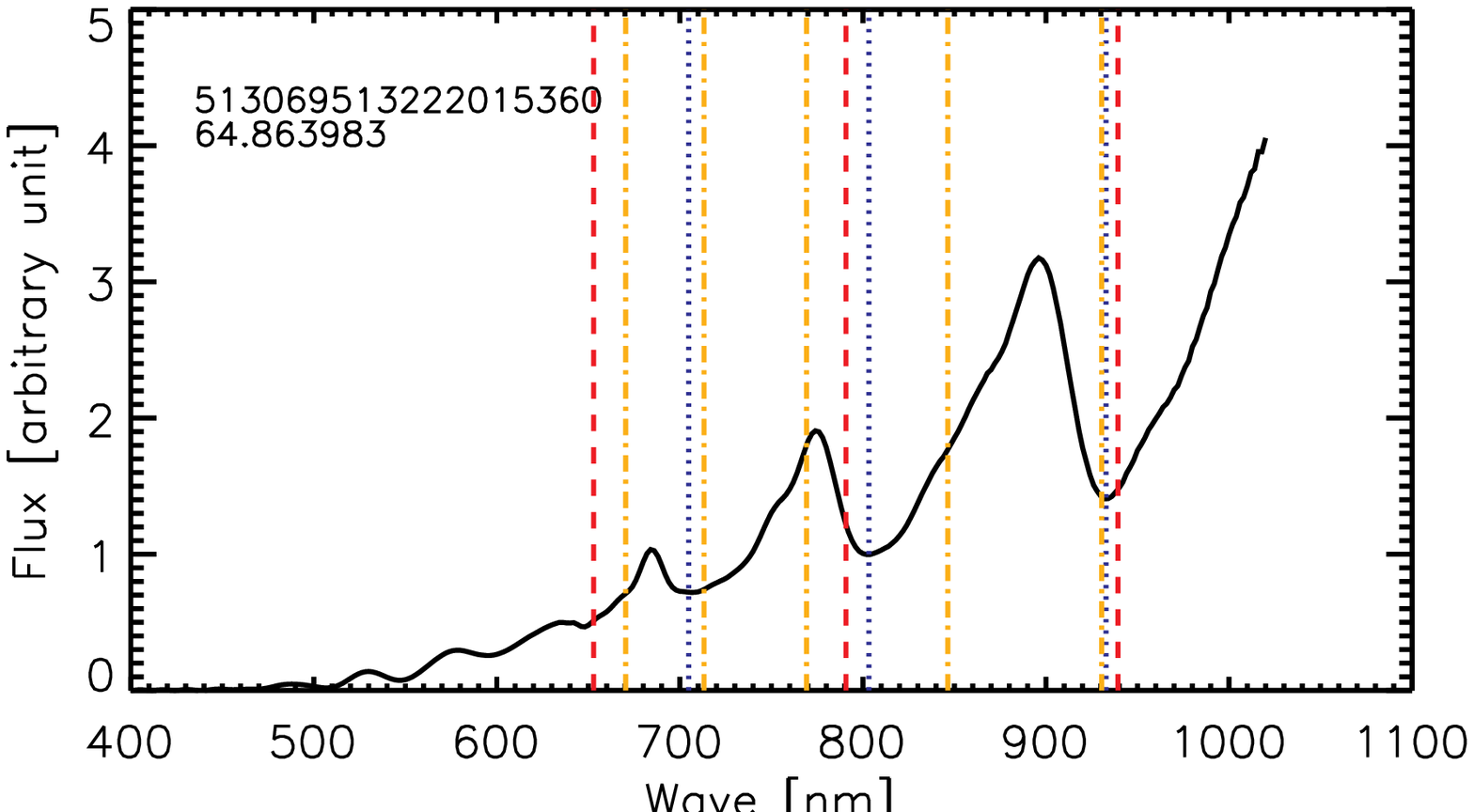}}
\resizebox{0.33\hsize}{!}{\includegraphics[angle=0]{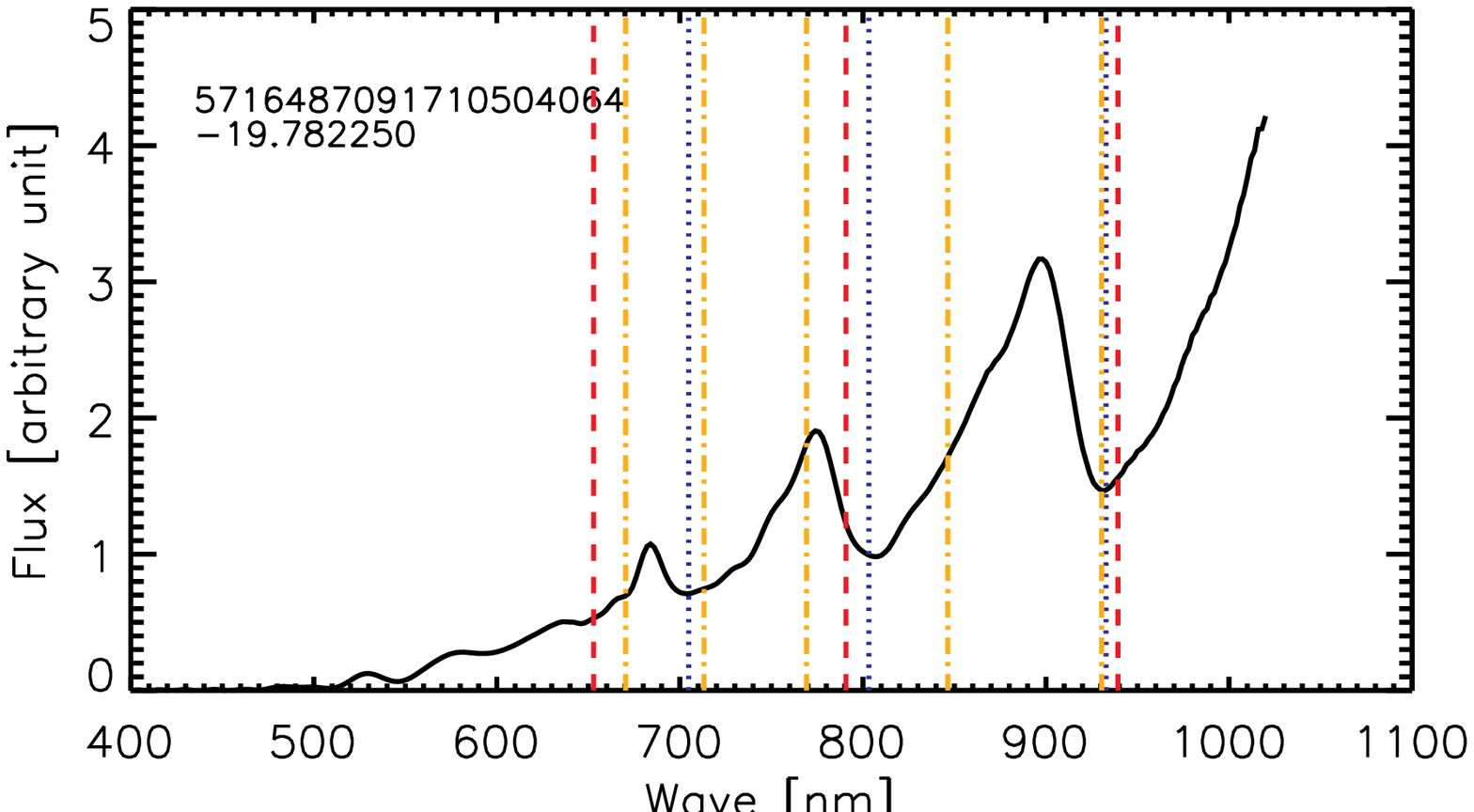}}
\resizebox{0.33\hsize}{!}{\includegraphics[angle=0]{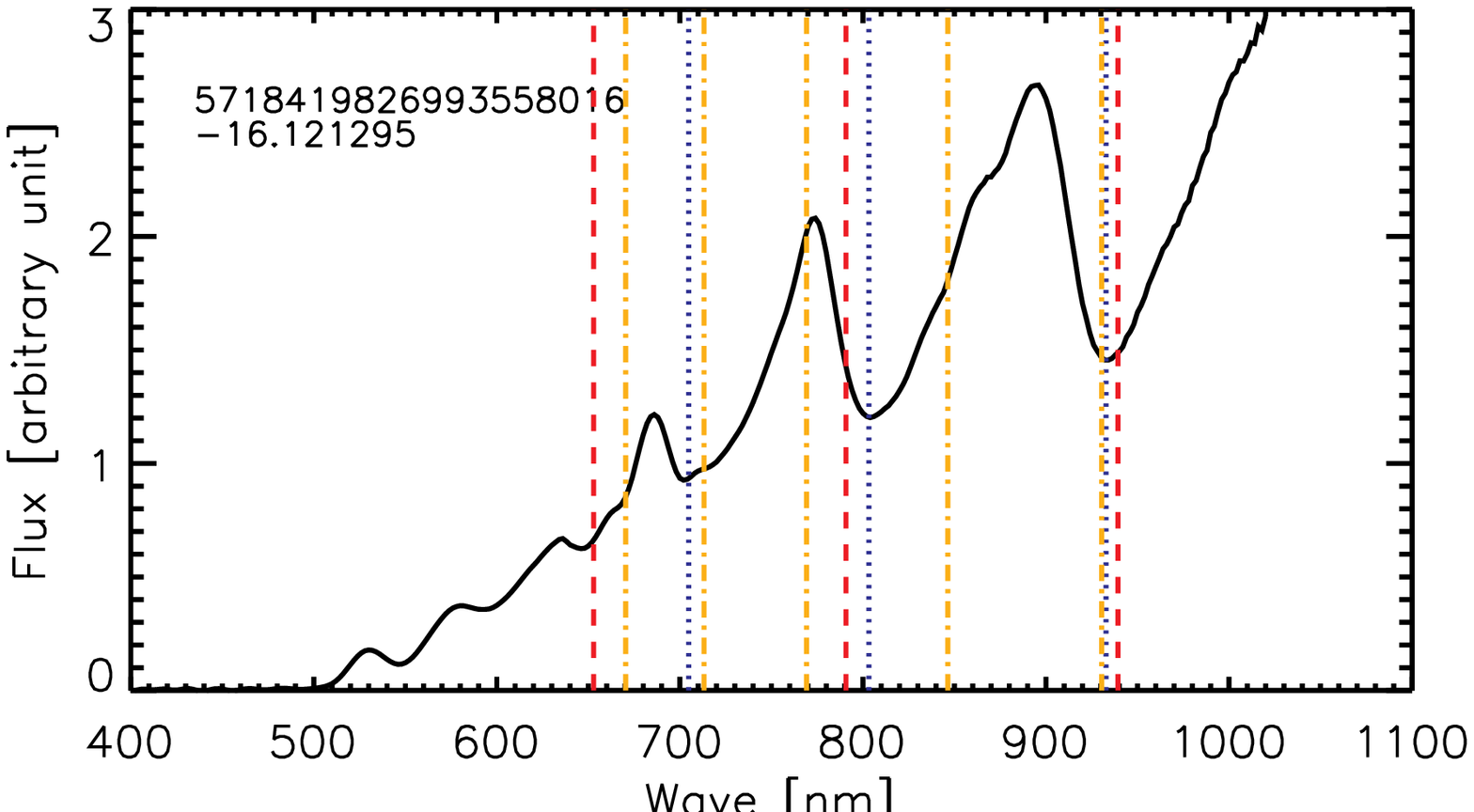}}
\resizebox{0.33\hsize}{!}{\includegraphics[angle=0]{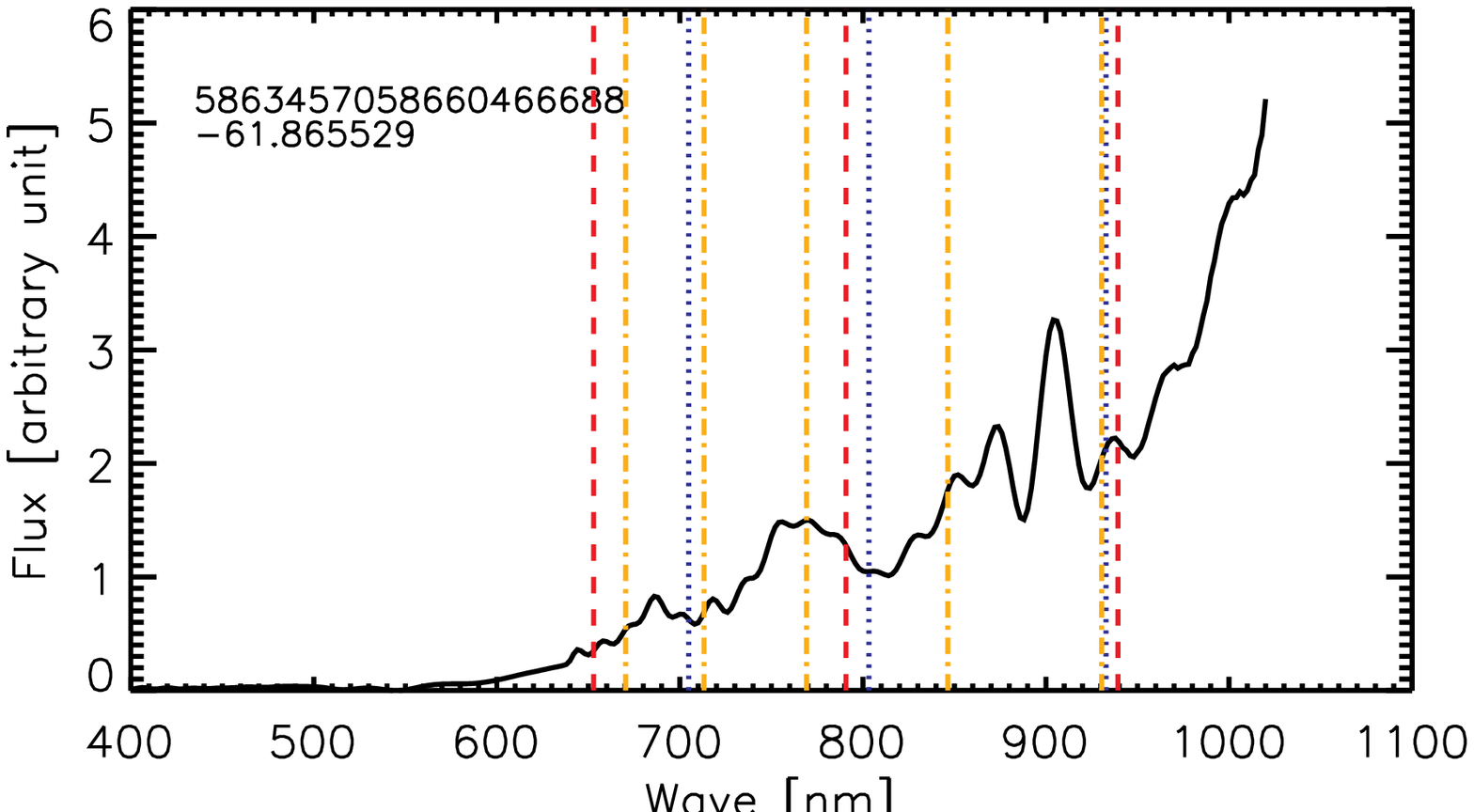}}
\end{center}
\caption{\label{Cstars} Gaia BP/RP spectra of known C-rich stars in the
new selected 203 bright late-type stars in Sect. \ref{detection}.
The vertical blue dotted lines  mark the locations of the main absorption lines
seen in C-rich stars.
For comparison, the centroids of the absorption seen in S-type (red) and  
 M1-M3 O-rich stars
(orange) are also shown.} 
\end{figure*}

In Figures   \ref{Sstars} and \ref{Cstars}, the BP/RP spectra of 
C-rich stars and S-type stars found among the 203 newly 
selected bright late-type stars
in Sect. \ref{detection}  are displayed.

\begin{figure*}
\begin{center}
\resizebox{0.33\hsize}{!}{\includegraphics[angle=0]{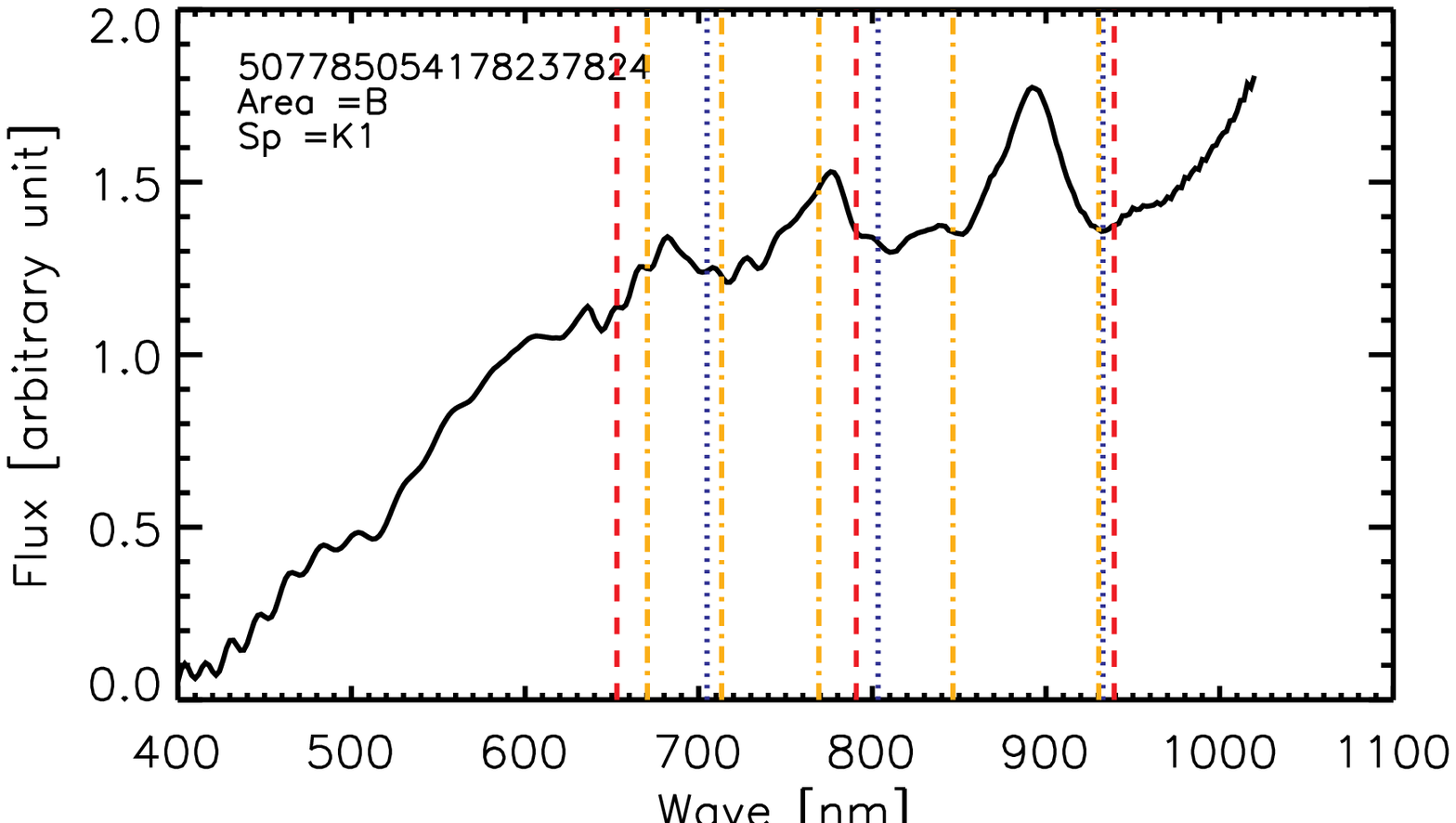}}
\resizebox{0.33\hsize}{!}{\includegraphics[angle=0]{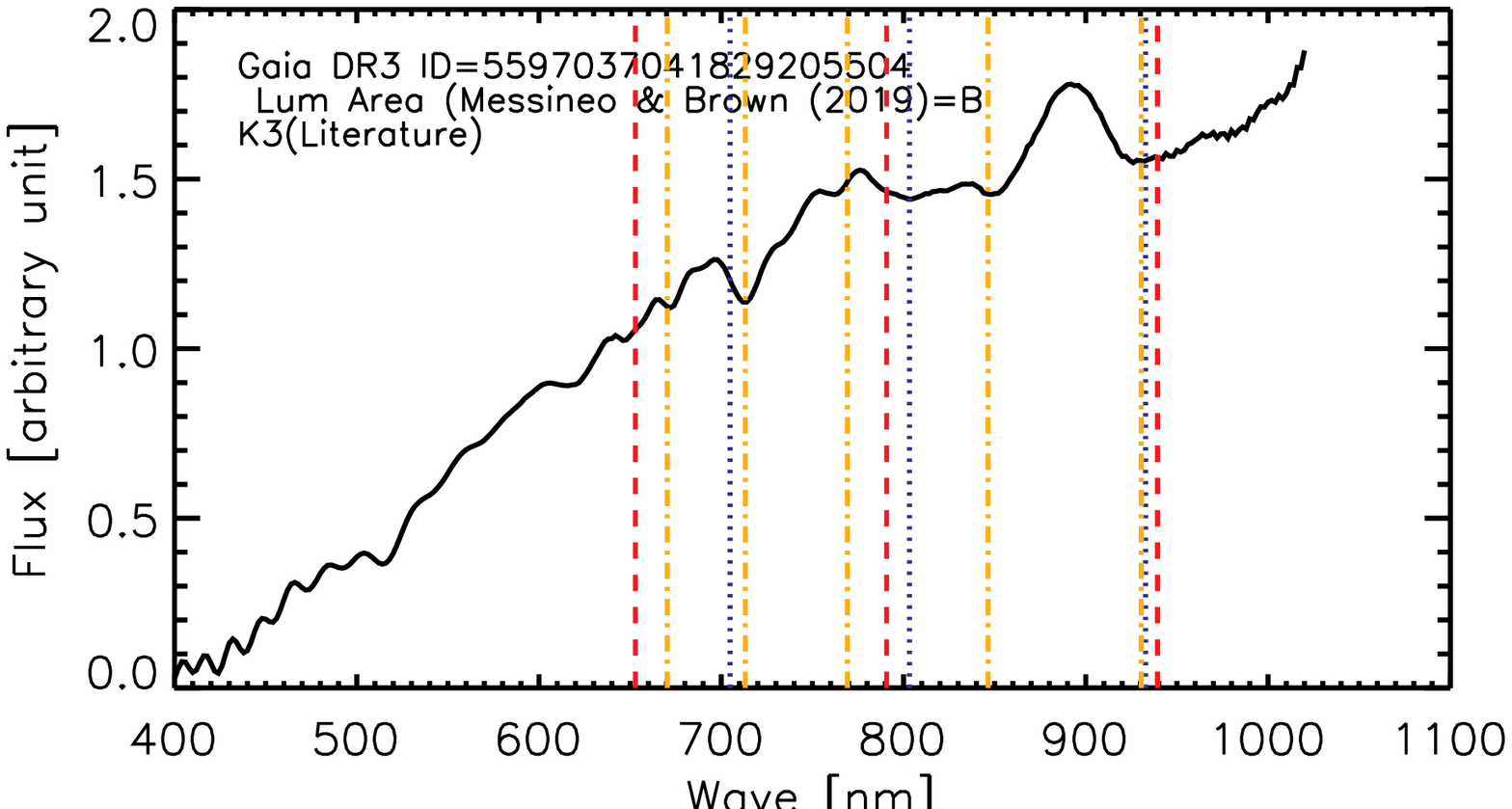}}
\resizebox{0.33\hsize}{!}{\includegraphics[angle=0]{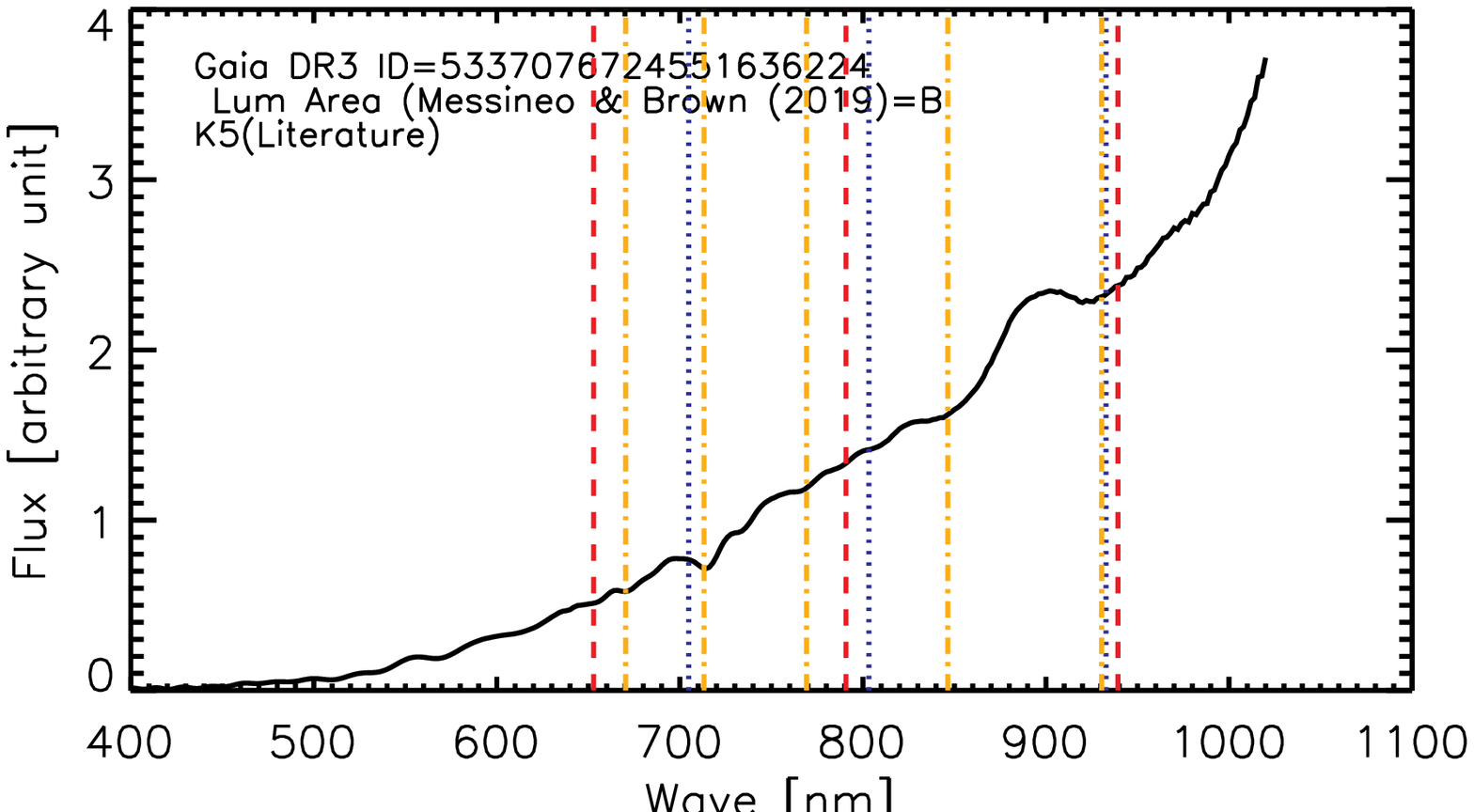}}
\end{center}
\begin{center}
\resizebox{0.33\hsize}{!}{\includegraphics[angle=0]{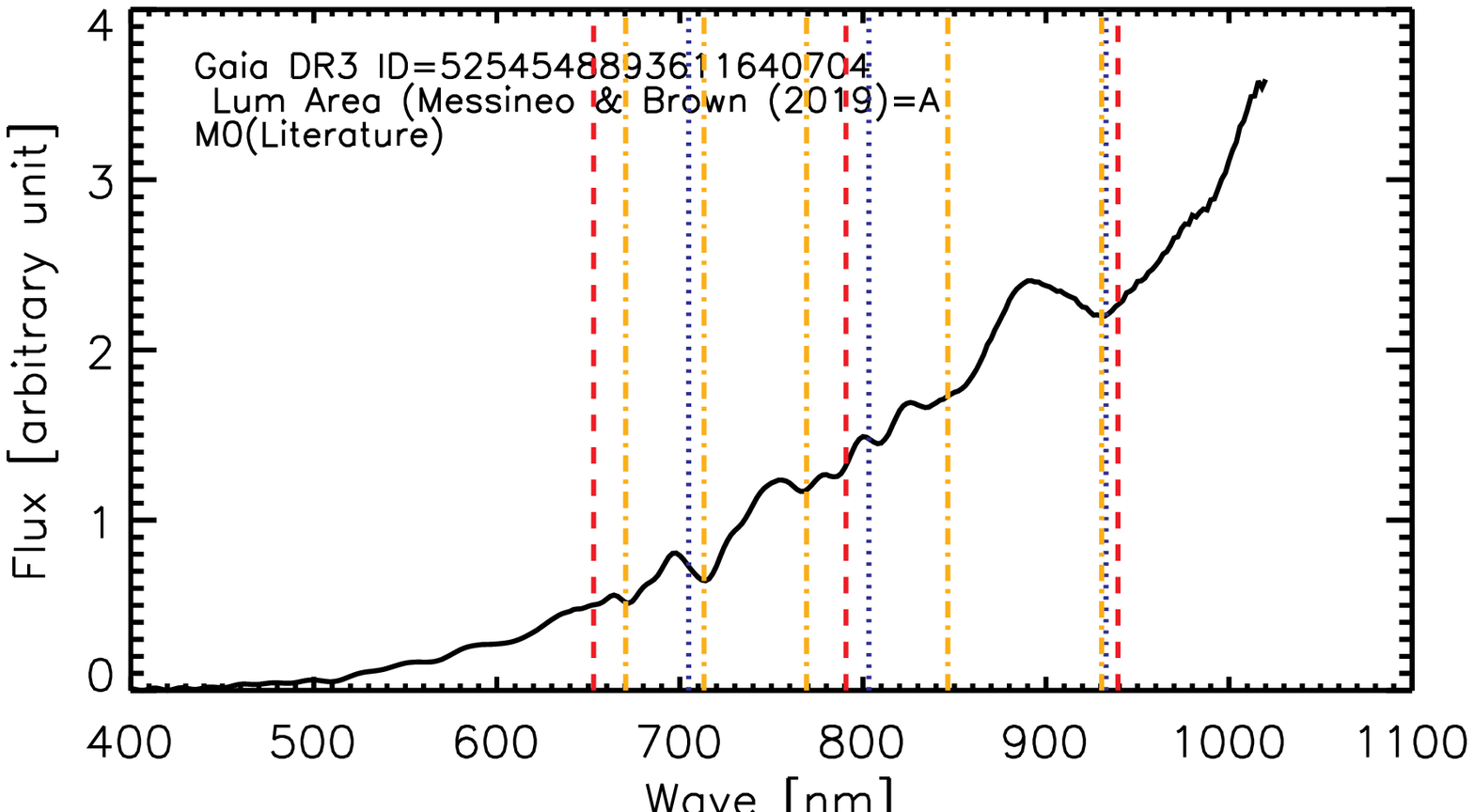}}
\resizebox{0.33\hsize}{!}{\includegraphics[angle=0]{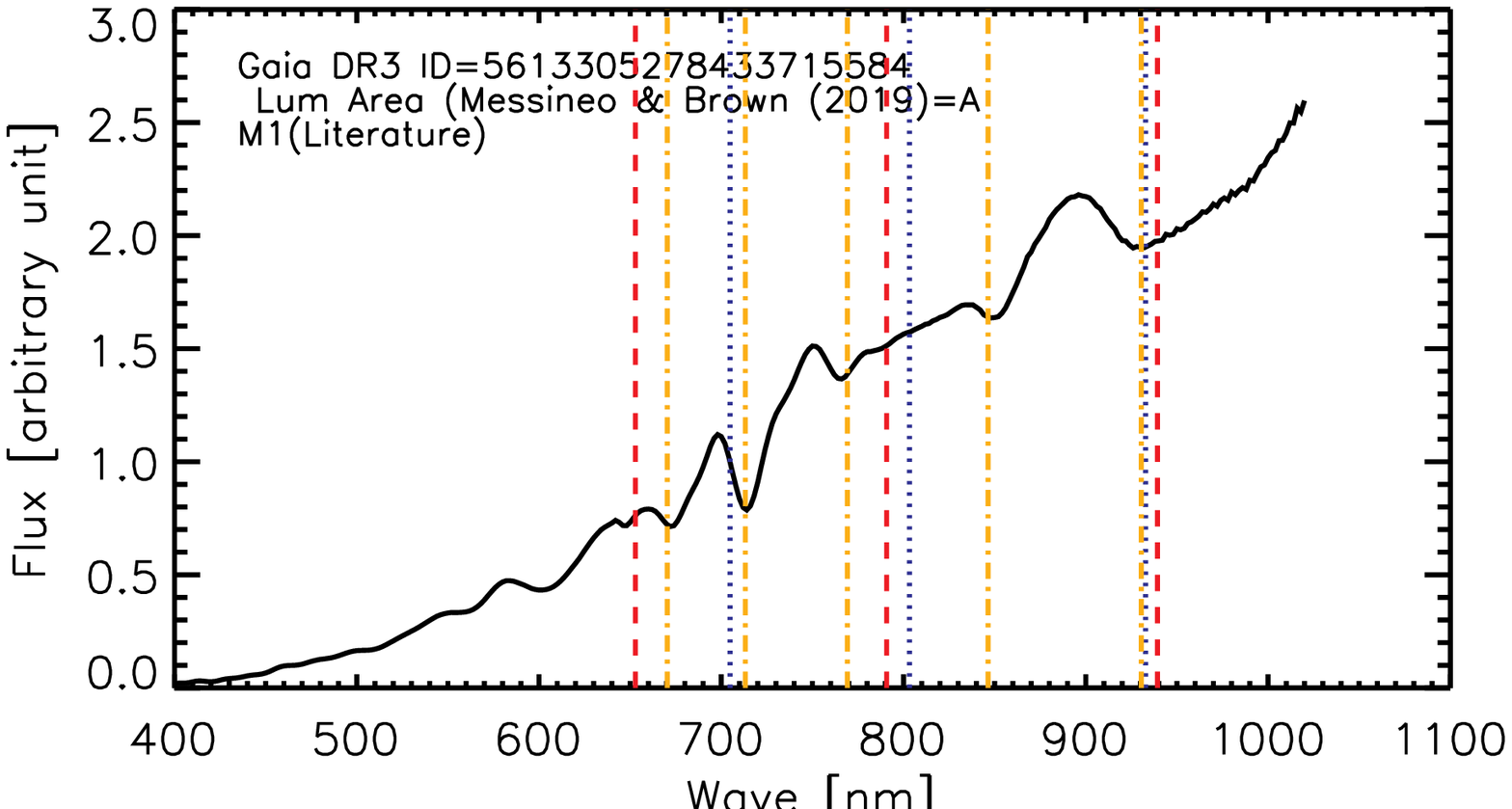}}
\resizebox{0.33\hsize}{!}{\includegraphics[angle=0]{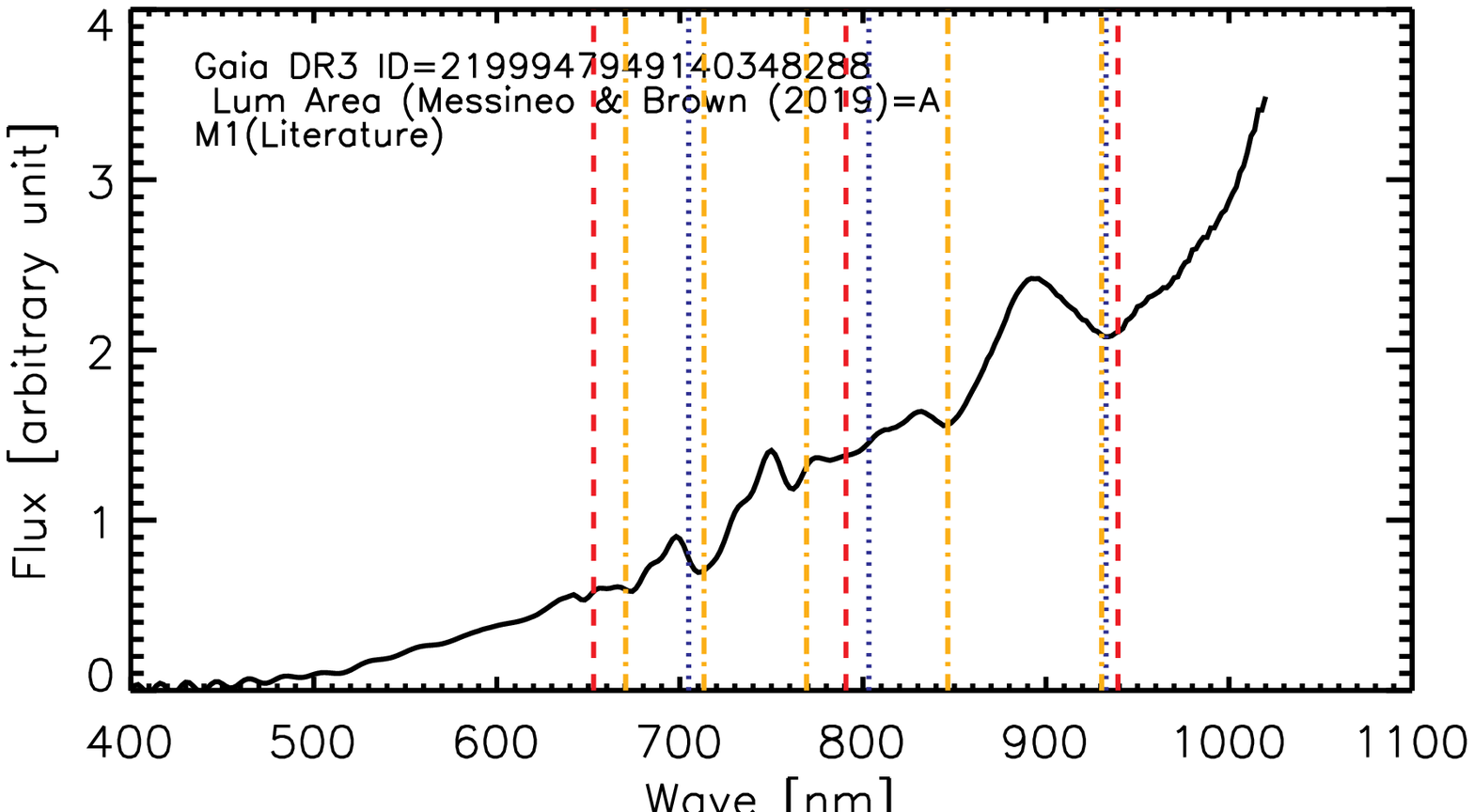}}
\end{center}
\begin{center}
\resizebox{0.33\hsize}{!}{\includegraphics[angle=0]{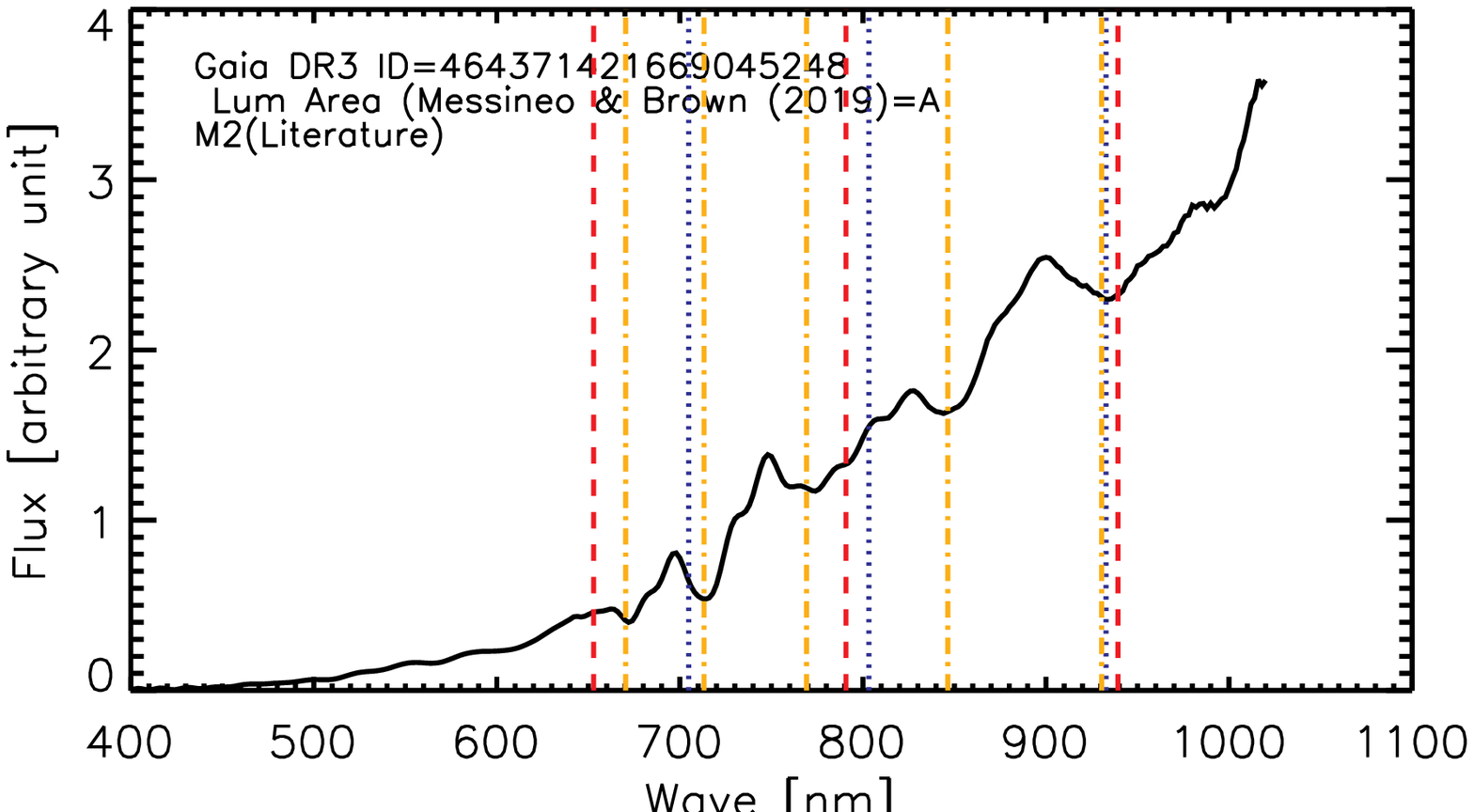}}
\resizebox{0.33\hsize}{!}{\includegraphics[angle=0]{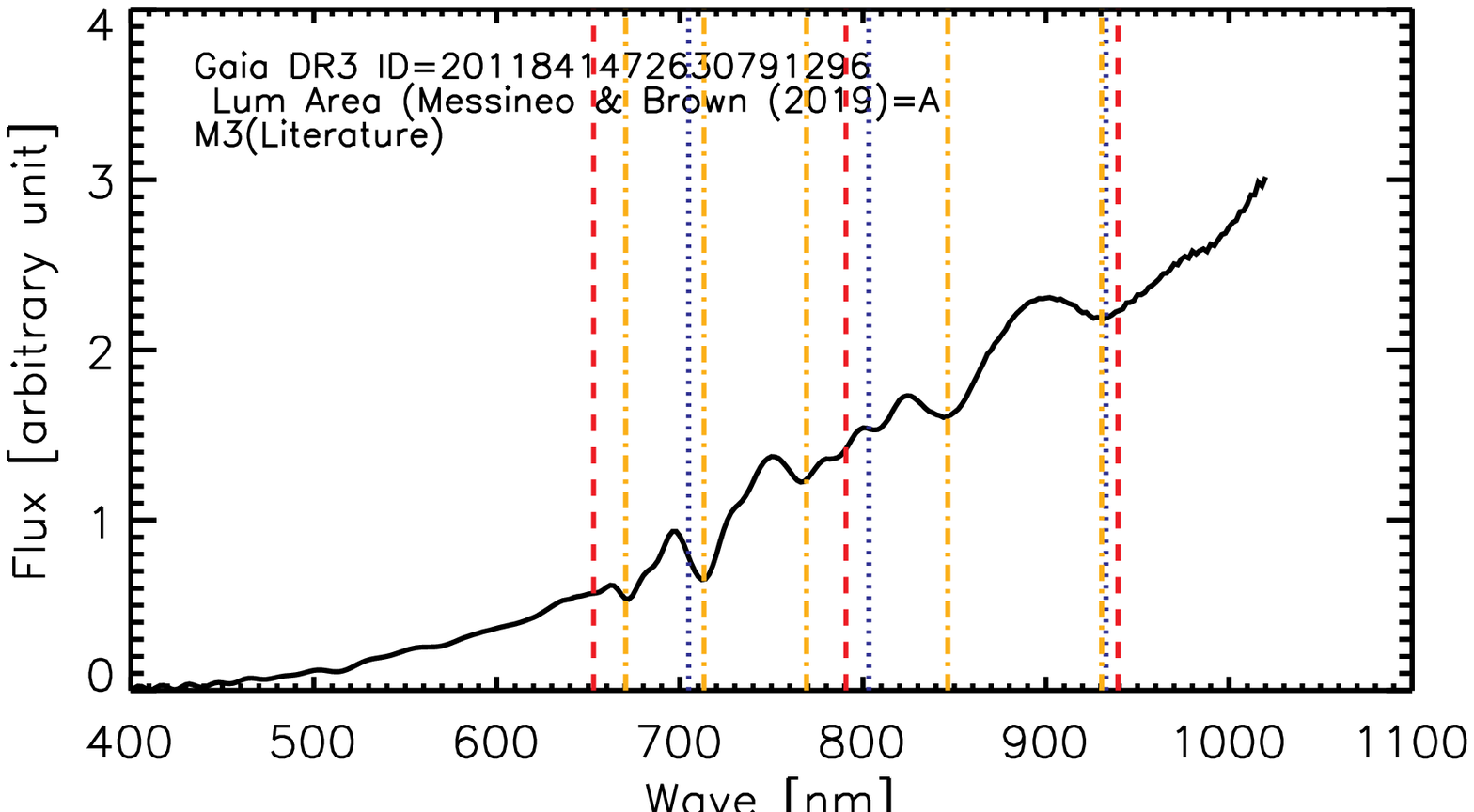}}
\resizebox{0.33\hsize}{!}{\includegraphics[angle=0]{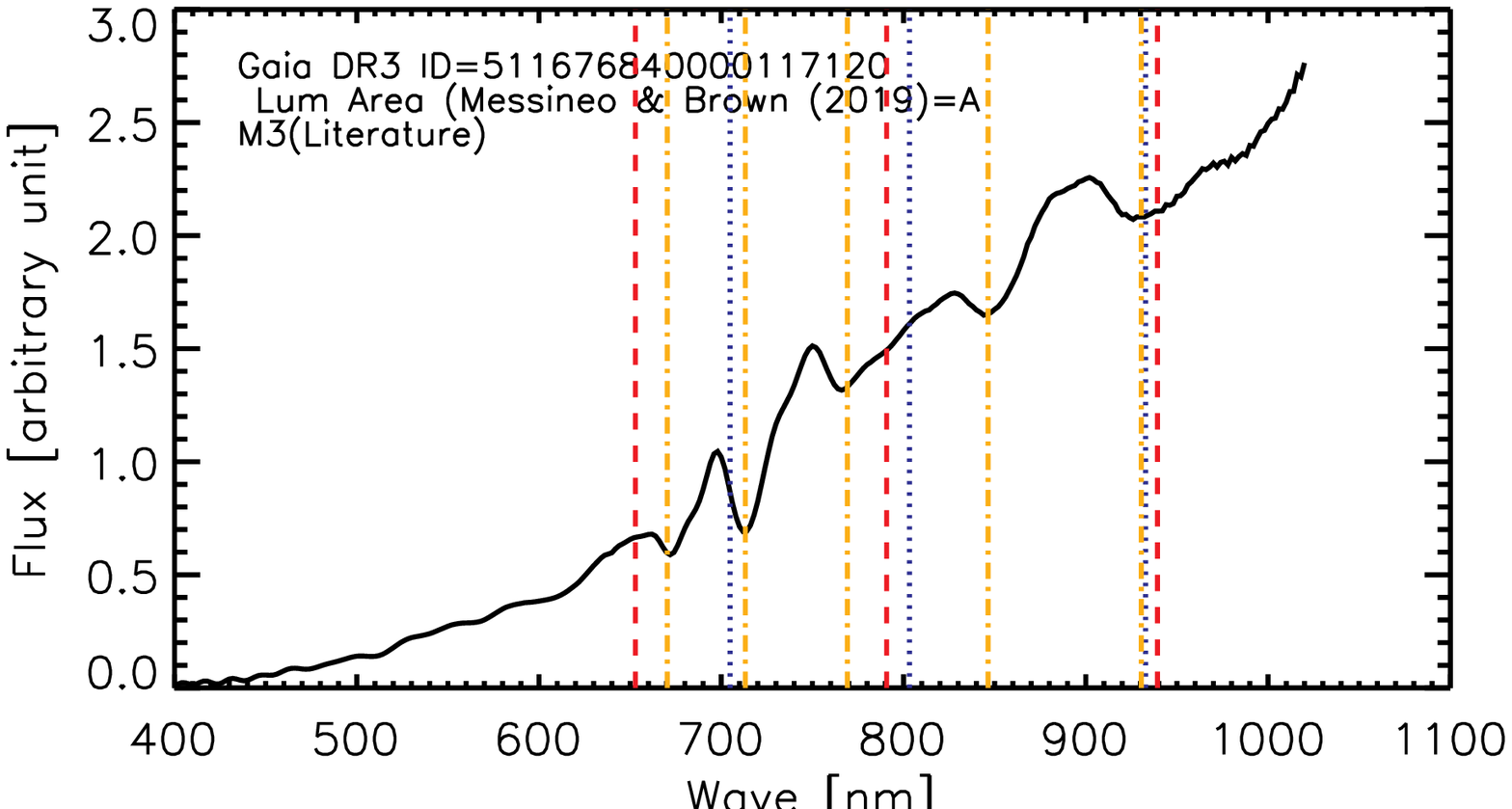}}
\end{center}
\begin{center}
\resizebox{0.33\hsize}{!}{\includegraphics[angle=0]{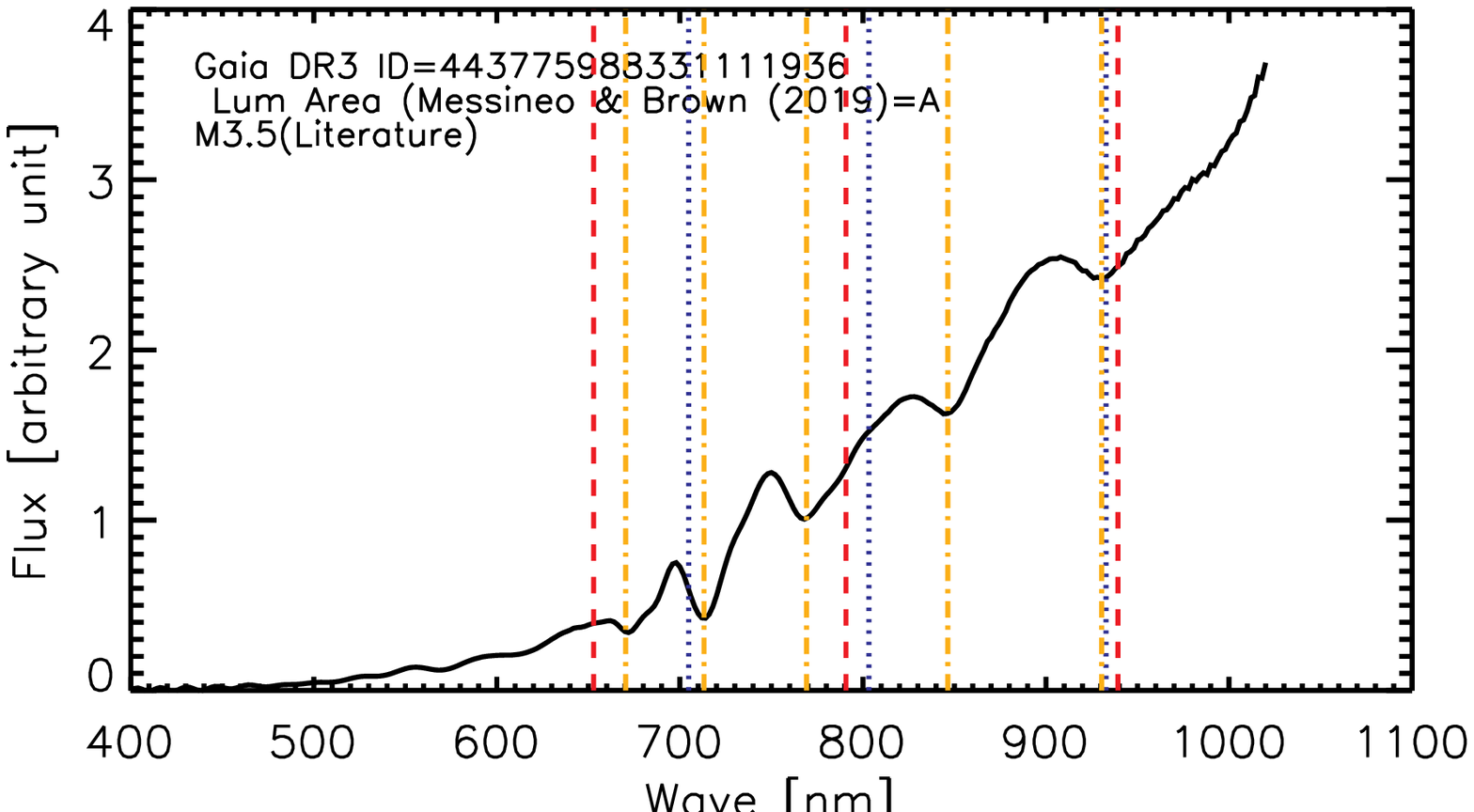}}
\resizebox{0.33\hsize}{!}{\includegraphics[angle=0]{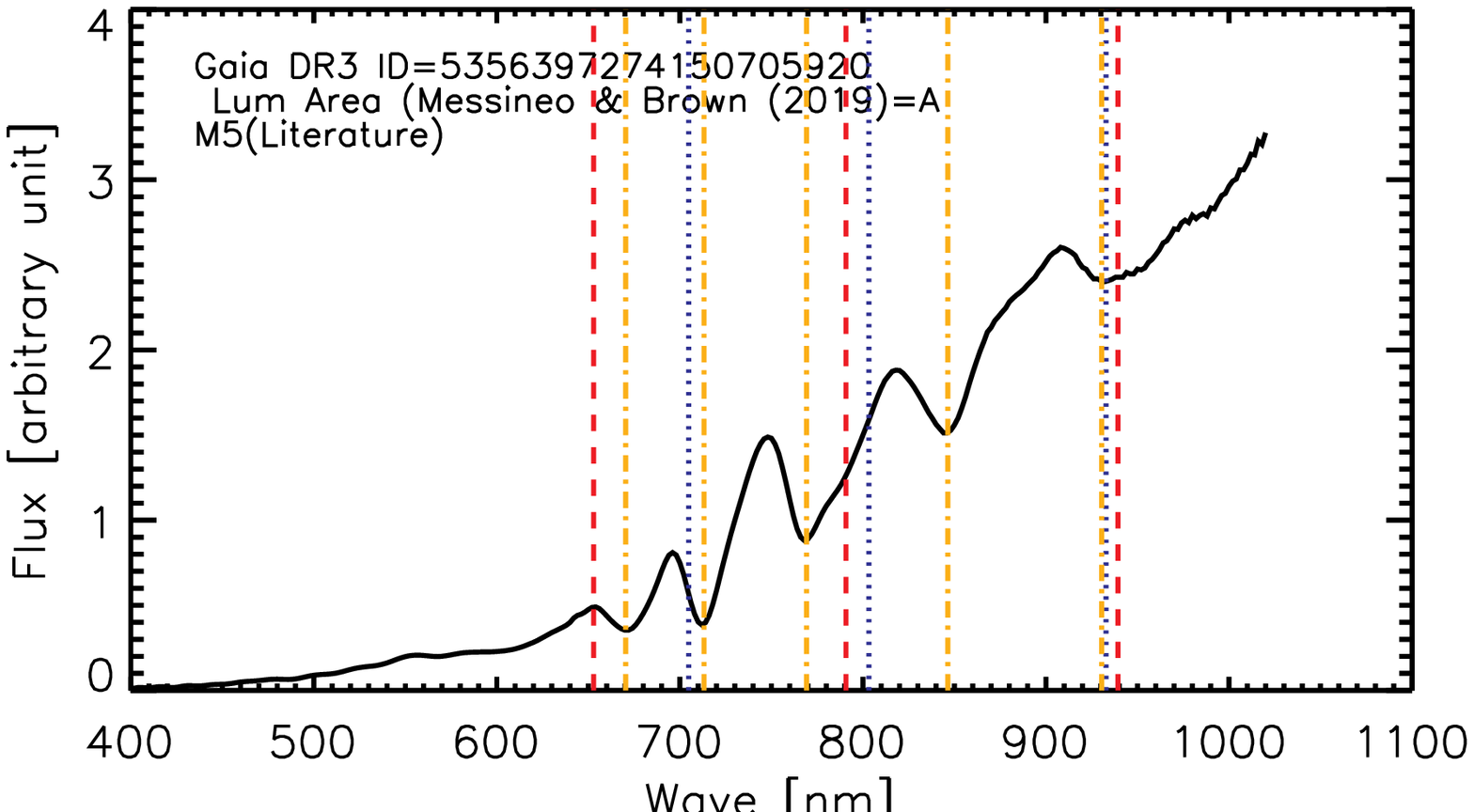}}
\resizebox{0.33\hsize}{!}{\includegraphics[angle=0]{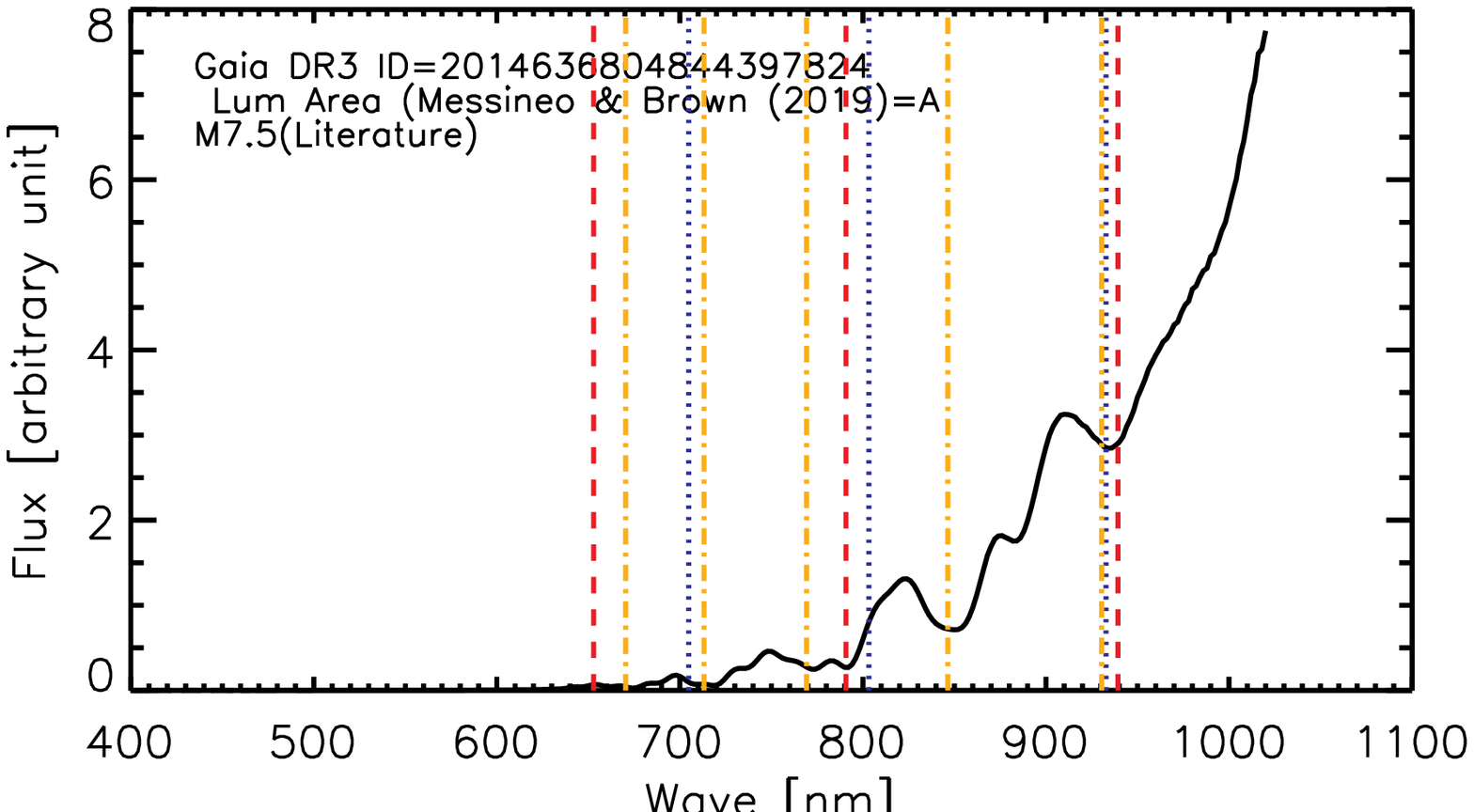}}
\end{center}
\caption{\label{RSGstarsREF} Gaia BP/RP spectra of well-known RSGs 
in the catalog of  \citet{messineo19}.
The vertical orange dotted-dashed lines mark the locations of the main 
absorption bands seen in M1-M3  RSGs and O-rich stars.
For comparison, the centroids of the absorption seen in S-type (red) 
and  C-rich stars (blue) are also shown
(see also \url{https://mariamessineo.github.io/KMclassI-DR3_AB_BP-RPspectra/}).
} 
\end{figure*}

In Figure \ref{RSGstarsREF} a small compiled 
library of spectra of known RSGs is displayed, while
the bright K-M stars erroneously flagged as  C-rich stars 
by the LPV pipeline are shown in Fig.\ \ref{RSGstars}.

\begin{figure*}
\begin{center}
\resizebox{0.33\hsize}{!}{\includegraphics[angle=0]{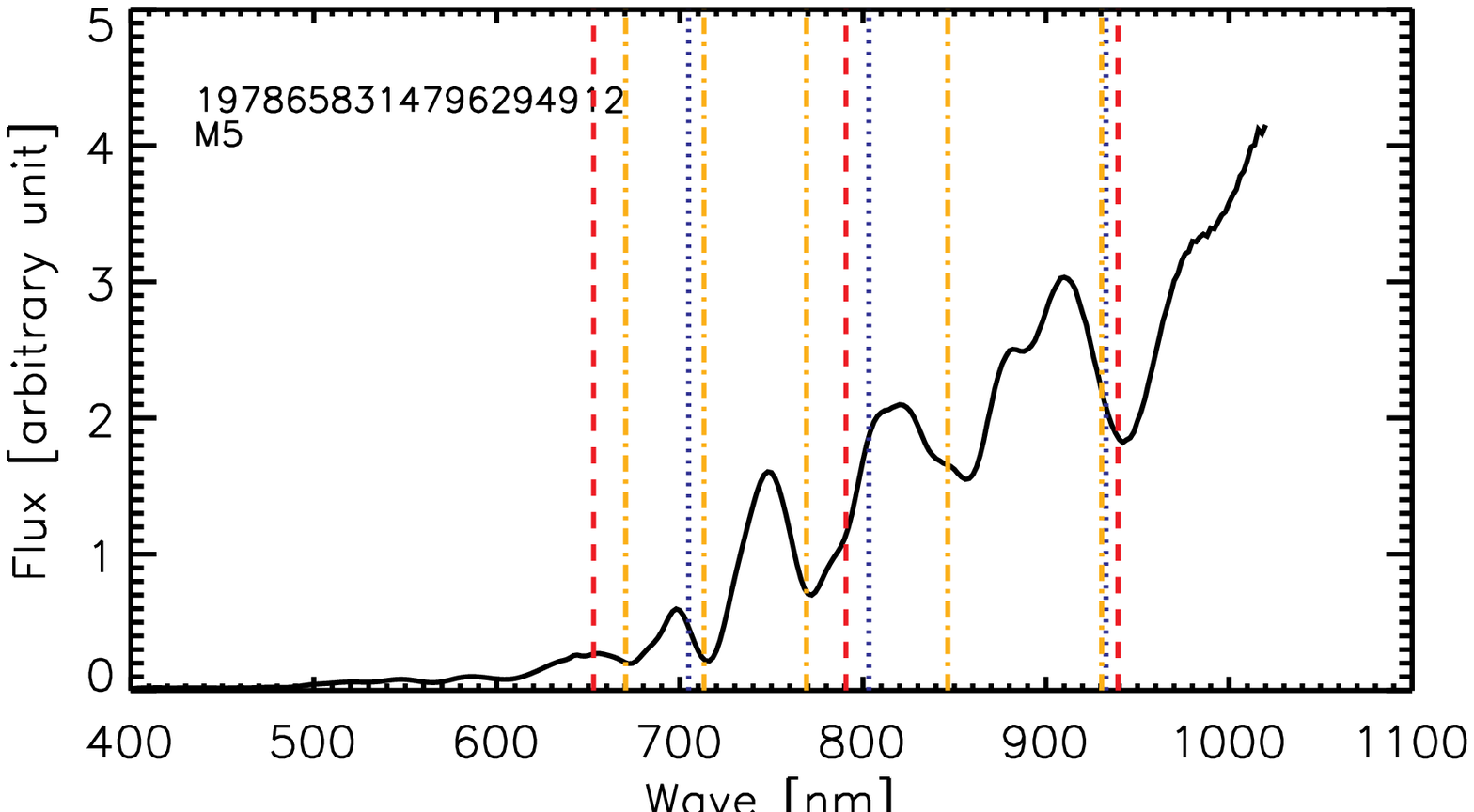}}
\resizebox{0.33\hsize}{!}{\includegraphics[angle=0]{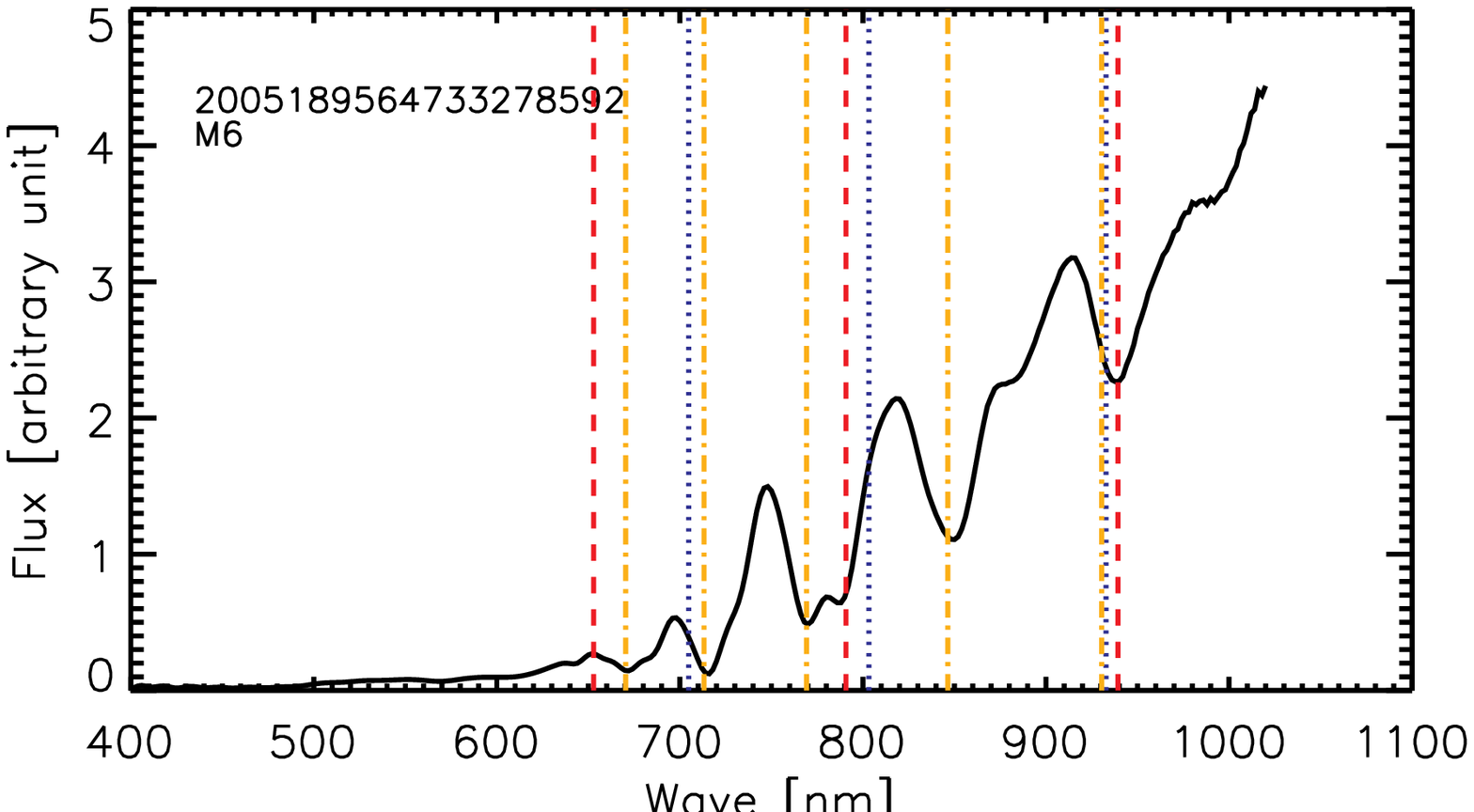}}
\resizebox{0.33\hsize}{!}{\includegraphics[angle=0]{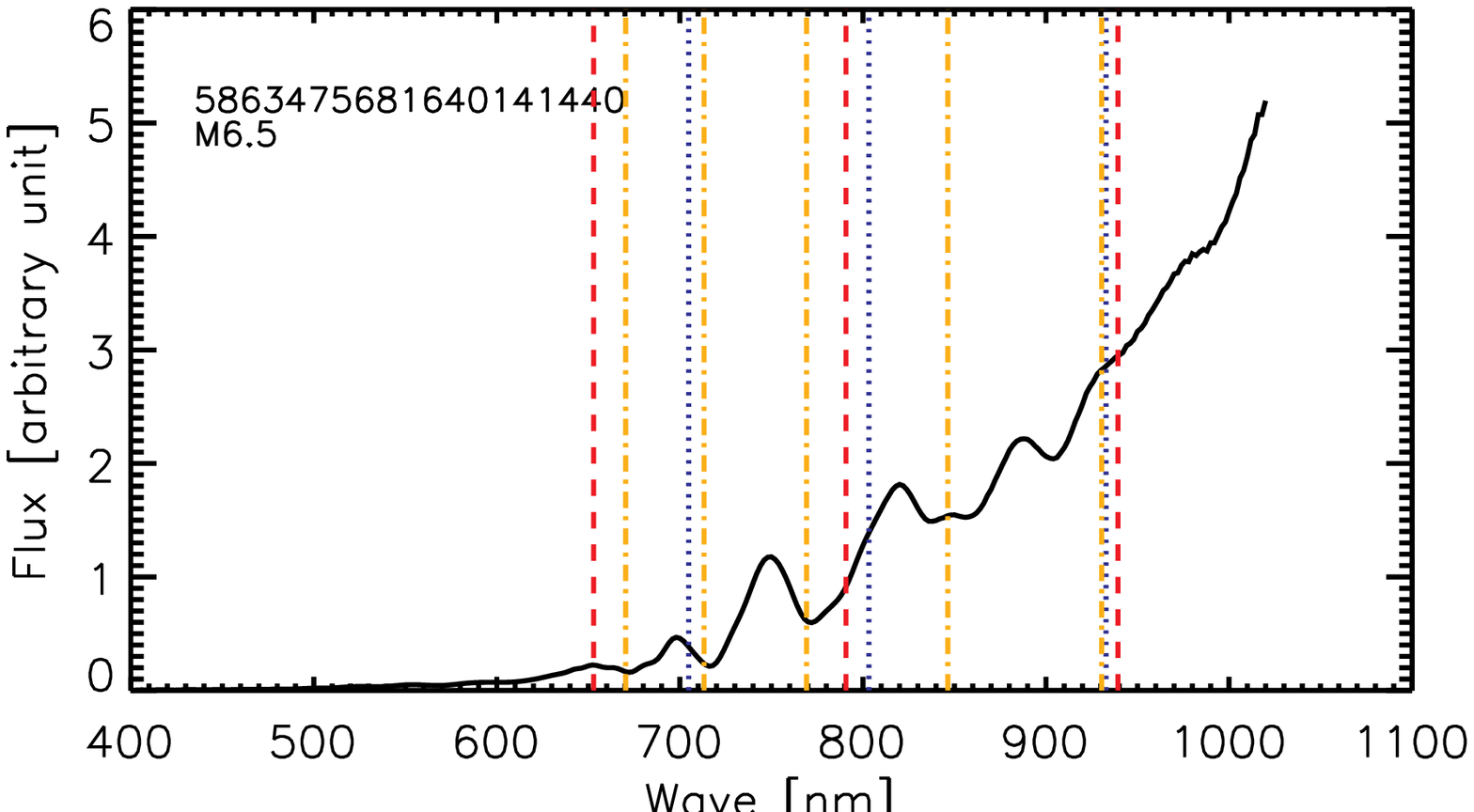}}
\end{center}
\begin{center}
\resizebox{0.33\hsize}{!}{\includegraphics[angle=0]{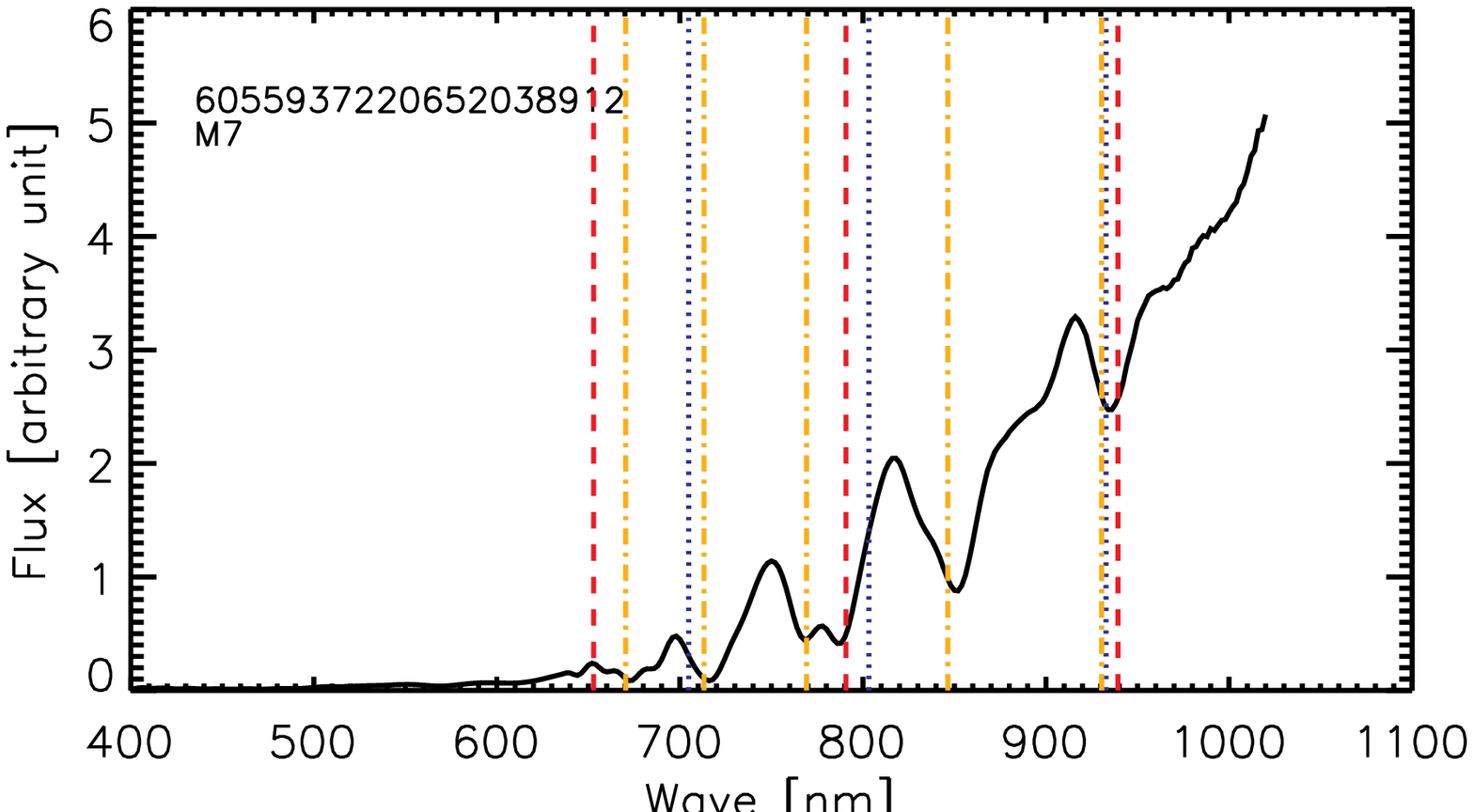}}
\resizebox{0.33\hsize}{!}{\includegraphics[angle=0]{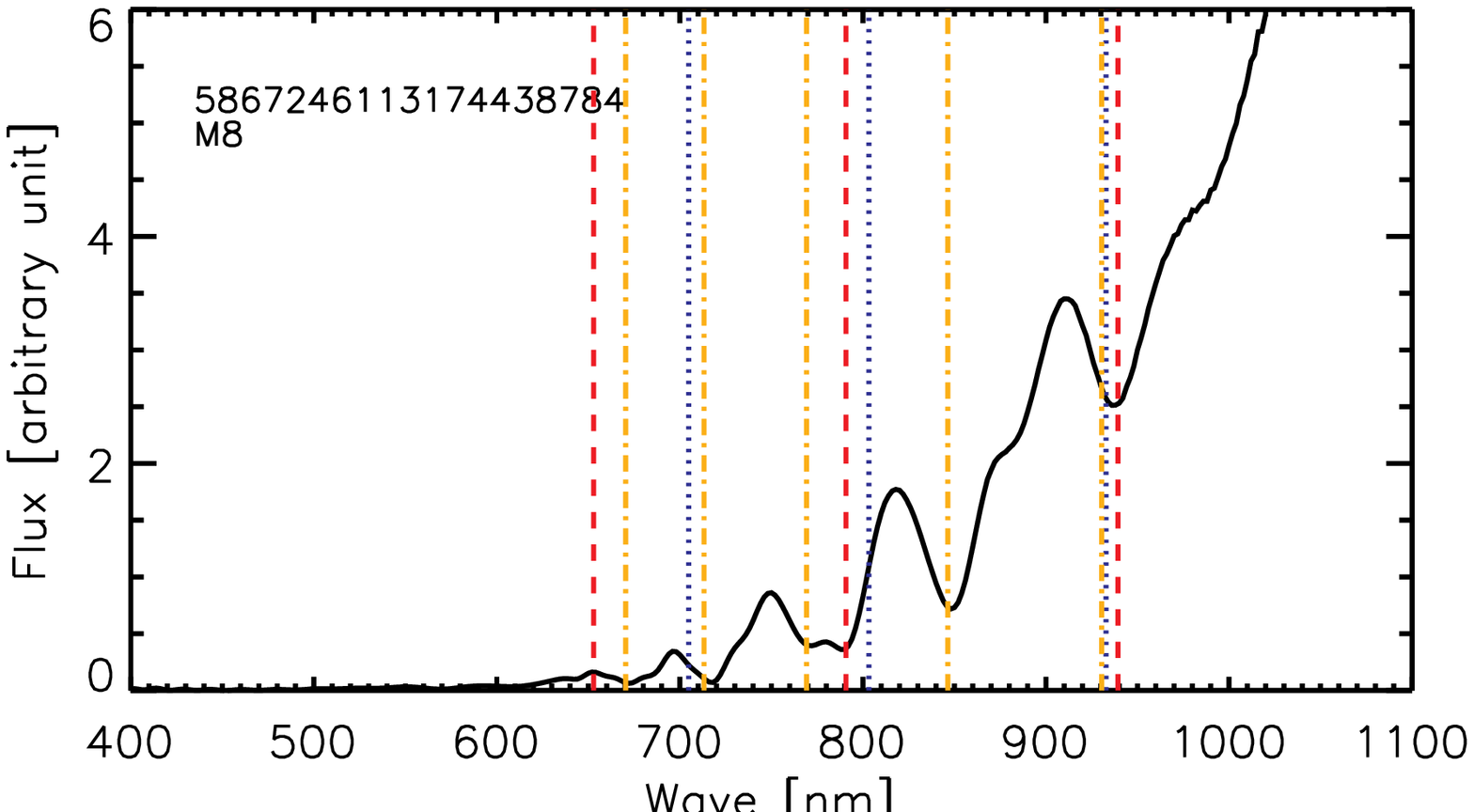}}
\resizebox{0.33\hsize}{!}{\includegraphics[angle=0]{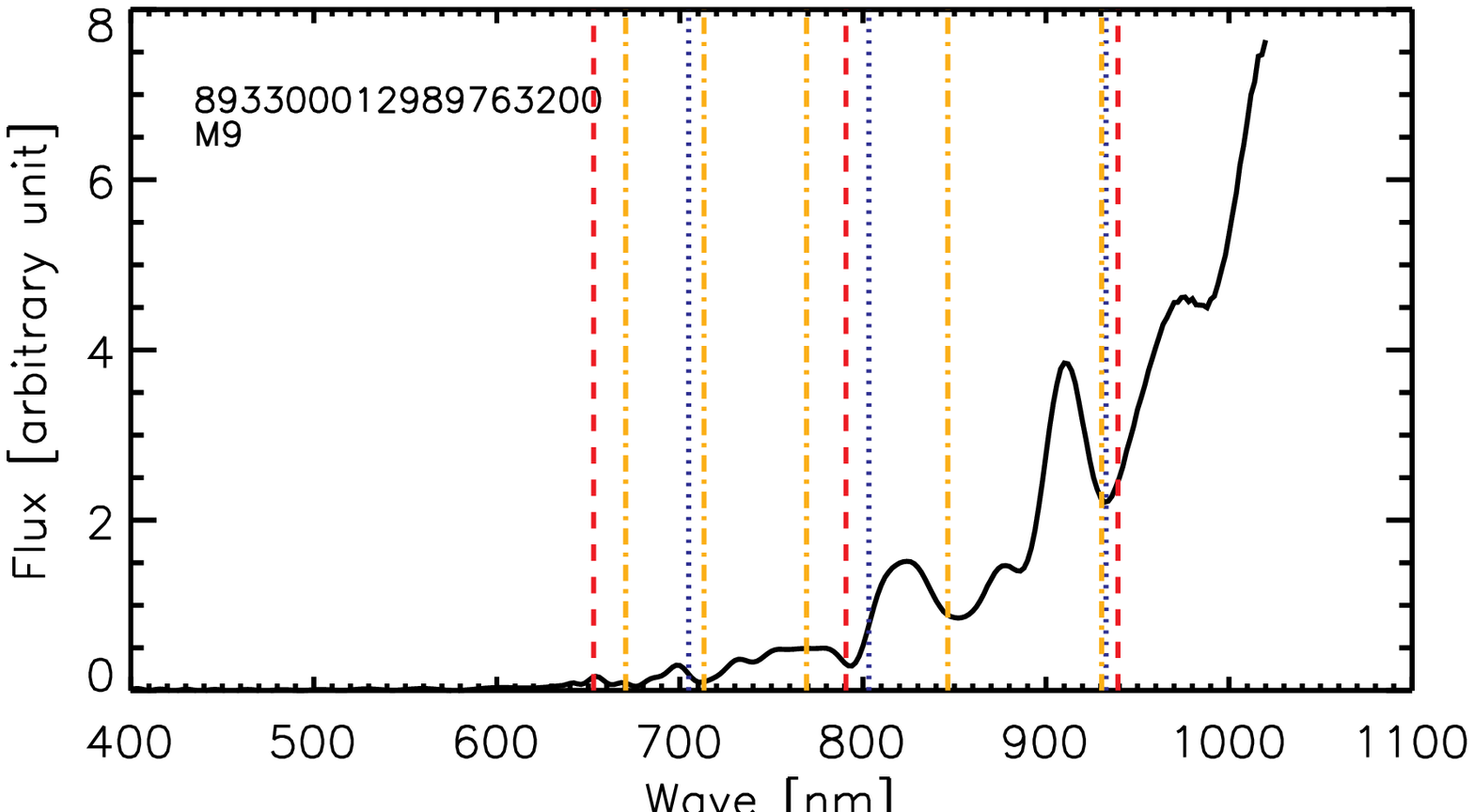}}
\end{center}
\caption{\label{RSGstarsREF2} Gaia BP/RP spectra of known late-M AGB stars found
among the 203 new Apsis selected bright late-type stars (see Sect.   \ref{detection}).
The vertical orange dotted-dashed lines mark the locations of the main 
absorption bands seen in M1-M3 RSGs and O-rich stars.
For comparison, the centroids of the absorption seen in S-type (red) 
and in C-rich stars (blue) are also shown.
} 
\end{figure*}

\begin{figure*}
\begin{center}
\resizebox{0.33\hsize}{!}{\includegraphics[angle=0]{507785054178237824.eps}}
\resizebox{0.33\hsize}{!}{\includegraphics[angle=0]{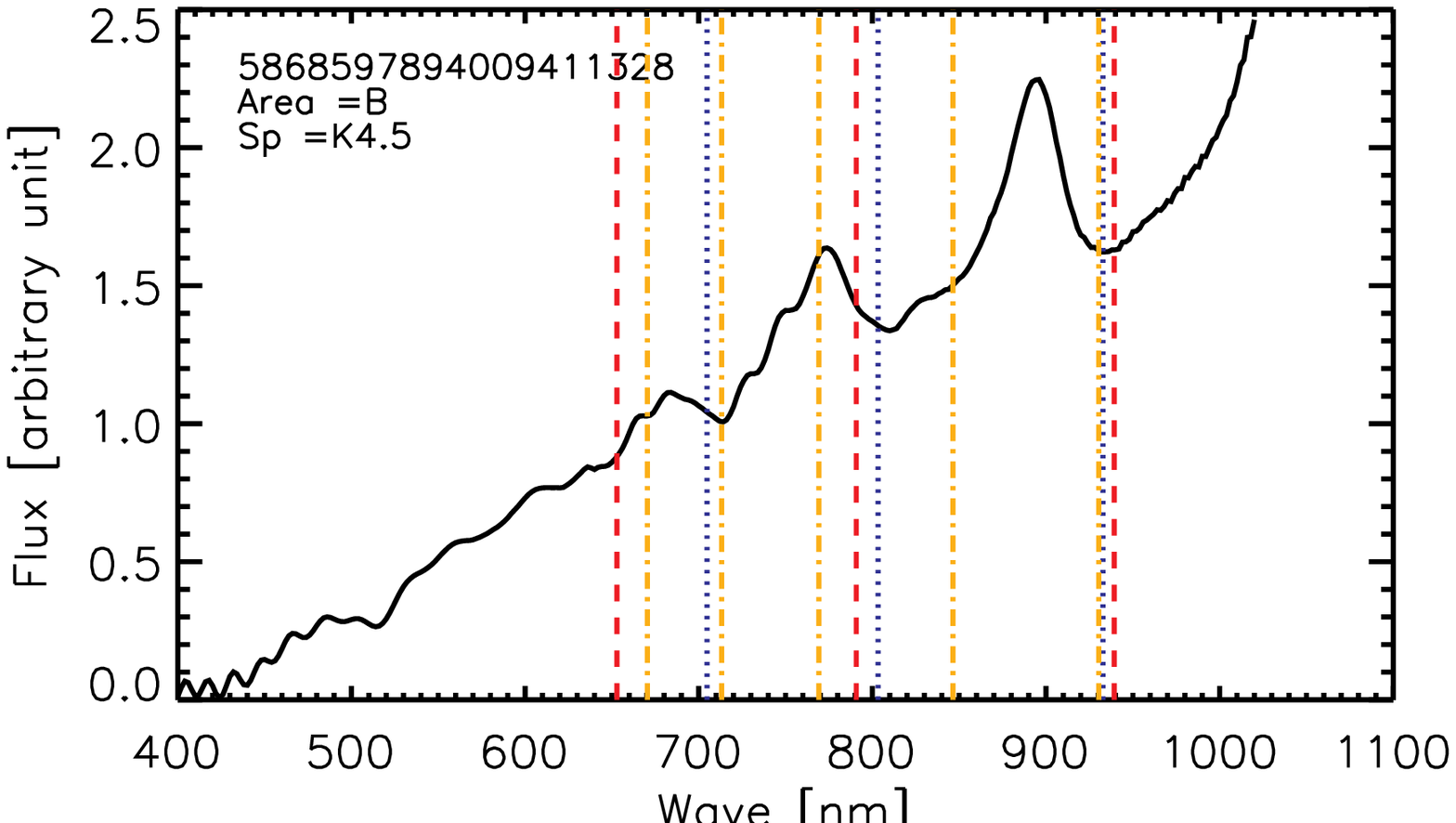}}
\resizebox{0.33\hsize}{!}{\includegraphics[angle=0]{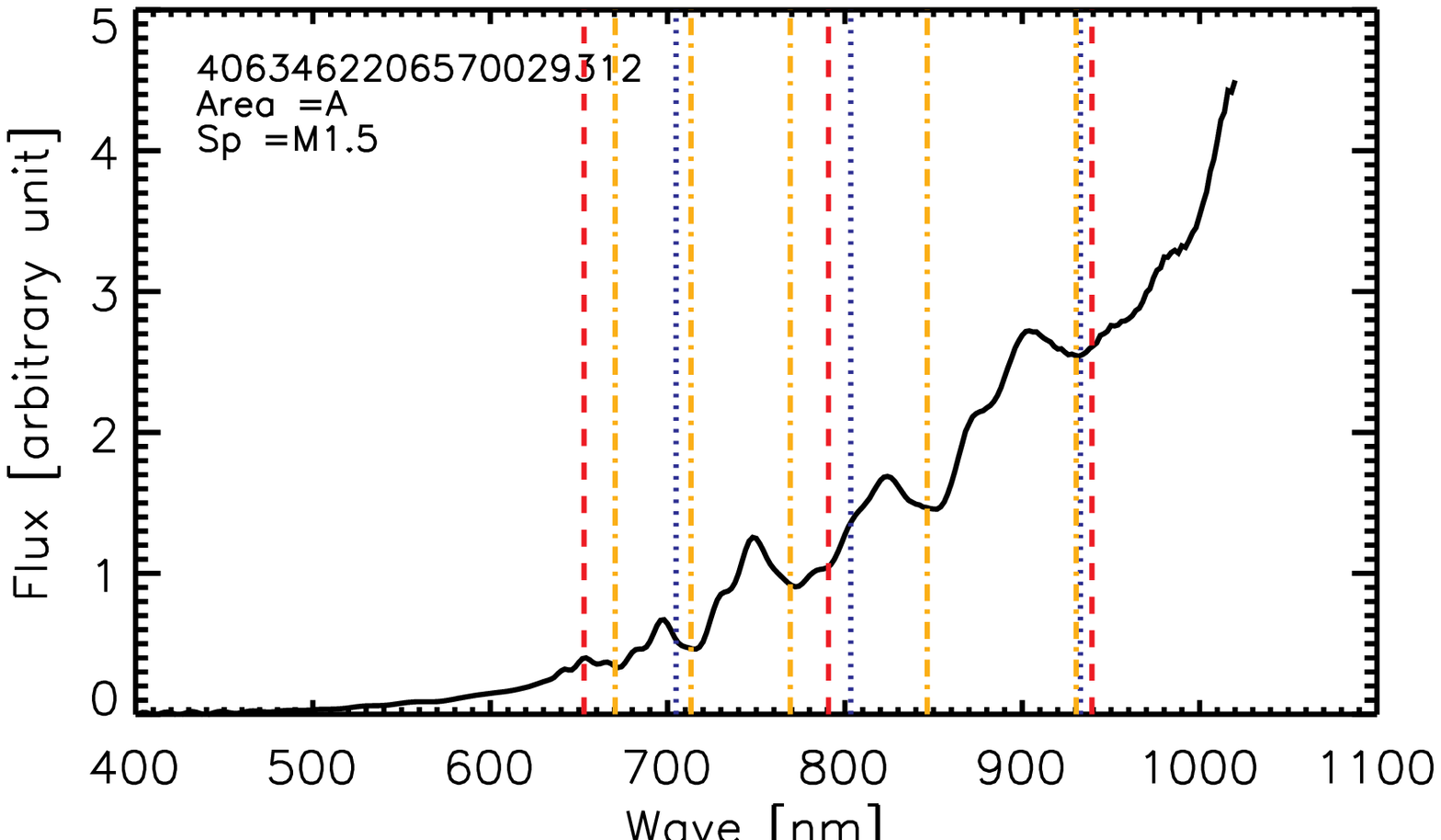}}
\end{center}
\begin{center}
\resizebox{0.33\hsize}{!}{\includegraphics[angle=0]{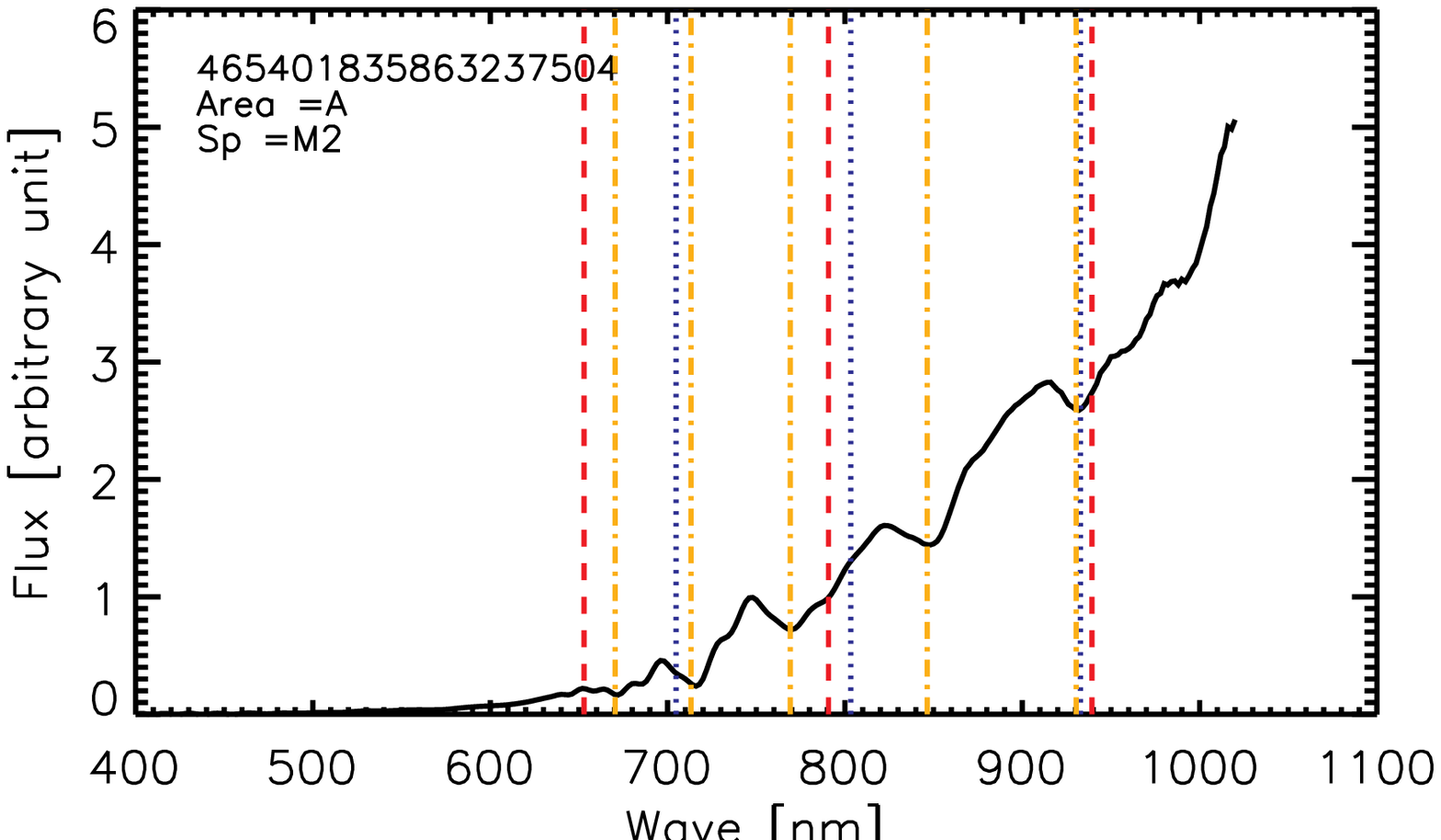}}
\resizebox{0.33\hsize}{!}{\includegraphics[angle=0]{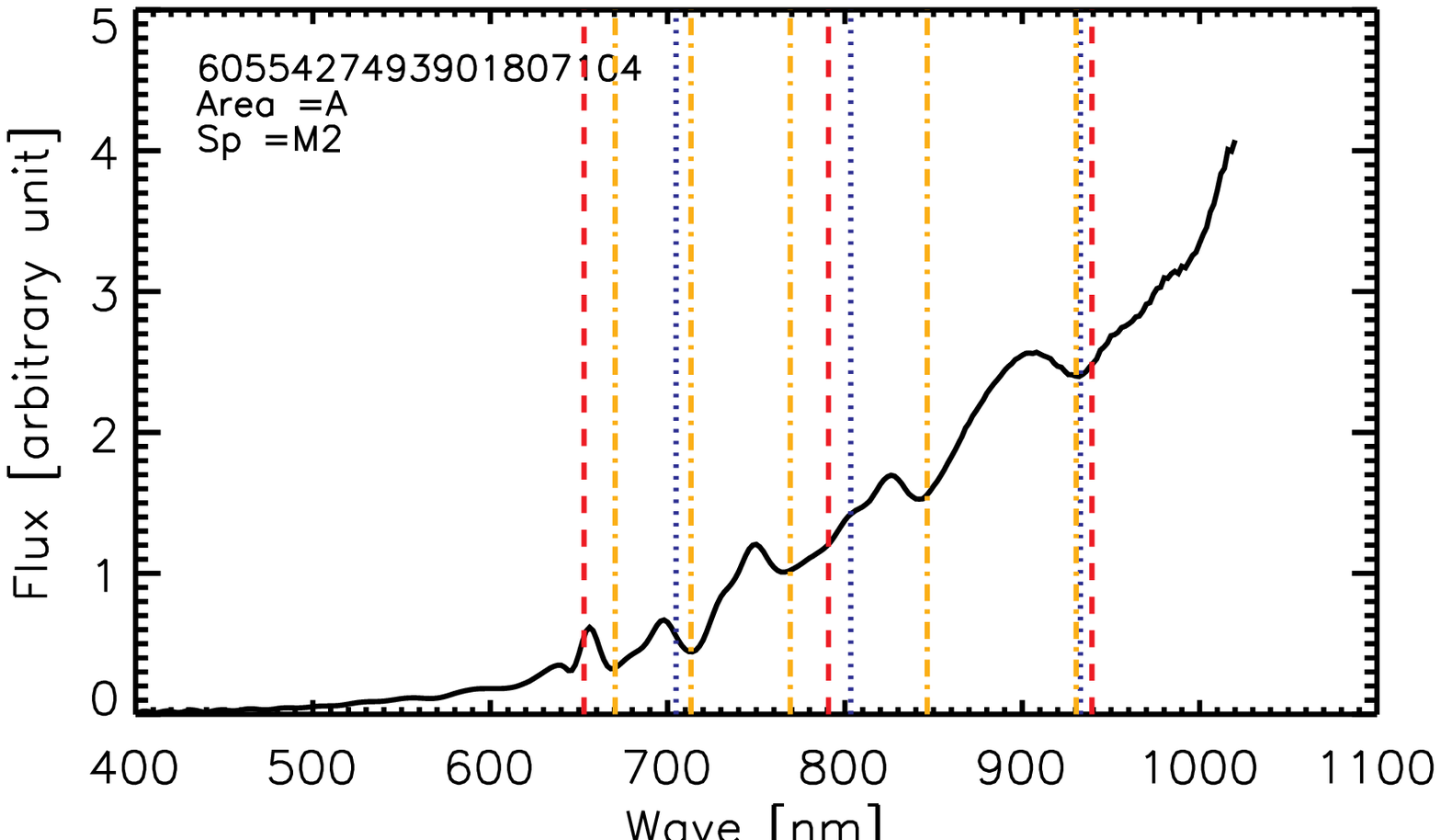}}
\resizebox{0.33\hsize}{!}{\includegraphics[angle=0]{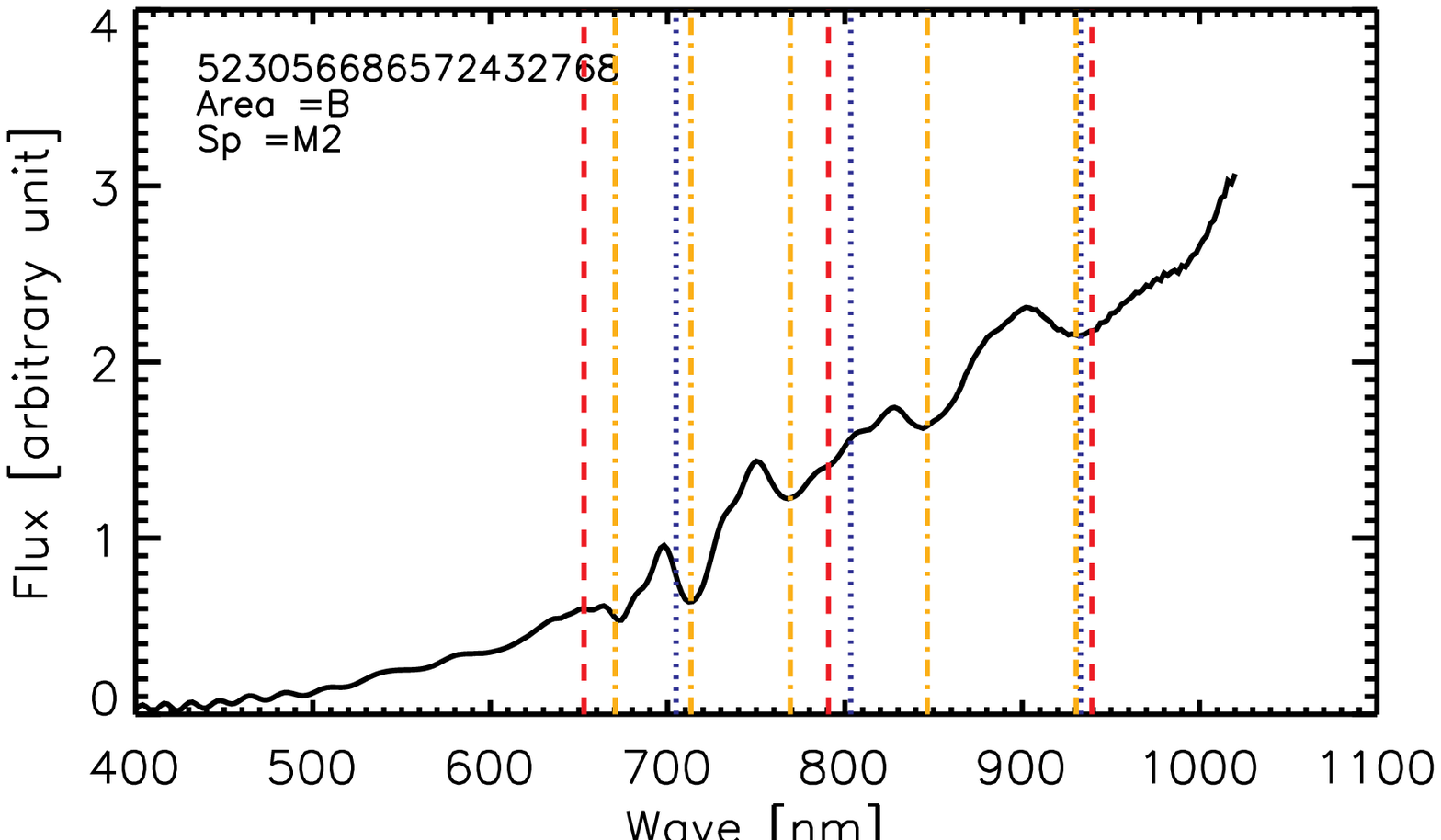}}
\end{center}
\begin{center}
\resizebox{0.33\hsize}{!}{\includegraphics[angle=0]{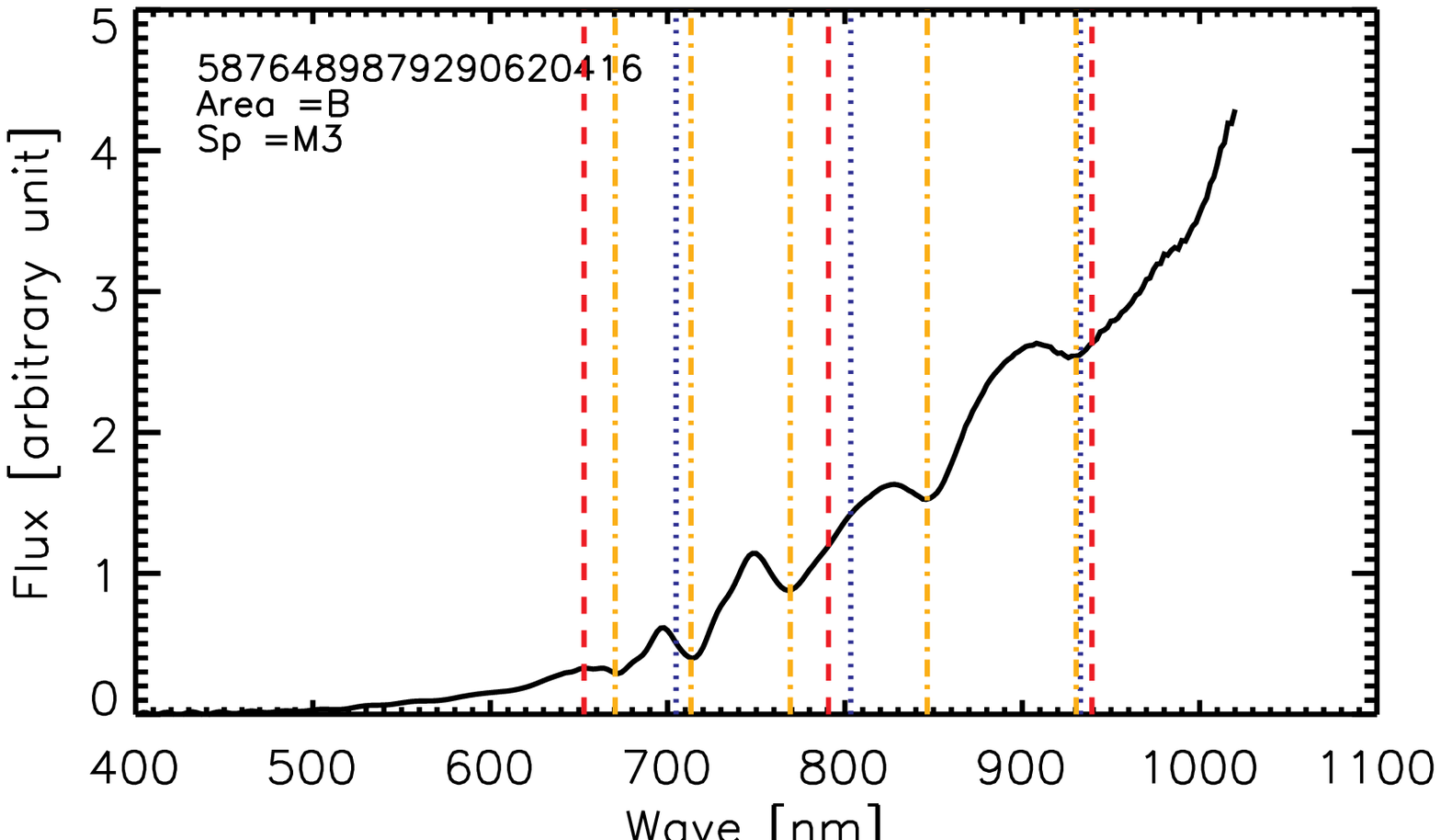}}
\resizebox{0.33\hsize}{!}{\includegraphics[angle=0]{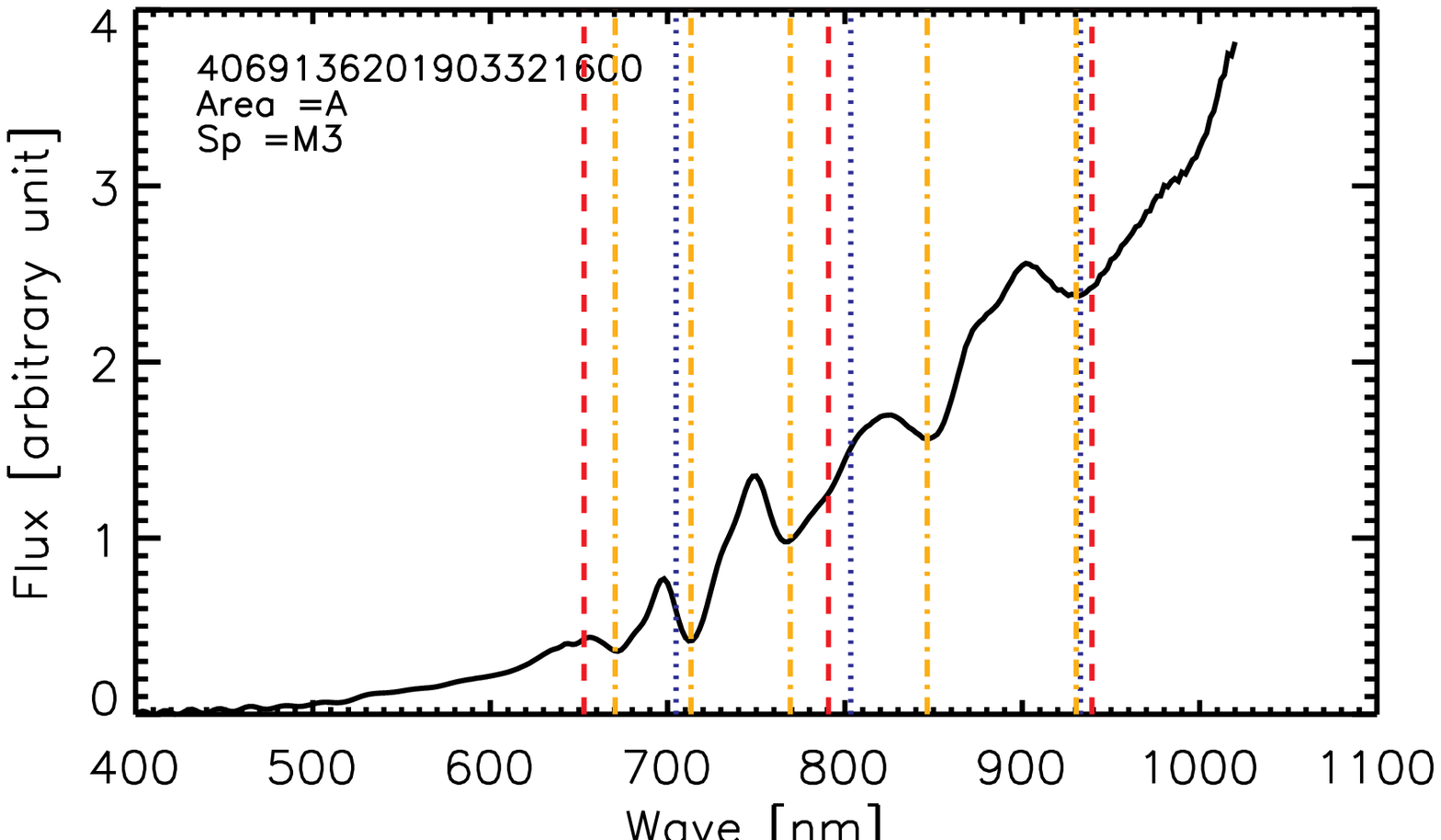}}
\resizebox{0.33\hsize}{!}{\includegraphics[angle=0]{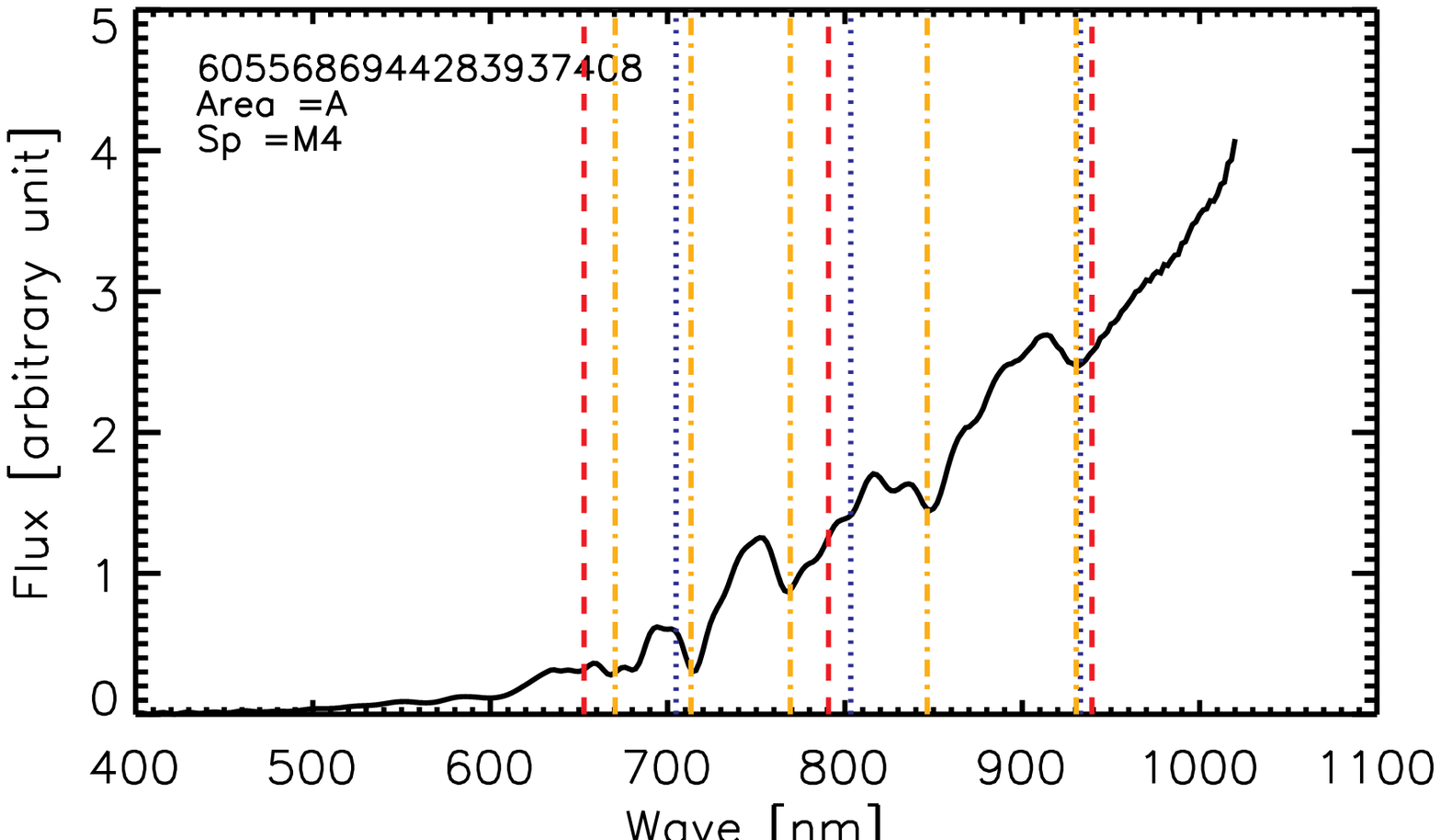}}
\end{center}
\caption{\label{RSGstars} Gaia BP/RP spectra of stars in the catalog of  
\citet{messineo19}, which are flagged as C-rich by the LPV pipeline.
The vertical orange dotted-dashed lines  mark the locations of the main 
absorption bands
seen  in M1-M3 RSGs and O-rich stars.
For comparison, the centroids of the absorption seen in S-type (red) 
and in C-rich stars (blue) are also shown.
} 
\end{figure*}

\begin{figure*}
\begin{center}

\resizebox{0.33\hsize}{!}{\includegraphics[angle=0]{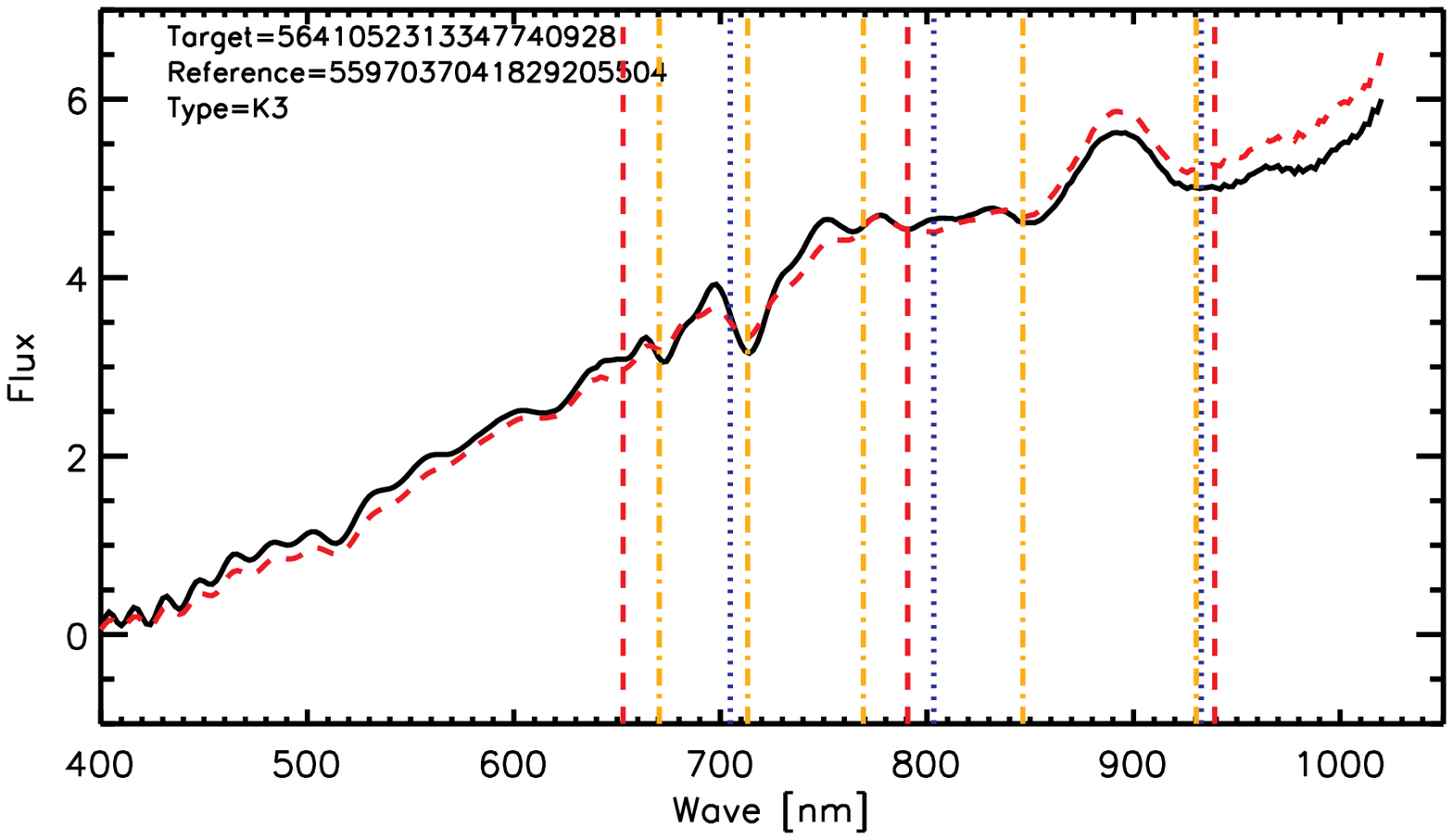}}
\resizebox{0.33\hsize}{!}{\includegraphics[angle=0]{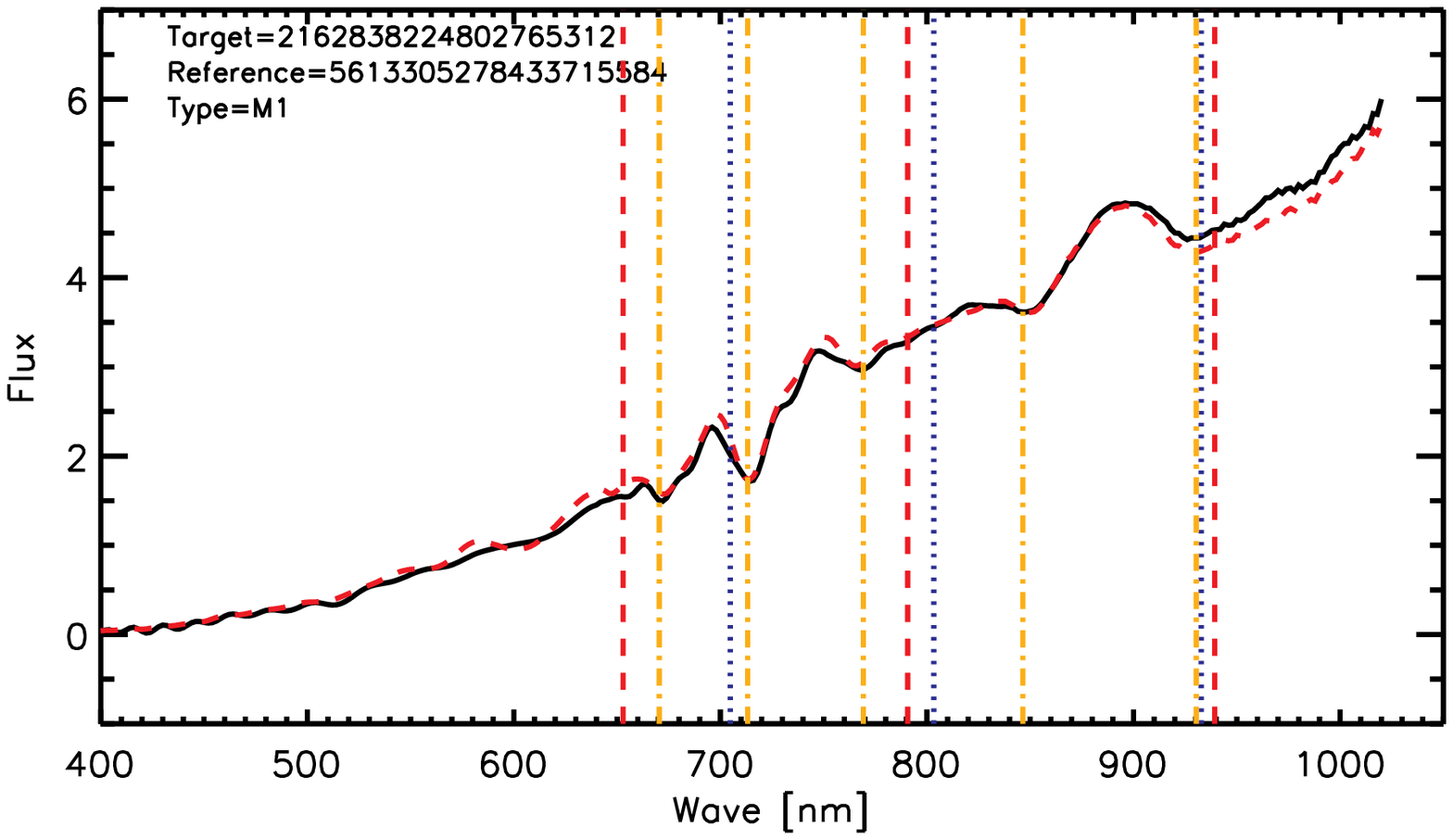}}
\resizebox{0.33\hsize}{!}{\includegraphics[angle=0]{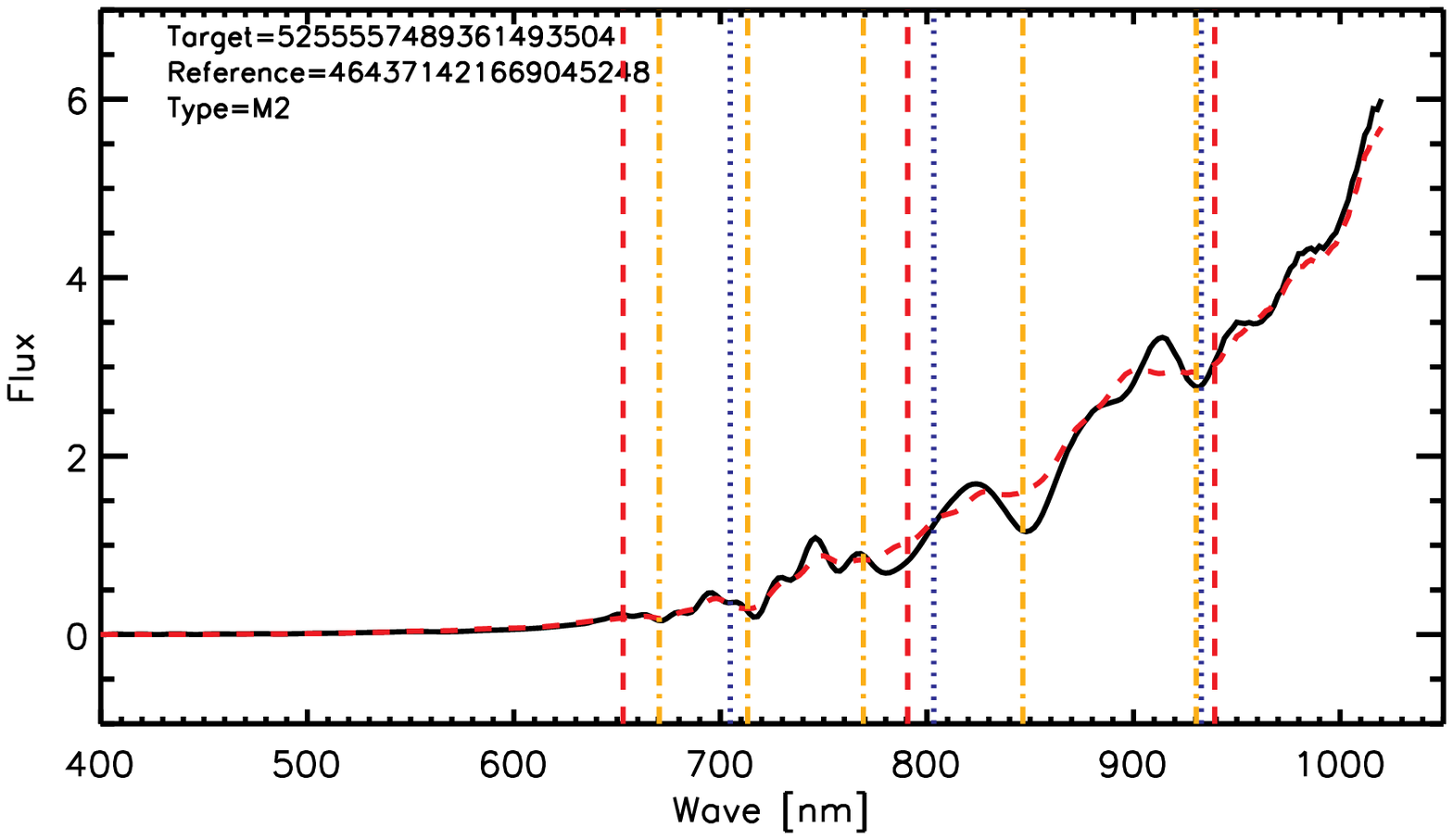}}
\end{center}
\begin{center}

\resizebox{0.33\hsize}{!}{\includegraphics[angle=0]{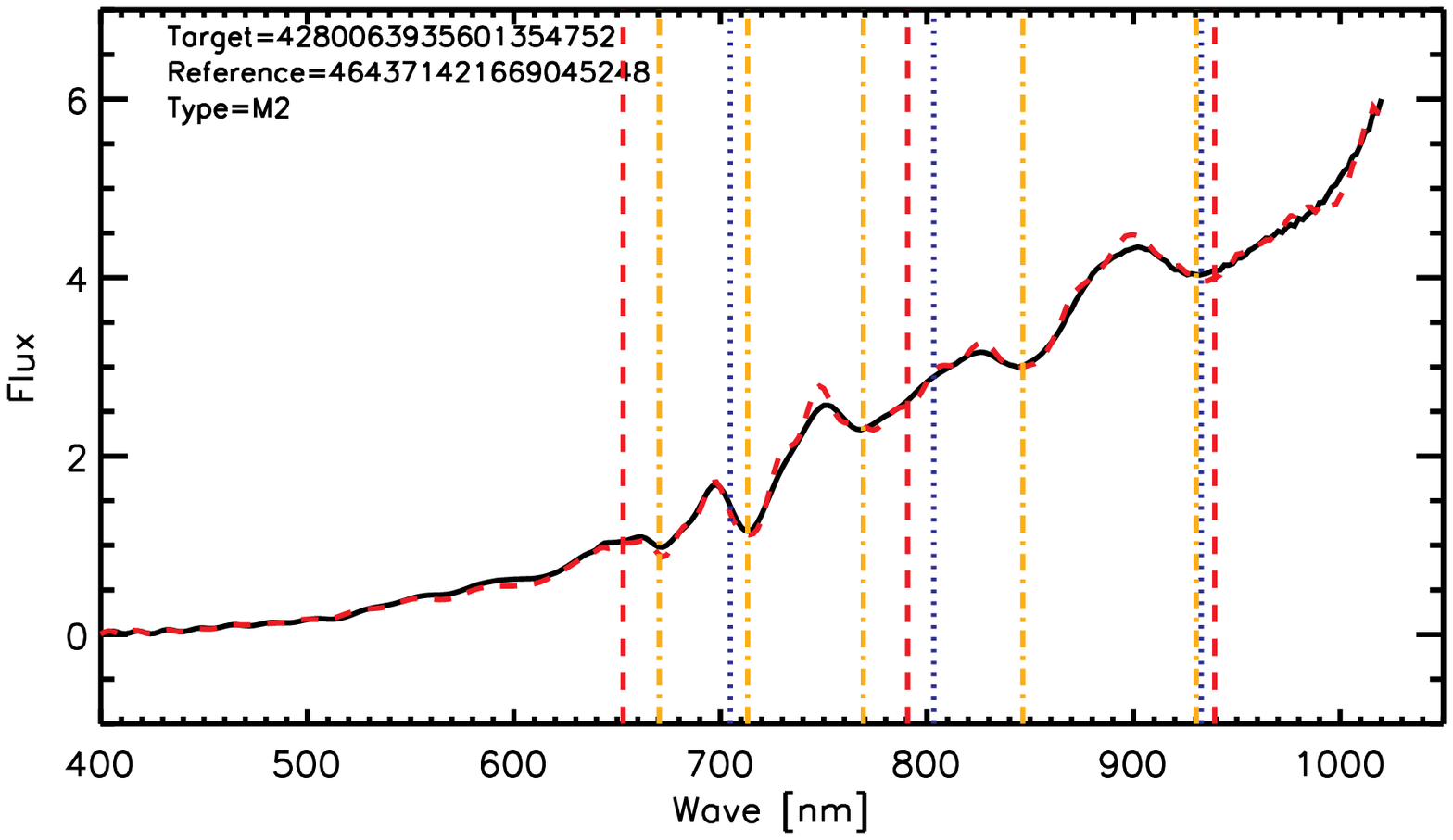}}
\resizebox{0.33\hsize}{!}{\includegraphics[angle=0]{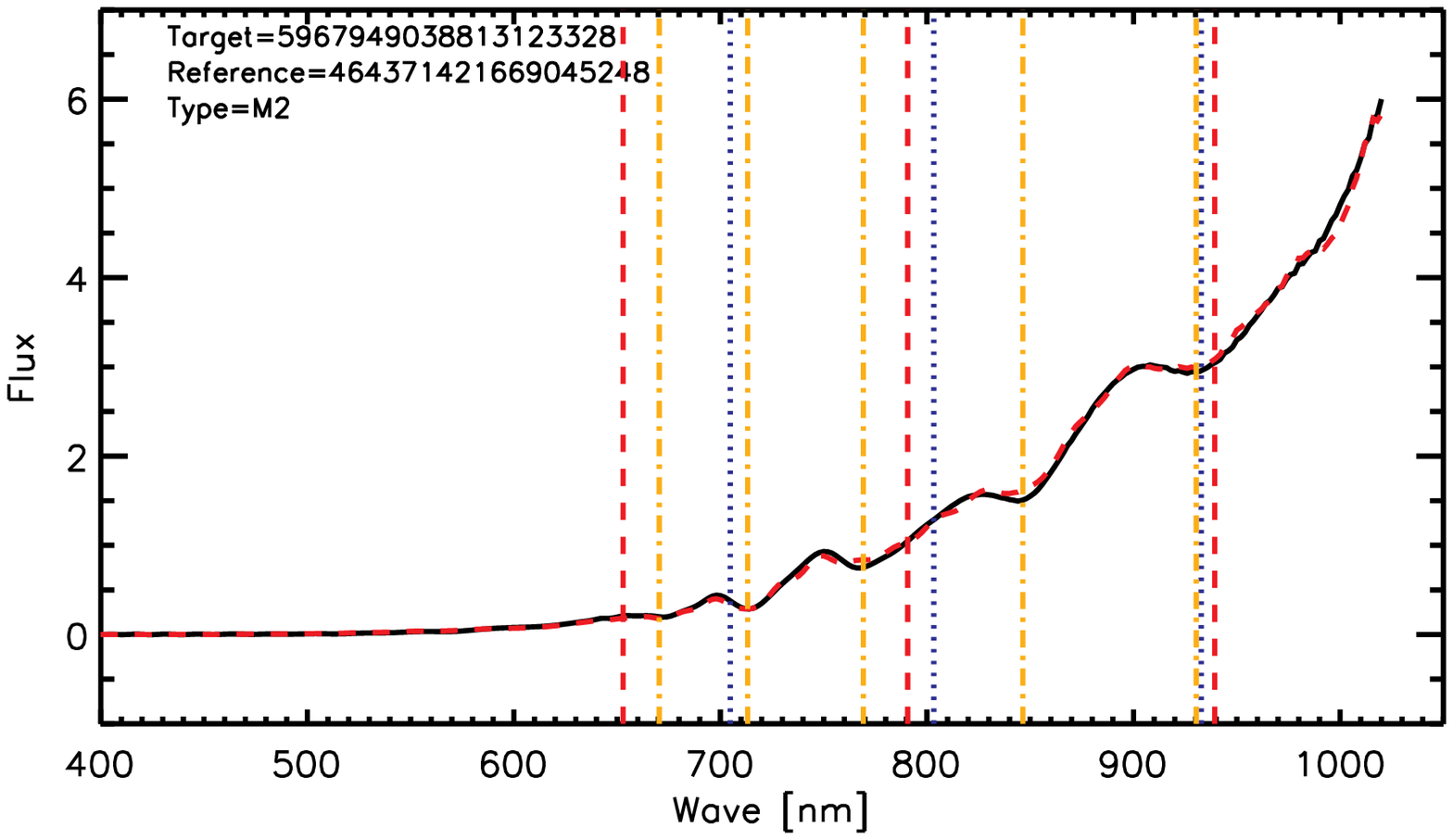}}
\resizebox{0.33\hsize}{!}{\includegraphics[angle=0]{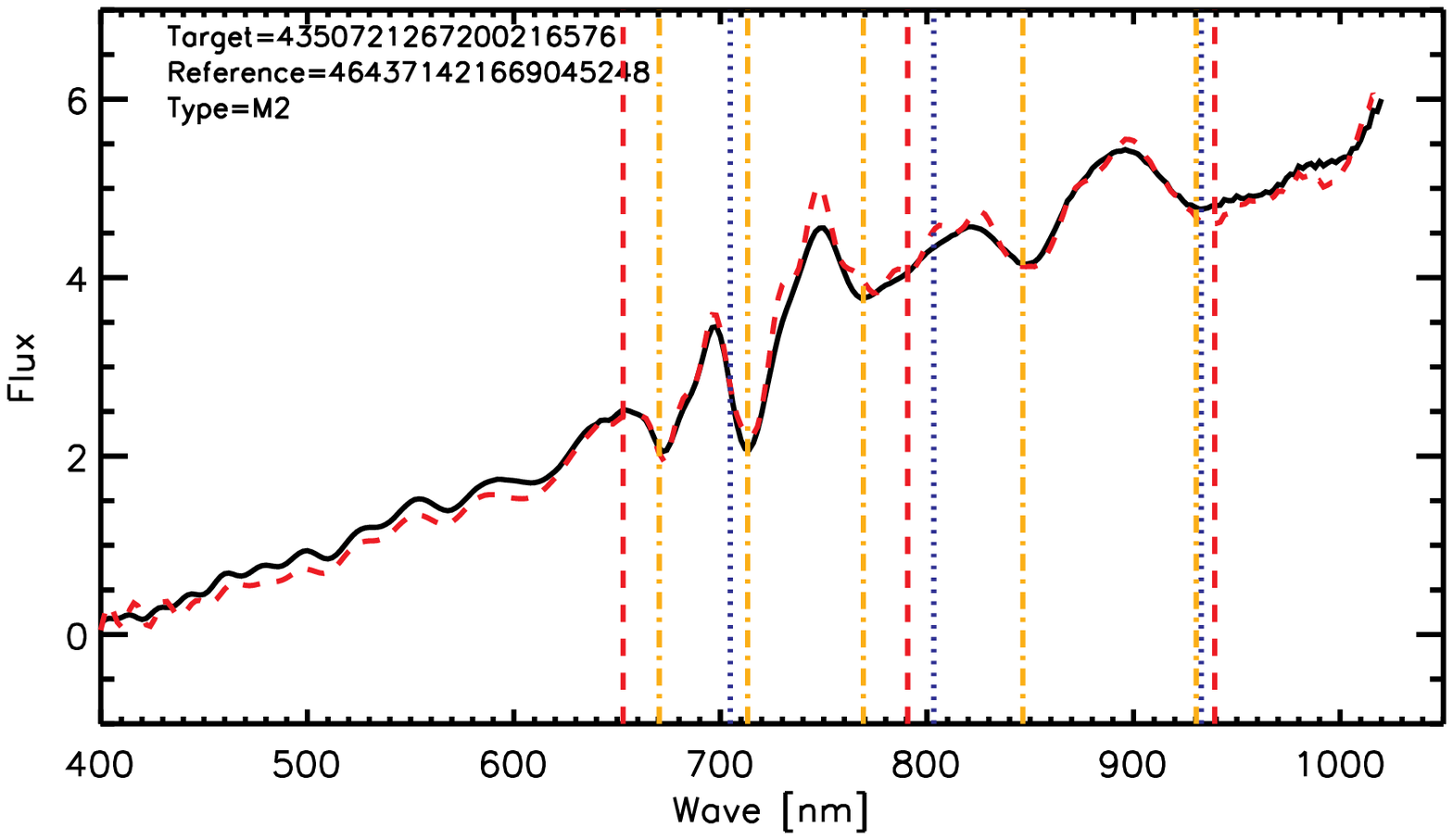}}
\end{center}

\begin{center}
\resizebox{0.33\hsize}{!}{\includegraphics[angle=0]{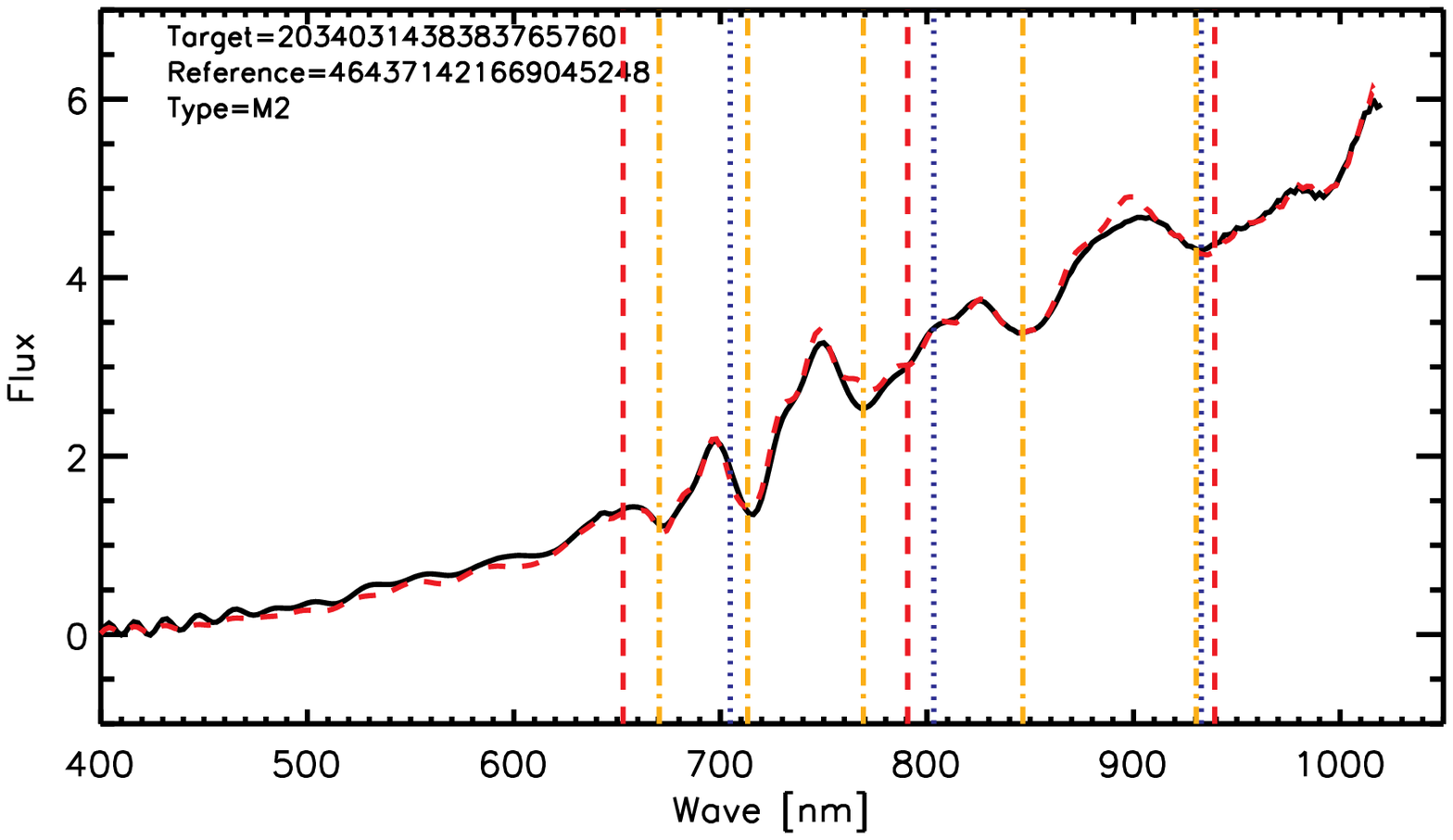}}
\resizebox{0.33\hsize}{!}{\includegraphics[angle=0]{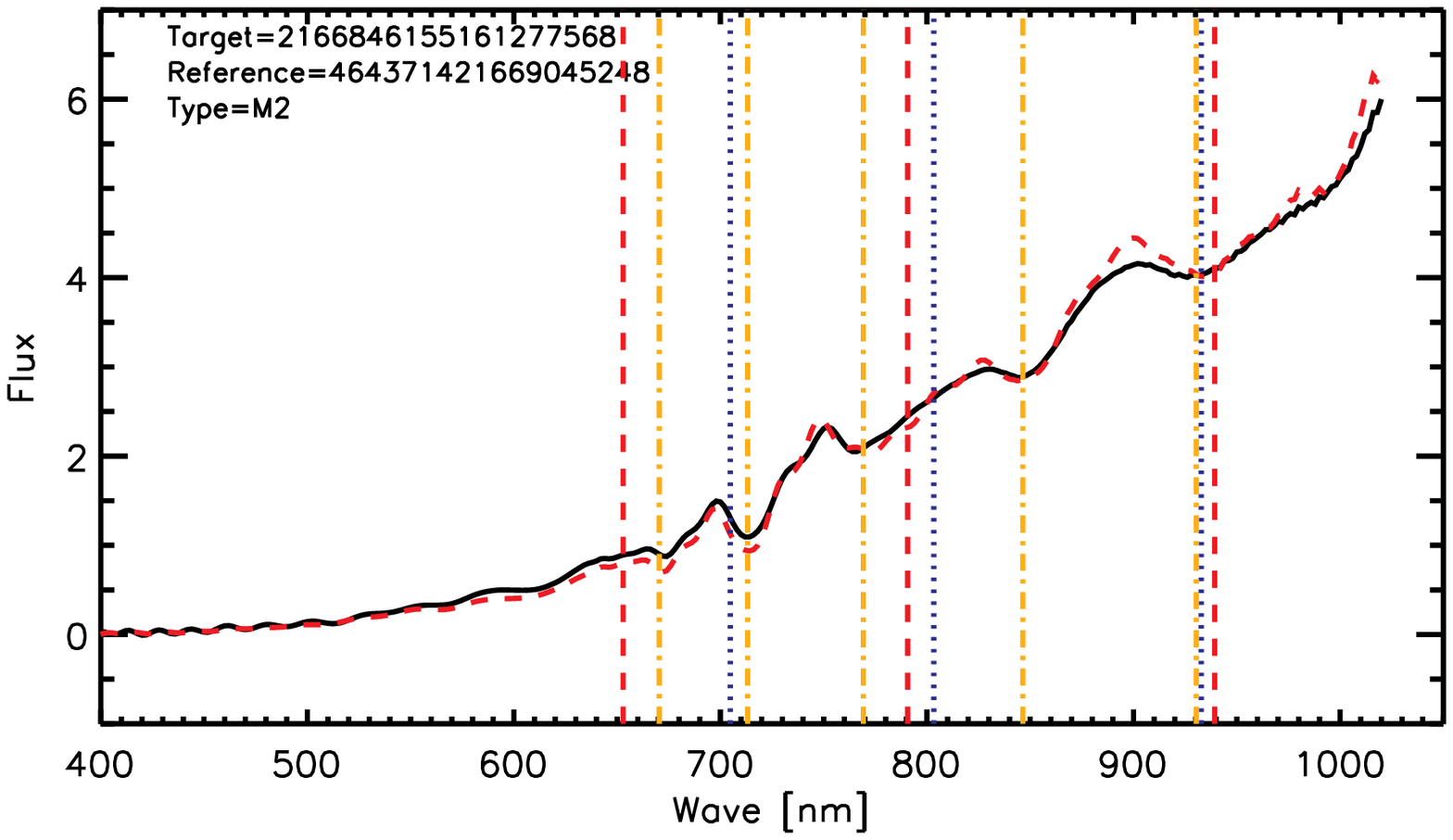}}
\resizebox{0.33\hsize}{!}{\includegraphics[angle=0]{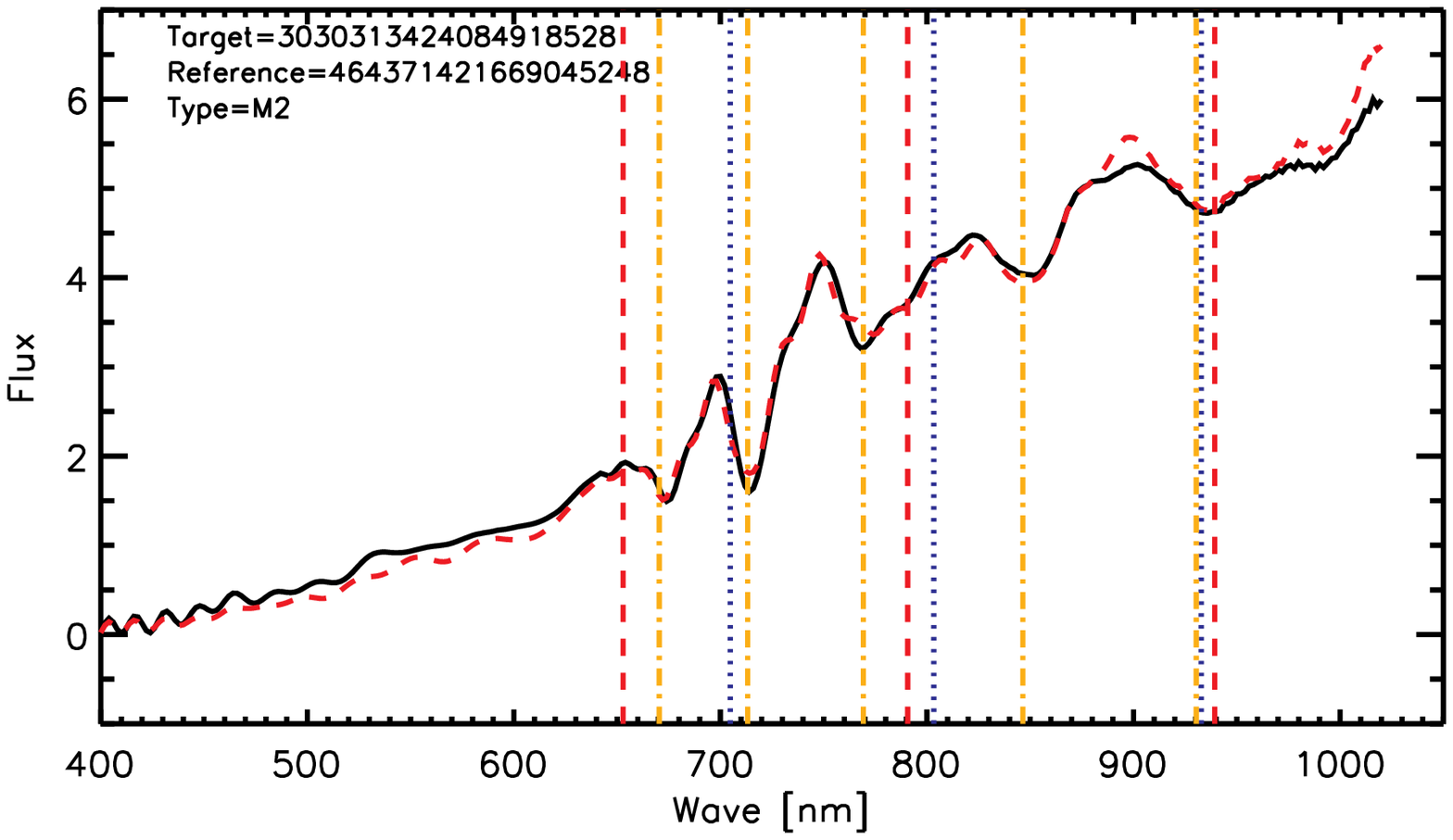}}
\end{center}
\begin{center}

\resizebox{0.33\hsize}{!}{\includegraphics[angle=0]{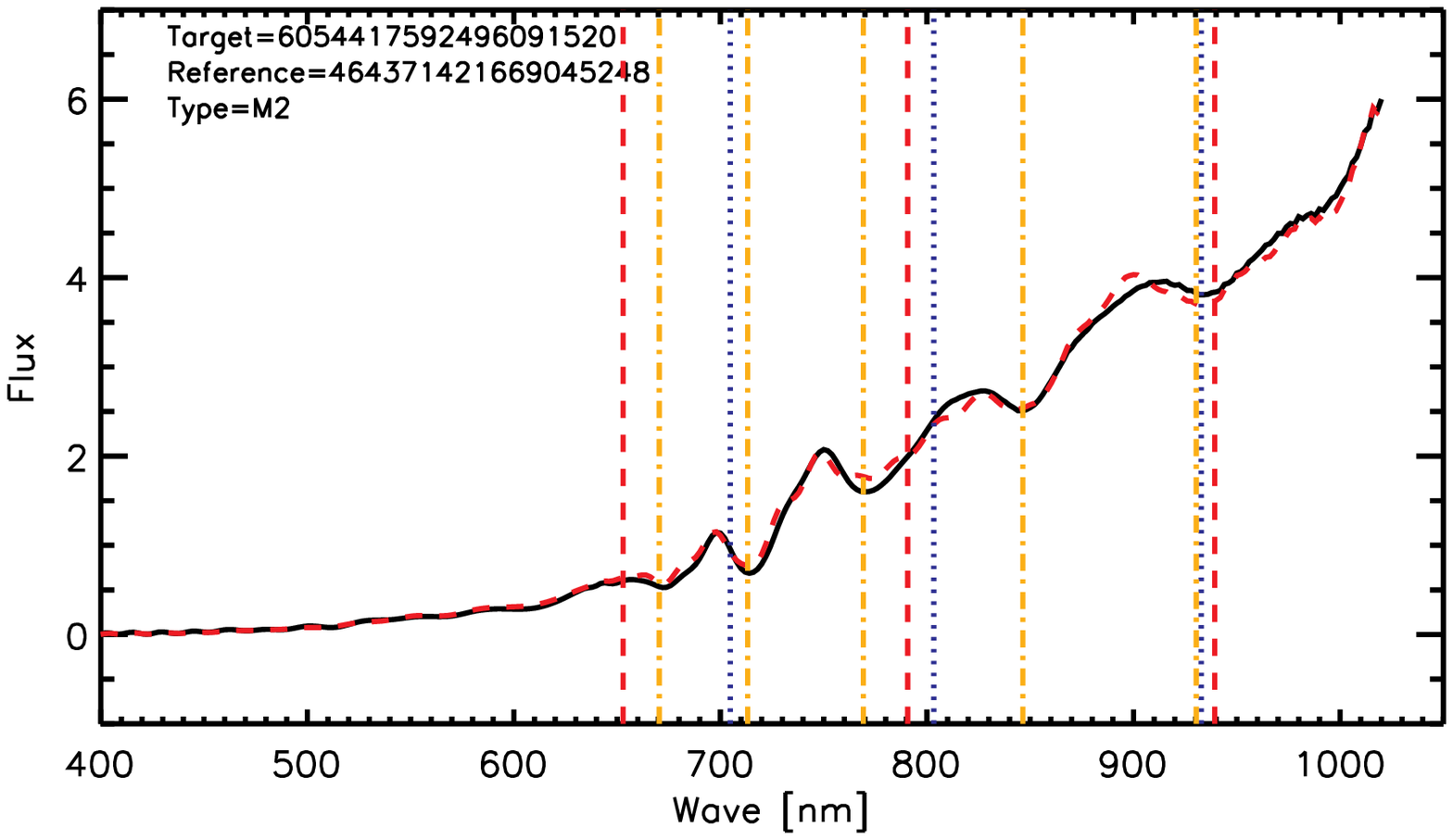}}
\resizebox{0.33\hsize}{!}{\includegraphics[angle=0]{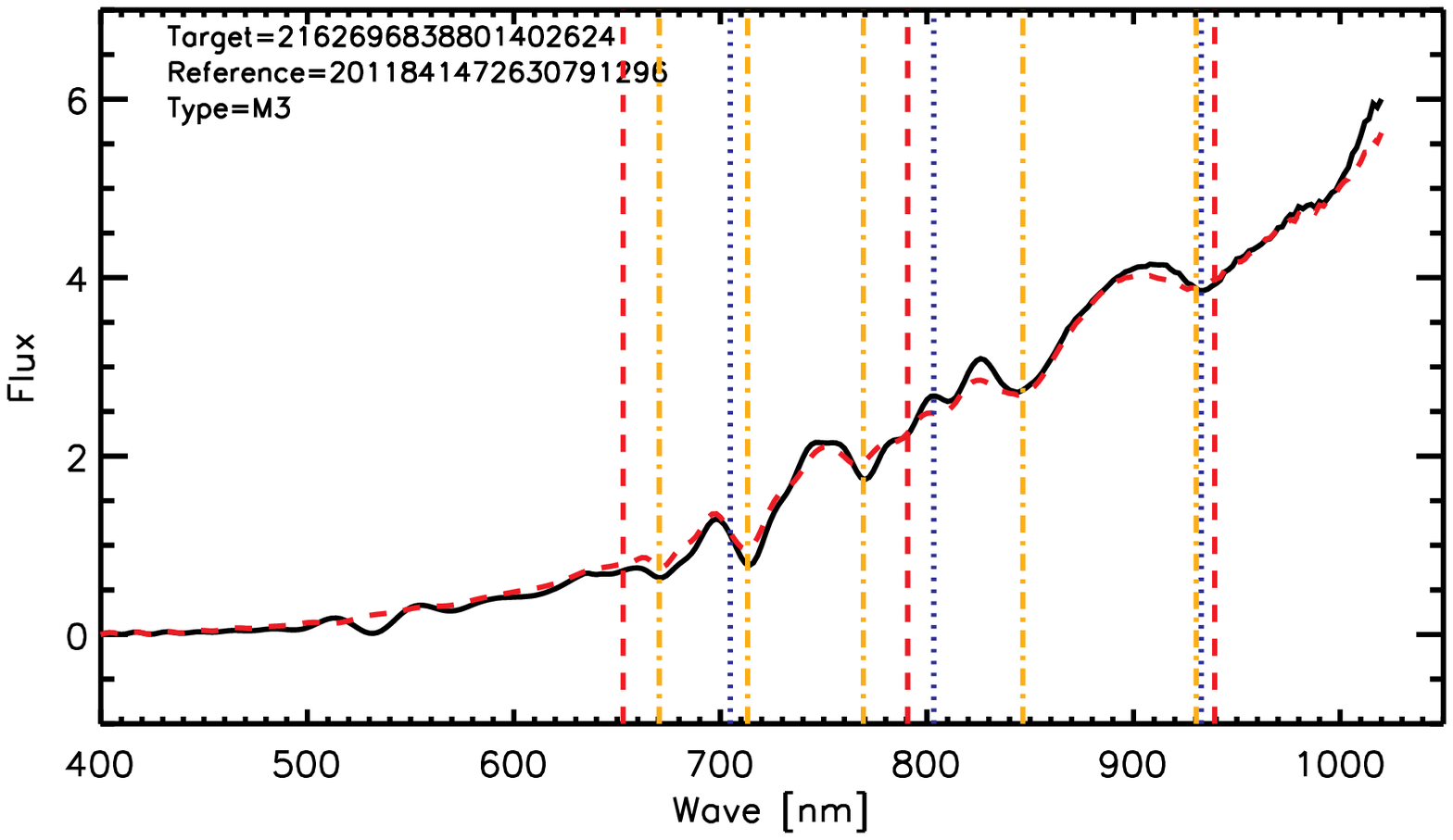}}
\resizebox{0.33\hsize}{!}{\includegraphics[angle=0]{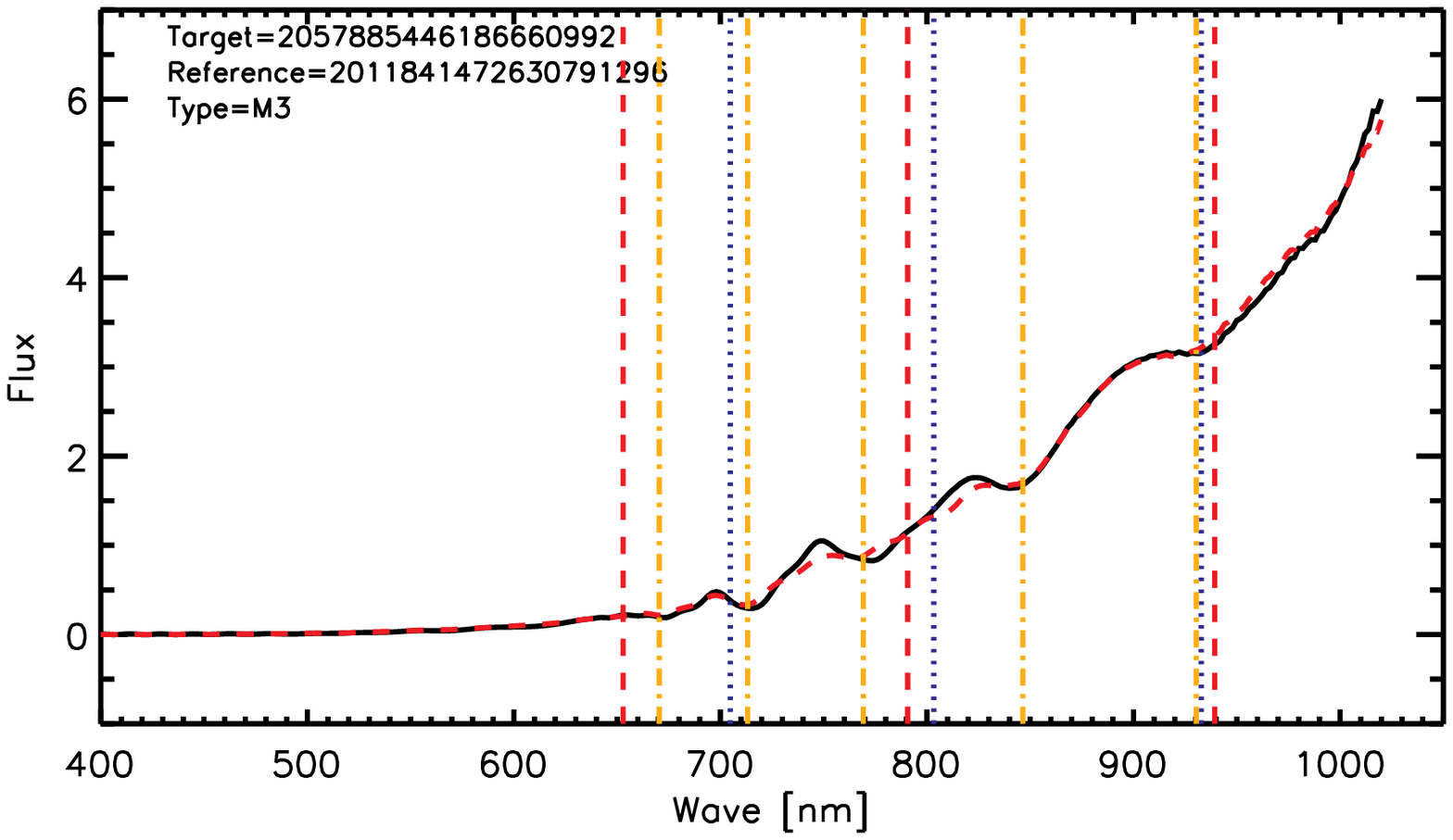}}
\end{center}
\begin{center}

\resizebox{0.33\hsize}{!}{\includegraphics[angle=0]{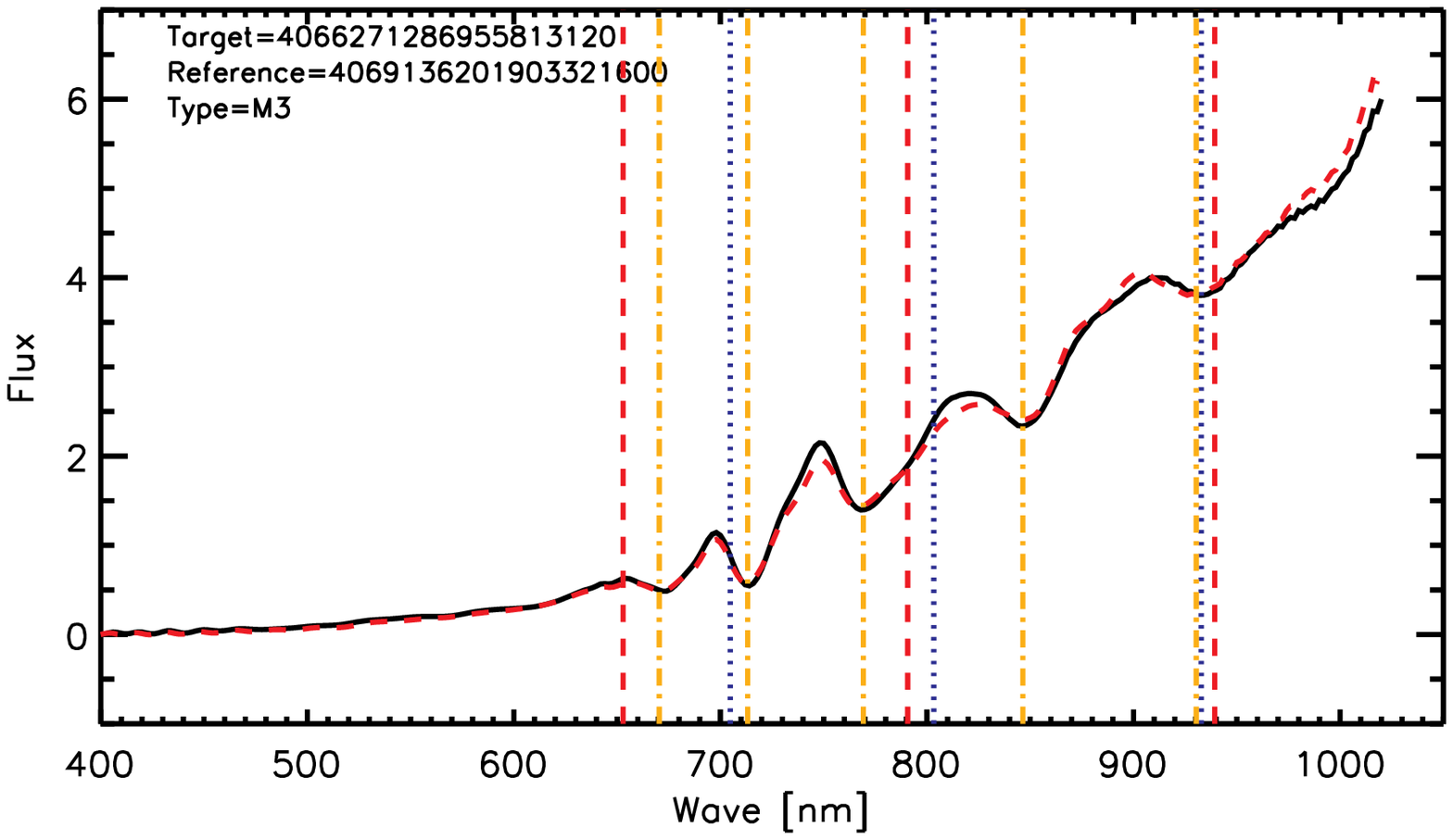}}
\resizebox{0.33\hsize}{!}{\includegraphics[angle=0]{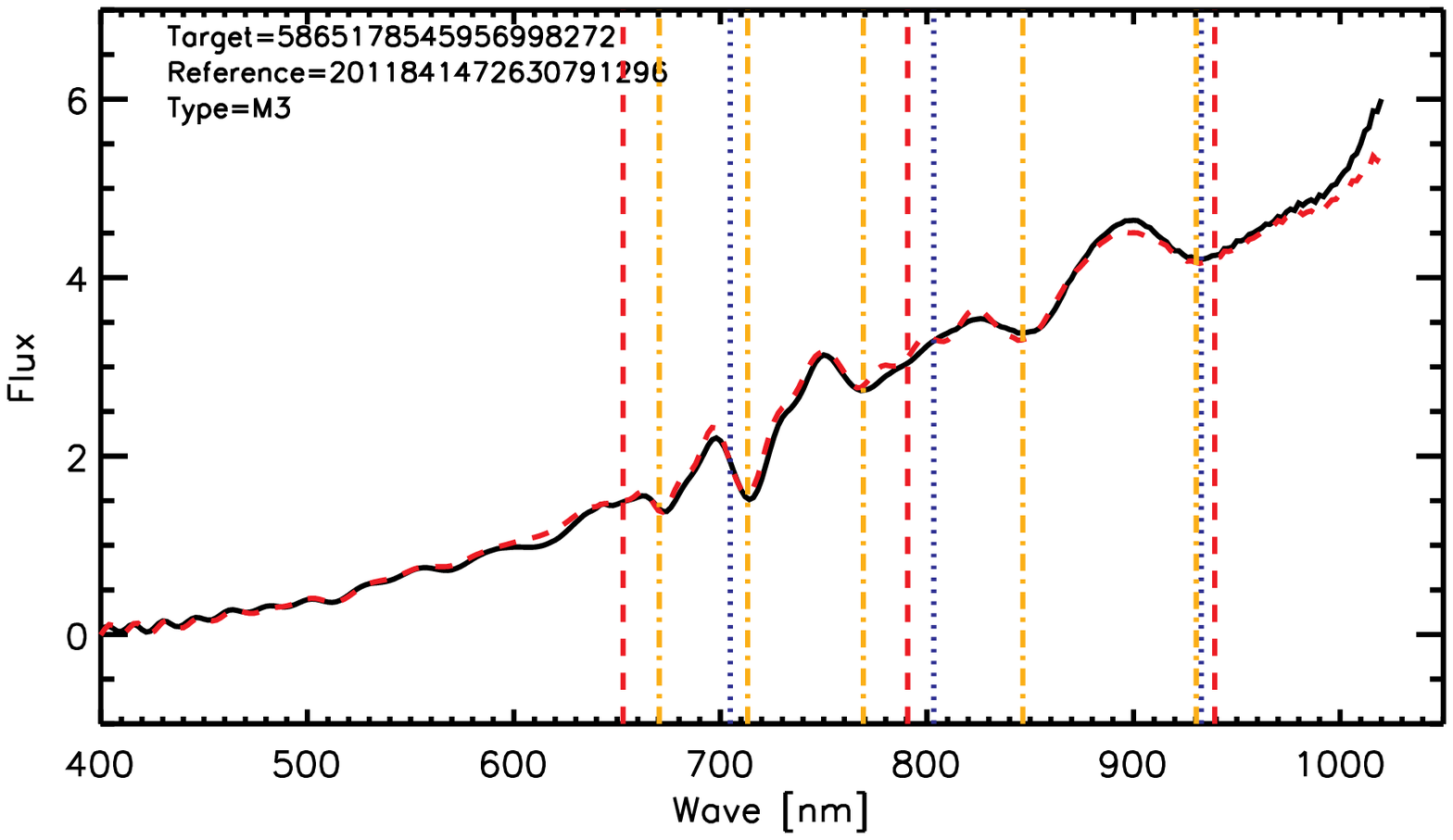}}
\resizebox{0.33\hsize}{!}{\includegraphics[angle=0]{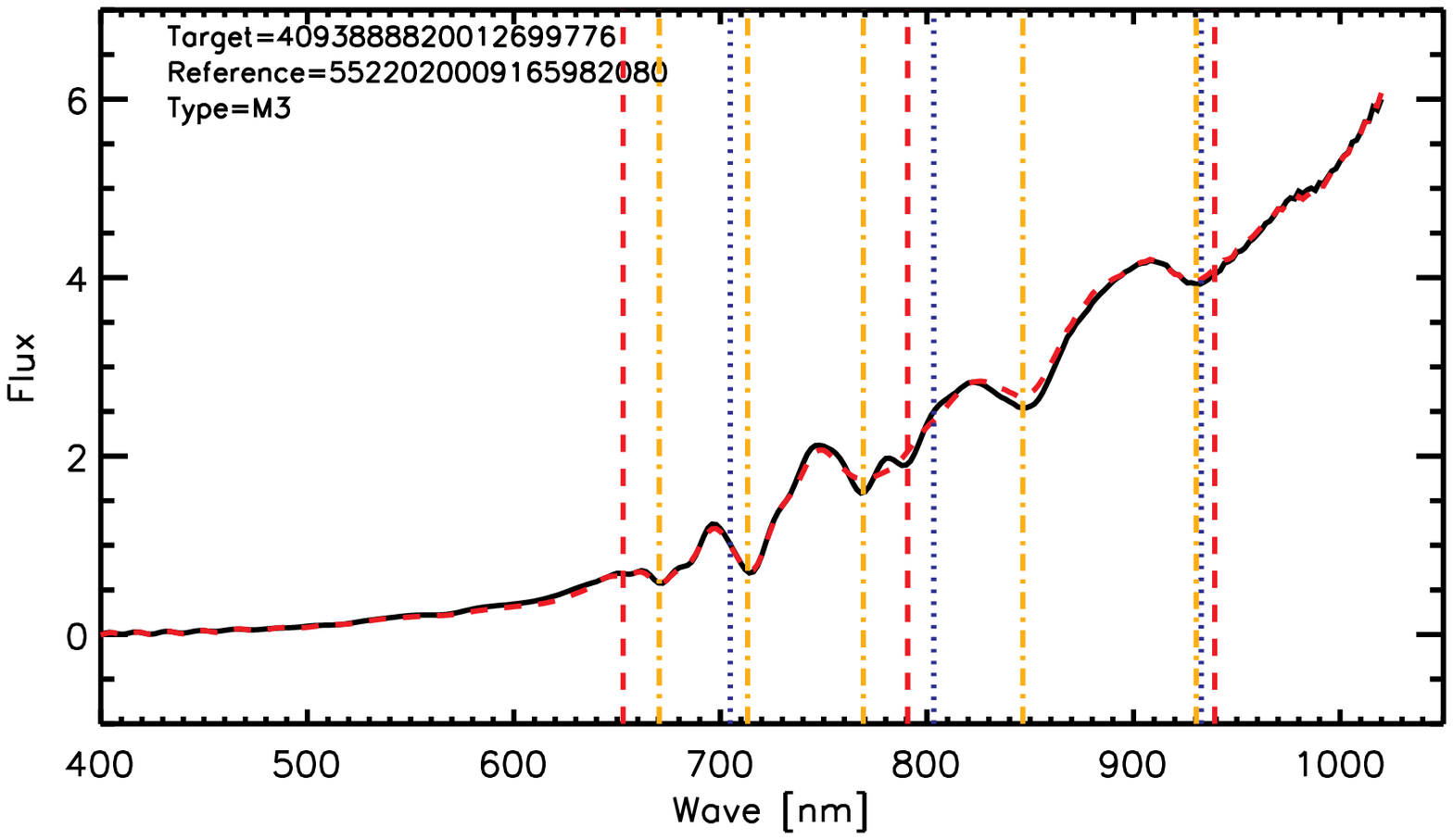}}
\end{center}
\begin{center}
\resizebox{0.33\hsize}{!}{\includegraphics[angle=0]{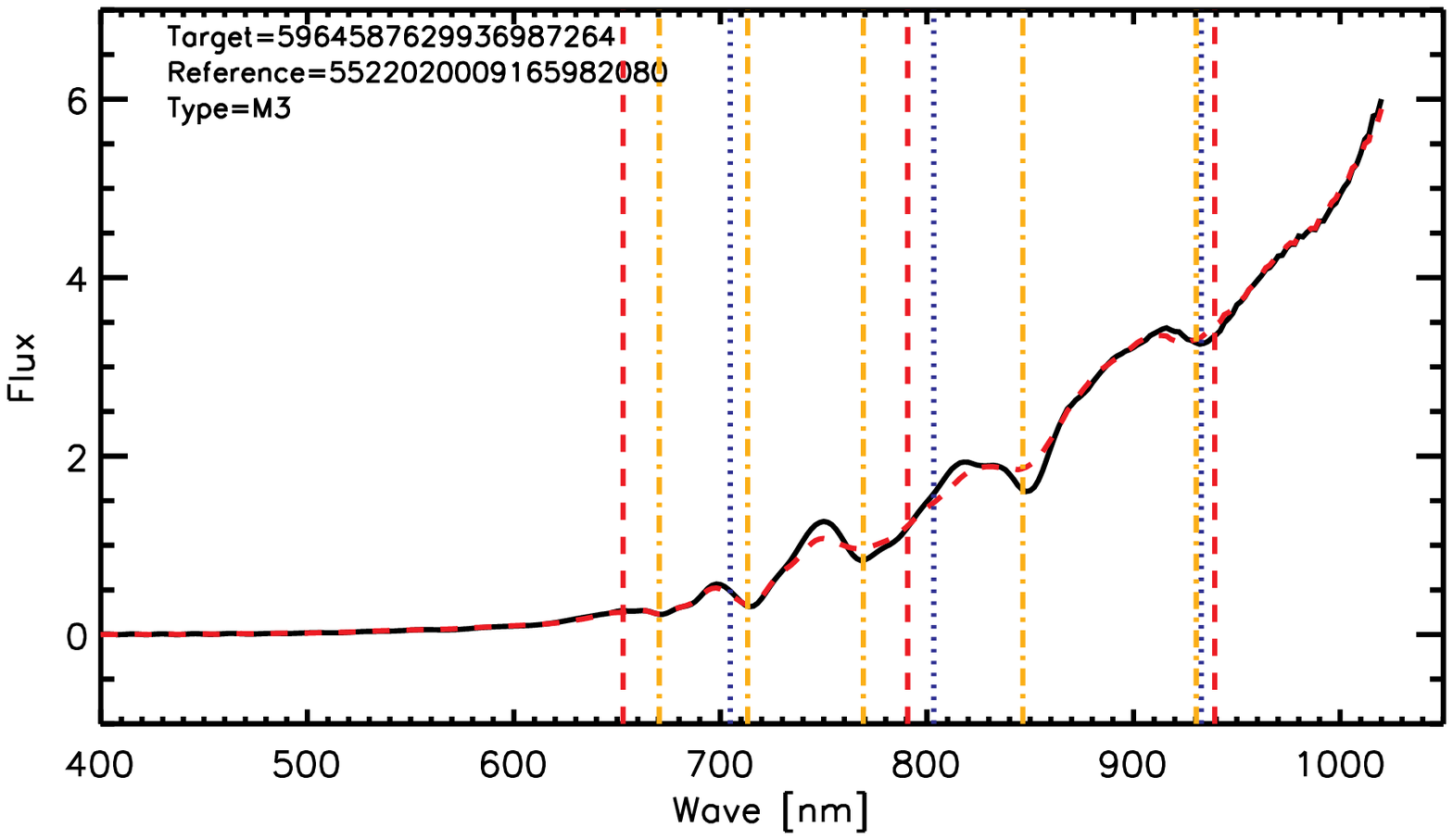}}
\resizebox{0.33\hsize}{!}{\includegraphics[angle=0]{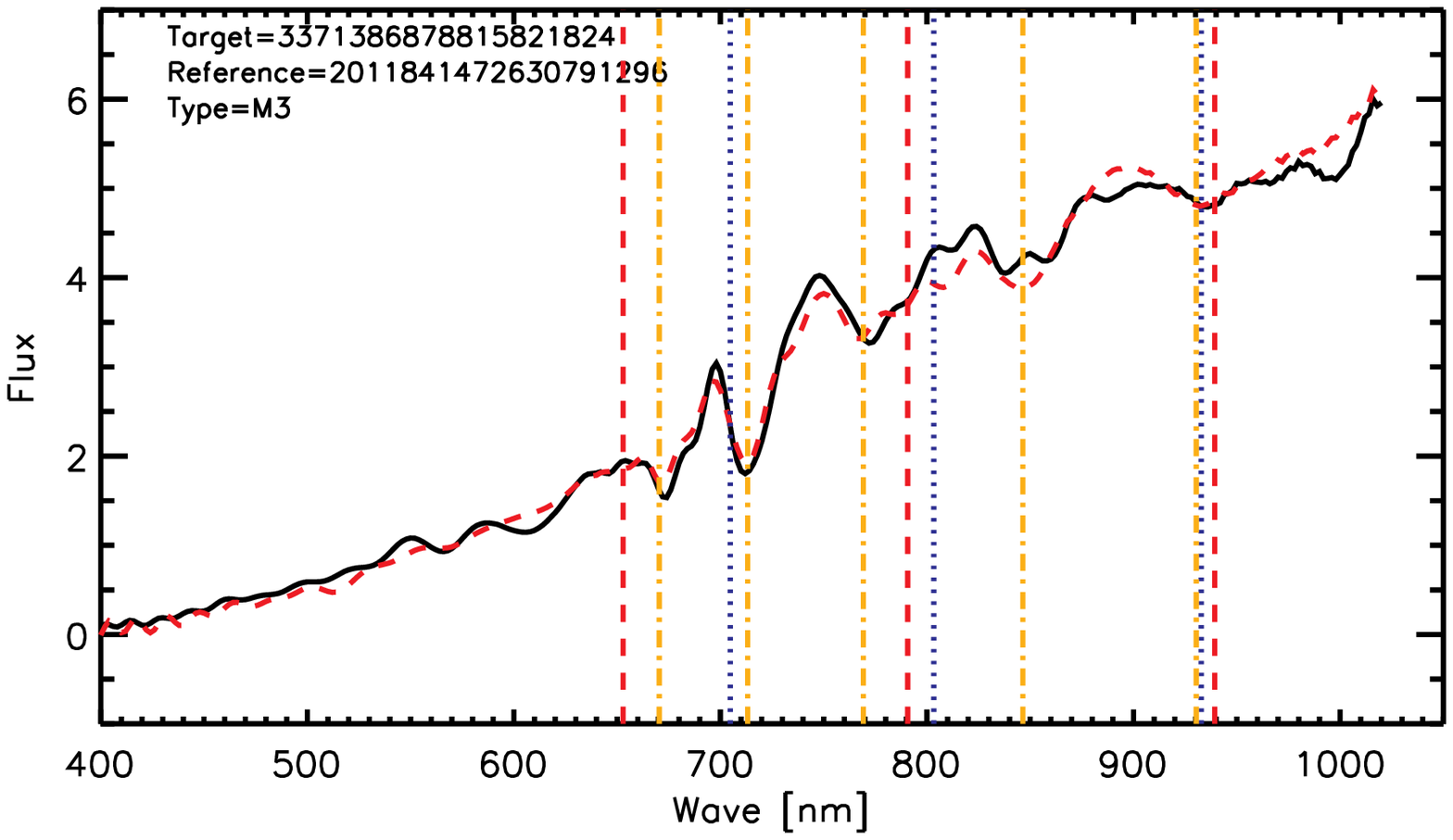}}
\resizebox{0.33\hsize}{!}{\includegraphics[angle=0]{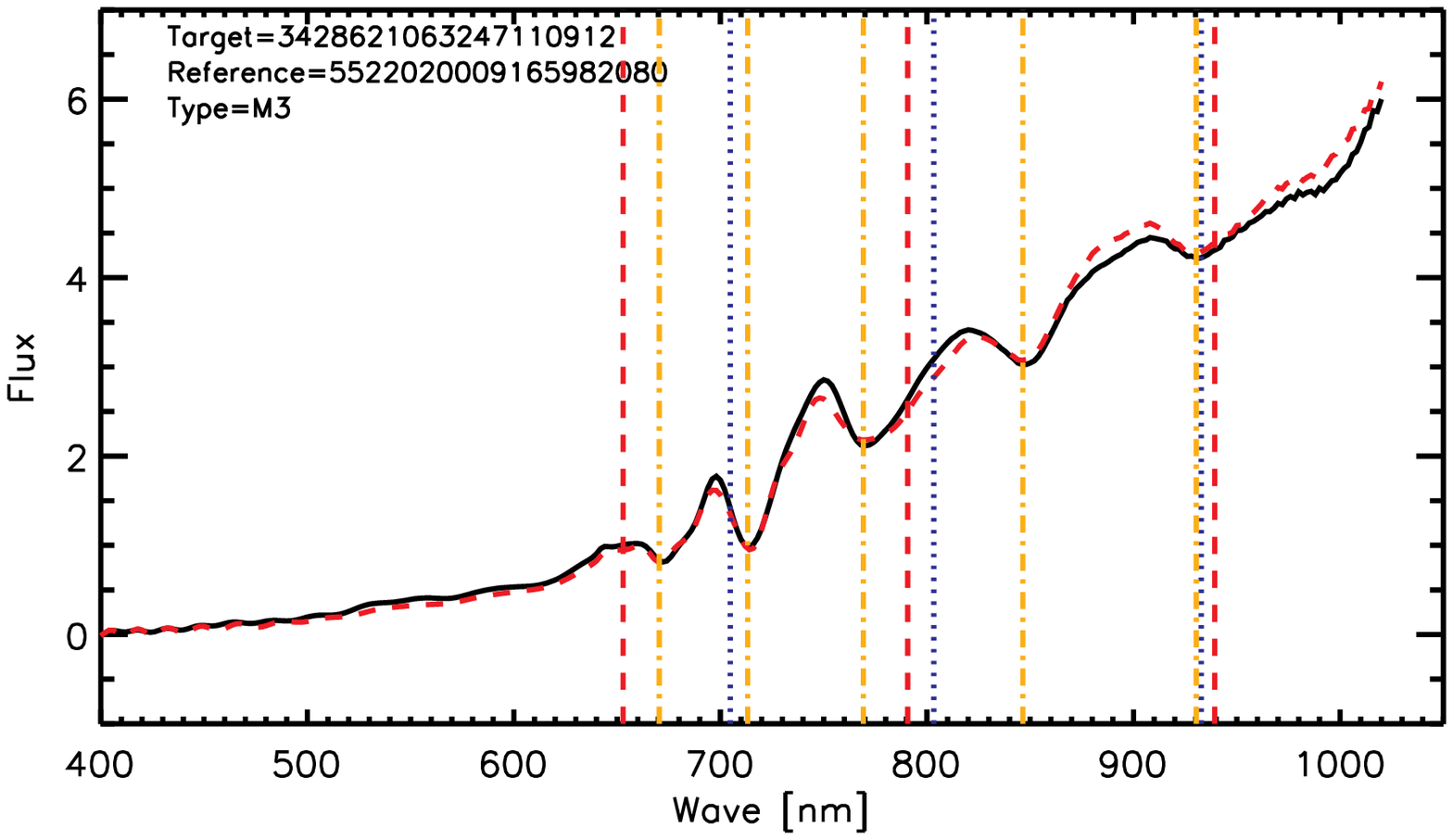}}
\end{center}
\begin{center}

\resizebox{0.33\hsize}{!}{\includegraphics[angle=0]{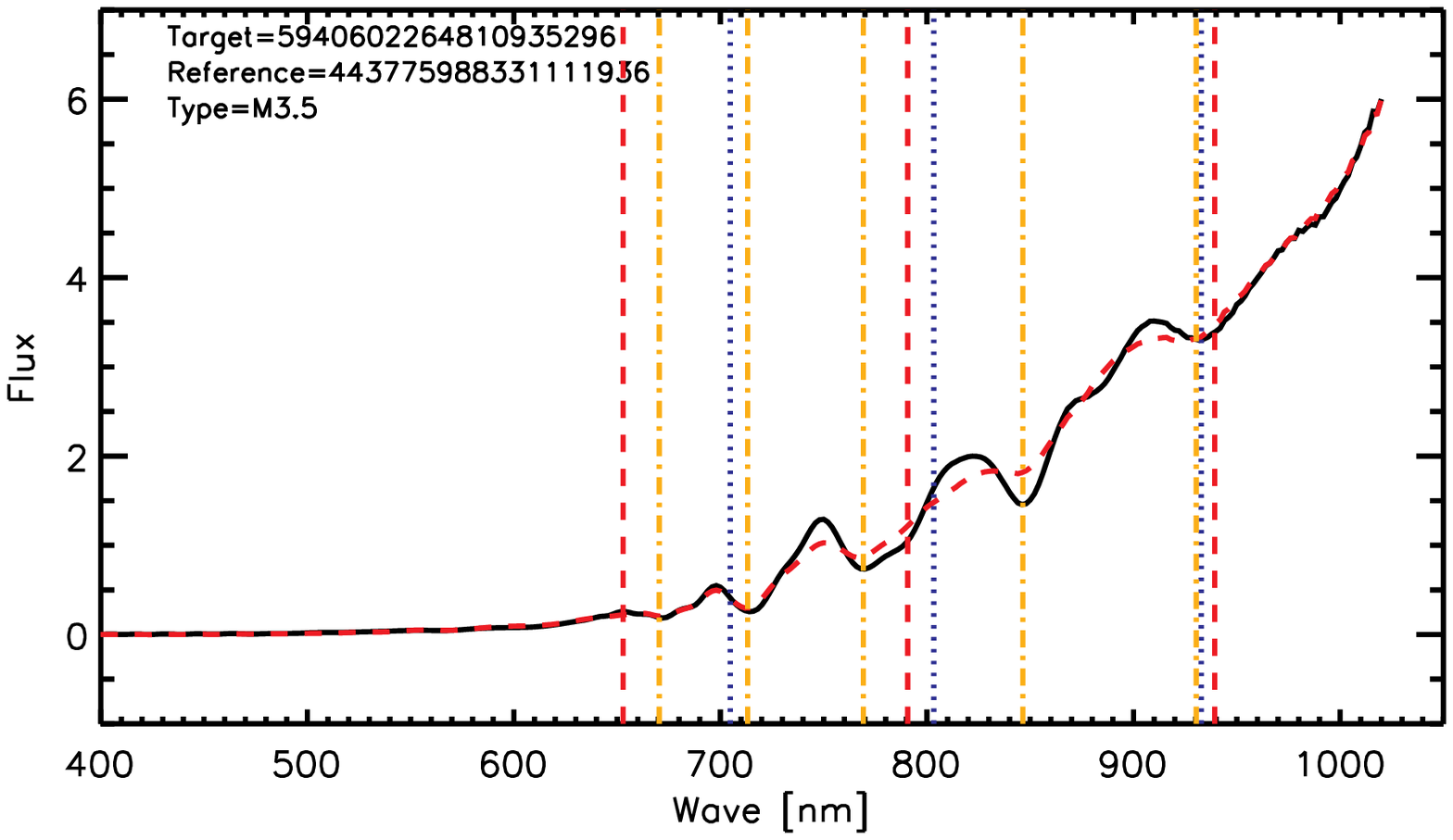}}
\resizebox{0.33\hsize}{!}{\includegraphics[angle=0]{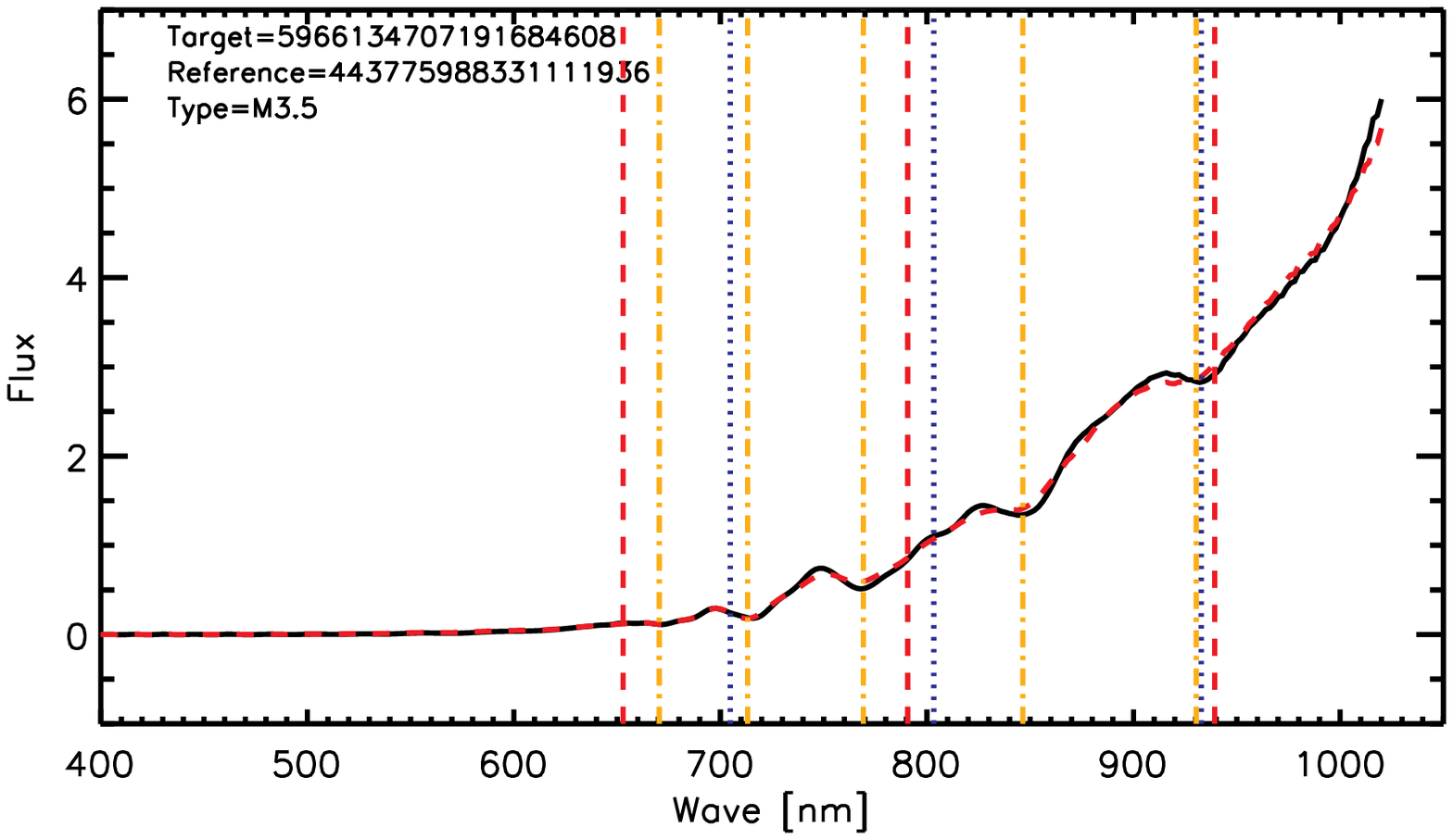}}

\end{center}
\caption{ \label{newrsg} The BP/RP spectra of the newly detected very-likely RSGs 
are shown with a black curve. The red-dashed curves show the  BP/RP spectra of stars
from \citet{messineo19} used as references. The reference spectra have been 
 brought to the target
extinction using the infrared-derived \Aks\ and the Galactic extinction curve 
of \citet{cardelli89} extended to infrared wavelengths with a power law with 
index $-1.9$ \citep{messineo05}.
Here again, the vertical orange dotted-dashed lines  mark the locations of the main 
absorption bands seen  in M1-M3 RSGs and O-rich stars; the vertical red lines 
mark the absorption seen in S-type, and the vertical blue lines are
those in C-rich stars.
} 
\end{figure*}

\acknowledgement
\addcontentsline{toc}{section}{Acknowledgements}

This work has made use of data from the European Space Agency (ESA) mission {\it Gaia}
($http://www.cosmos.esa.int/gaia$), processed by the {\it Gaia} Data Processing and Analysis
Consortium (DPAC, $http://www.cosmos.esa.int/web/gaia/dpac/consortium$). Funding for the DPAC
has been provided by national institutions, in particular the institutions participating in the {\it
Gaia} Multilateral Agreement. 
This publication makes use of data products from the Two Micron All
Sky Survey, which is a joint project of the University of Massachusetts and the Infrared Processing
and Analysis Center / California Institute of Technology, funded by the National Aeronautics and Space
Administration and the National Science Foundation. 
This work is based on observations made with
the Spitzer Space Telescope, which is operated by the Jet Propulsion Laboratory, California
Institute of Technology under a contract with NASA.
This research made use of data products from the Midcourse Space Experiment, the processing of which
was funded by the Ballistic Missile Defense Organization with additional support from the NASA
office of Space Science. 
This publication makes use of data products from WISE, which is a joint
project of the University of California, Los Angeles, and the Jet Propulsion Laboratory / California
Institute of Technology, funded by the National Aeronautics and Space Administration. 
This research
has made use of the VizieR catalogue access tool, CDS, Strasbourg, France, and SIMBAD database.
This research utilized  the NASA’s Astrophysics Data System Bibliographic Services. 
This work uses the RSG catalog by Messineo M. \& Brown A. (2019) which was
supported by the National Natural Science Foundation of China
(NSFC-11773025, 11421303), and USTC grant KY2030000054. 
MM thanks Prof. Anthony Brown for helping her to realize the 2019 catalog.
MM thanks the referee for his careful reading and constructive comments.


\end{document}